\documentclass[twocolumn,groupedaddress,assymb,amsmath]{revtex4}
\usepackage{}
\usepackage{graphicx}
\usepackage{amssymb}
\usepackage{amsmath}

\usepackage{extarrows} %$\xlongrightarrow{\Omega _1(t)}$
\usepackage[bookmarks=false]{hyperref}
\hypersetup{colorlinks=true,citecolor=blue,linkcolor=blue,urlcolor=blue,pdfstartview=FitH,bookmarksopen=true}

\usepackage{color}
\definecolor{zzz}{rgb}{0.9,0.0,0.4}
\begin{document}
\title{Optomechanical second-order sidebands and group delays in a spinning resonator with parametric amplifier and non-Markovian effects}
\author{Wei Zhang$^{1}$ and H. Z. Shen$^{1,2,}$\footnote{ \textcolor{zzz}{Corresponding author: shenhz458@nenu.edu.cn }}}
\affiliation{$^1$Center for Quantum Sciences and School of Physics,
Northeast Normal University, Changchun 130024, China\\
$^2$Center for Advanced Optoelectronic Functional  Materials
Research, and Key Laboratory for UV Light-Emitting Materials and
Technology of Ministry of Education, Northeast Normal  University,
Changchun 130024, China}
\date{\today}

\begin{abstract}
We investigate the generation of the frequency components at the second-order sidebands based on a spinning resonator containing a degenerate optical parametric amplifier (OPA). We show an OPA driven by different pumping frequencies inside a cavity can enhance and modulate the amplitude of the second-order sideband with different influences. We find that both the second-order sideband amplitude and its associated group delay sensitively depend on the nonlinear gain of the OPA, the phase of the field driving the OPA, the rotation speed of the resonator, and the incident direction of the input fields. Tuning the pumping frequency of the OPA can remain the localization of the maximum value of the sideband efficiency and nonreciprocal behavior due to the optical Sagnac effect, which also can adjust the linewidth of the suppressive window of the second-order sideband. Furthermore, we extend the study of second-order sideband to the non-Markovian bath which consists of a collection of infinite oscillators (bosonic photonic modes). We illustrate the second-order sidebands in a spinning resonator exhibit a transition from the non-Markovian to Markovian regime by controlling environmental spectral width. \textbf{We also study the influences of the decay from the non-Markovian environment coupling to an external reservoir on the efficiency of second-order upper sidebands.} This indicates a promising new way to enhance or steer optomechanically induced transparency devices in nonlinear optical cavities and provides potential applications for precision measurement, optical communications, and quantum sensing.
\end{abstract}

%²©

%\pacs{42.50.Pq, 42.50.Ct, 42.50.Ar, 42.50.Dv} \maketitle
%42.50.pq: Cavity quantum electrodynamics; micromasers
%42.50.Ct: Quantum description of interaction of light and matter; related %experiments
%42.50.Ar: Photon statistics and coherence theory
%42.50.Dv: Quantum state engineering and measurements

\maketitle
\section{Introduction}
In recent years, quantities of attention have been paid to the field of optomechanics \cite{Aspelmeyer861391,Aspelmeyer6529,Kippenberg3211172,Marquardt240,Sainadh92033824}, in which different considerable phenomena have been met. There are different applications such as cooling of a mechanical resonator \cite{Metzger4321002,Gigan44467,Arcizet44471,Schliesser5509,Meystre525215}, gravitational wave detection \cite{Caves4575,Abramovici256325,Braginsky293228}, optical bistability \cite{Nejad562816,Sarma331335,Shahidani311087}, optomechanical mass sensors \cite{Jiang2213773}, quantum measurement \cite{Thompson45272}, and detection of weak microwave signals \cite{Bagci50781,Andrews10321,Nejad97053839} in merged quantum mechanical systems with nano and micro mechanics. The recent advance in connection with the present study closely is optomechanically induced transparency (OMIT) \cite{Weis3301520,Safavi47269,Jia91043843,Jing59663,Wang90023817}. In OMIT, the intense red-detuned optical control field produces anti-Stokes scattering, which alters the optical response of the optomechanical cavity, making it transparent in a narrow bandwidth around the cavity resonance for a probe beam \cite{Karuza88013804}. As an analog of electromagnetically induced transparency \cite{Fleischhauer77633,Agarwal81041803}, OMIT plays an essential role in optical storage and optical telecommunication \cite{Chang13023003,Fiore107133601,Zhou9179,Hill31196}. In the last several years, the main progress has concentrated on the linearization of the optomechanical interaction, where we properly explain OMIT by linearizing the optomechanical interaction in the case of ignoring the intrinsic nonlinear nature of the optomechanical coupling \cite{Agarwal81041803,Huang83043826}. In recent years, nonlinear optical interactions in materials can increase the photons circulating in microcavities, such as parametric amplification and optical Kerr effect \cite{Xiong58050302,Bartolo94033841,Zhou421289,Ilchenko92043903}, which has emerged as an important new frontier in cavity optomechanics. In the classical mechanism, nonlinear optomechanical interaction brings about unconventional photon blockade \cite{Rabl107063601,Flayac96053810,Lemonde90063824}, optomechanical chaos \cite{Marino87052906}, and sideband generation \cite{Xiong86013815}.

Nonreciprocal transmission plays a very important role in the process of quantum information \cite{Bi5758,Aleahmad72129,AbdelMalek528560,Bernier8604} due to the characteristics of unidirectional transmission. The nonreciprocal transmission of the optical signal allows the flow of light from one side but blocks it from the other, which resembles the traditional semiconductor p-n junction. Recently, OMIT has been demonstrated in a rotating optomechanical system with a whispering-gallery-mode (WGM) microresonator \cite{Schliesser10095015,H5367,Jiang3371}. The experiment \cite{Maayani558569} shows that optical nonreciprocal devices can be achieved by spinning an optomechanical resonator. In such a spinning resonator, due to the Sagnac effect, the frequencies of the clockwise and counterclockwise modes experience Sagnac-Fizeau shifts. Additionally, it also suggests a new scheme to achieve optical nonreciprocity that the optical sidebands strongly rely on the rotary direction of the resonator, which is different from the nonlinearity-based schemes demonstrated \cite{Shen10657,Ruesink713662,Fang7465,Cao118033901,Li102033526}. The spinning resonator systems have developed rapidly, including nanoparticles sensing \cite{Jing51424}, mass sensing \cite{Chen14082005}, nonreciprocal photon blockades \cite{Huang121153601,Li7630}, nonreciprocal phonon lasers \cite{Jiang10064037}, unidirectional signal amplification \cite{Peng5332000405}, breaking anti-PT symmetry \cite{Zhang207594}, and optical solitons \cite{Li103053522}.

It has been shown that combining nonlinear optics and optomechanics has resulted in many kinds of physical phenomena to enhance quantum effects \cite{Otey93033835,Coillet9828}. An optical parametric amplifier (OPA) inside the optomechanical cavity, which is pumped by an external laser, can directly lead to optical amplification and modulate the optomechanical coupling in a way analogous to periodic cavity driving \cite{Adamyan92053818,Hu100043824,Lu114093602}. The OPA is able to generate pairs of down-converted photons, which shows nearly perfect single or dual squeezing. Therefore, the OPA can modify the dynamical instabilities and nonlinear dynamics of the system \cite{Mi67115,Hu7124,Xuereb86013809}. Numerous applications have been studied owing to these features, such as the realization of strong mechanical squeezing \cite{Agarwal93043844}, enhancing optomechanical cooling \cite{Huang79013821}, the normal-mode splitting \cite{Huang80033807}, controlling the photon blockade \cite{Sarma96053827,Shen98023856,Shen101013826}, and the increase of atom-cavity coupling \cite{Qin120093601}.

Recently, studying the nonlinear optomechanical interactions in the presence of a coherent mechanical pump has emerged as an important frontier \cite{Lemonde111053602,Liu111083601,Mikkelsen96043832,Ferretti85033303}. Due to the existence of nonlinear optomechanical interactions, second-order and higher-order sidebands are generated in optomechanical systems \cite{Xiong86013815,Kronwald111133601,Suzuki92033823,Jiao18083034,Liu717637,Fan65850}. Generation of spectral components at high-order OMIT sidebands is demonstrated analytically, which may have great potential in precise sensing of charges \cite{Xiong423630,Kong95033820}, phonon number \cite{Cohen520522}, weak forces \cite{Nunnenkamp111053603,Zhao63224211}, single-particle detection \cite{Li116}, magnetometer \cite{Liu712521}, mass sensor \cite{Liu99033822,Wang106803908}, and high-order squeezed frequency combs \cite{Liu439}. But actually, high-order OMIT sidebands are generally much weaker than the probe signal, which imposes many difficulties in detecting and utilizing the second-order sideband. Therefore, the enhancement and control of second-order sidebands have attracted much interest. Moreover, by controlling the group delay of the output light field, which is caused by rapid phase dispersion, slow light or fast light effects can be achieved \cite{Boyd3261074,Jiao97013843,H5367,He351649,Li635090,Mirza2725515,Liao11698}. The fast and slow light effects of the optomechanical system have a wide range of applications in optical communication and interferometry \cite{Zimmer92253201,Shahriar75053807}. The hybrid nonlinear optomechanical system provides an important platform for further study of the tunable slow and fast effect.

For open systems \cite{breuer2002,Weiss2008}, only if the coupling between the system and environment is weak, where the characteristic times of the bath are sufficiently smaller than those of the quantum system under study, the Markovian approximation is valid. This means that the Markovian approximation may fail in some cases, e.g., two-state systems, harmonic oscillators, coupled cavities, etc \cite{Chang052105,Tan032102,Longhi063826,Leggett5911987,breuer1032104012009,laine810621152010,addis900521032014,wibmann860621082012,wibmann920421082015,shen960338052017,lorenzo880201022013,rivas1050504032010,luo860441012012,wolf1011504022008,lu820421032010,chruscinski1121204042014,Zhang063853,Zhang19083022,Xiong436053,Zhao2729082,Triana116183602,Cheng430385,Cheng623678,Mu46270,Li361363,Ding111814,Sinha124043603,Wu103010601,Mu94012334}, where we need to consider the influences of non-Markovian effects on the system dynamics. Moreover, we show that the non-Markovian process proves to be useful in quantum information processing including quantum state engineering, quantum control, quantum channel capacity \cite{caruso8612032014,darrigo3502112014,lofranco900543042014,bylicka457202014,xue860523042012}, and has been realized in experiment \cite{Xiong2019100,Cialdi2019100,Tang201297,Groblacher20156,Liu20117,Hoeppe2012108,Xu201082,Madsen2011106,Guo2021126,Khurana201999,Uriri2020101,Liu2020102,Anderson199347,Li2022129,breuer880210022016,Vega015001}.

The above two considerations motivate us to explore that how to enhance and control the second-order OMIT sidebands and group delays in a spinning resonator with parametric amplifier and non-Markovian effects.

In this paper, we consider the influence of the OPA driven with different pumping frequencies on the second-order sideband generation in a rotating optomechanical system, which is coherently driven by a control field and a probe field. The results show that the second-order sidebands in the rotating resonator can be greatly enhanced in the presence of the OPA and meanwhile, remain the nonreciprocal behavior due to the optical Sagnac effect. The second-order sidebands can be adjusted simultaneously by the pumping frequency and phase of the field driving the OPA, the gain coefficient of the OPA, the rotation speed of the resonator, and the incident direction of the input fields. We compare the differences in efficiency of the second-order sideband generation when the OPA is driven by different pumping frequencies. Due to the Sagnac transformation and presence of the OPA, we find that the group delay of the second-order upper sideband can be tuned by adjusting the nonlinear gain and phase of the field driving the OPA, the rotation speed of the resonator, and the incident direction of the input fields in the spinning optomechanical system. The second-order OMIT sidebands in the spinning resonator are then generalized to the non-Markovian regimes and compared with the Markovian approximation in the wideband limit. The influences of the decay from the non-Markovian environment coupling to an external reservoir on the efficiency of second-order upper sidebands are also investigated. Our paper indicates the advantage of using a hybrid nonlinear system, which provides an effective way to further control and enhance second-order and higher-order sidebands in a nonreciprocal optical device.

The rest of this paper is organized as follows. In Sec. II, we give the efficiency of the second-order sideband and its group delay by solving the Heisenberg-Langevin equations. In Sec. III, we discuss the influence of the OPA excited by a pump driving with the frequency being the sum of the frequencies of the strong control field and the weak probe field driving the resonator on the second-order upper and lower sidebands generation in the spinning resonator. In Sec. IV, we study the group delay of the second-order upper sideband. In Sec. V, we show the influence of the OPA on the second-order sideband generation when the OPA is excited by a pump driving with the frequency setting to twice the frequency of the strong control field. In Sec. VI, we extend nonreciprocal second-order sidebands in the spinning resonator to a non-Markovian bath and compare it with that in the Markovian regime. Moreover, we also study the influences of the decay from the non-Markovian environment coupling to an external reservoir on the efficiency of second-order upper sidebands. Sec.~{\rm VII} is devoted to conclusions.

%figure1
\begin{figure}[h]
\centerline{
\includegraphics[width=8.0cm, height=12.5cm, clip]{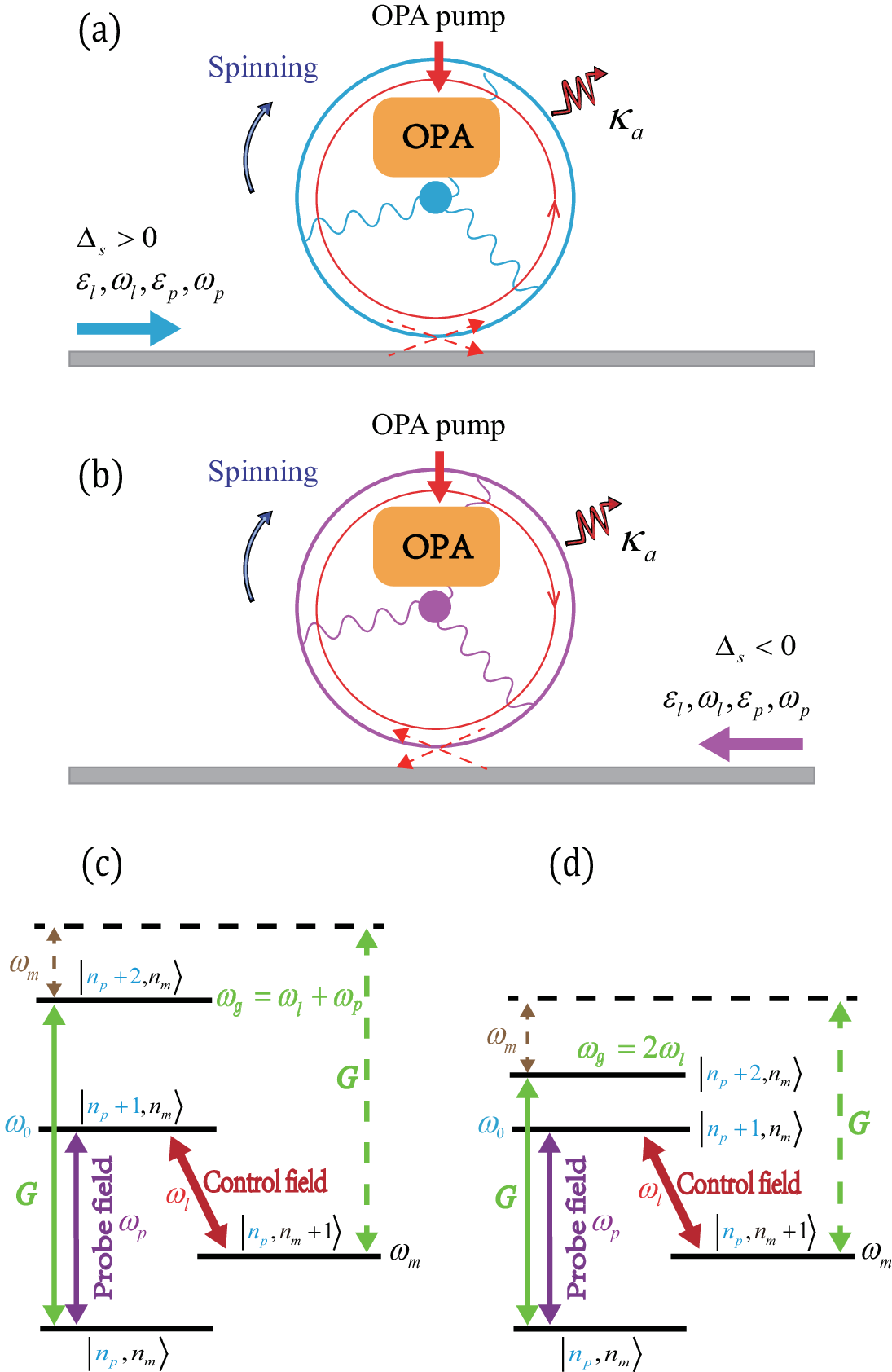}}
\caption{Schematic diagram of the spinning optomechanical system. A rotating whispering-gallery-mode (WGM) microresonator (containing an OPA \cite{Gerry,Clerk821155,Nation841,Leghtas347853,Shen100023814,Li100023838} with the frequency $\omega _g$) is coupled to a stationary tapered fiber. The resonator supports a mechanical mode at frequency ${{\omega _m}}$. We fix the clockwise rotation of the resonator, which leads to that the light circulating in the resonator experiences a Sagnac-Fizeau shift. (a) ${\Delta _s} > 0$ and (b) ${\Delta _s} < 0$ respond to the control-probe fields come from the left side and right side, respectively. The nonlinear crystal is pumped by an additional laser beam to produce parametric amplification. (c) with pump frequency $\omega_g=\omega_l+\omega_p$ and (d) with pump frequency $\omega_g=2\omega_l$ show the level schematic of the optomechanical system with OPA, where $| {{n_p}} \rangle $ and $| {{n_m}} \rangle $ denote the number states of the cavity and the mechanical mode, respectively.} \label{model}
\end{figure}

\section{The Model}
As schematically shown in Fig.~\ref{model}(a) and (b), the model we consider is a rotating whispering-gallery-mode (WGM) microresonator (containing an optical parametric amplifier), which is coupled to a stationary tapered fiber. The resonator (driven by a strong control field at frequency ${\omega _l}$ and a weak probe field at frequency ${\omega _p}$), with optical resonance frequency ${{\omega _0}}$ and intrinsic loss ${\kappa _a} = {{{\omega _0}} \mathord{\left/{\vphantom {{{\omega _0}} Q}} \right. \kern-\nulldelimiterspace} Q}$ ($Q$ is the optical quality factor), supports a mechanical breathing mode (frequency ${{\omega _m}}$ and effective mass $m$). A control laser and a probe laser are applied to the system via the evanescent coupling of the optical fiber and resonator, and the field amplitudes are given by ${\varepsilon _l} = \sqrt {{P_l}/\hbar {\omega _l}} $ and ${\varepsilon _p} = \sqrt {{P_p}/\hbar {\omega _p}} $, where ${{P_l}}$ and ${{P_p}}$ are the control and probe powers, respectively. It is well-known that due to the rotation, optical mode frequency experiences Sagnac-Fizeau shift \cite{Maayani558569,Post39475,Malykin431229}, which transforms
\begin{align}
&{\omega _0} \to {\omega _0} + {\Delta _s},\label{omega}\\
&{\Delta _s} = \frac{{nR\Omega {\omega _0}}}{c}\left( {1 - \frac{1}{{{n^2}}} - \frac{\lambda }{n}\frac{{dn}}{{d\lambda }}} \right),\label{deltas}
\end{align}
where $\Omega  = \dot \phi $ is the angular velocity of the spinning resonator. $n$ and $R$ are the refractive index and radius of the resonator, respectively. $c$ and $\lambda $ are the speed of light and the light wavelength in a vacuum, respectively. The dispersion term ${{dn} \mathord{\left/ {\vphantom {{dn} {d\lambda }}} \right. \kern-\nulldelimiterspace} {d\lambda }}$ represents a negligibly small relativistic (dispersion) correction in the Sagnac-Fizeau shift \cite{Maayani558569,Jiang10064037}. In Eq.~(\ref{deltas}), the first term in the parenthesis shows the Sagnac contribution which arises from the rotation of the resonators, while the two last terms with negative signs take into account the Fizeau drag due to the light propagation through a moving resonator medium. As shown in Refs.\cite{Agarwal93043844,Huang95023844,Scully}, the operating mechanism of the OPA is standard two-photon squeezing. Embedding the OPA in an optomechanical cavity makes the squeezed state transfer between a photon of a cavity field and a phonon of mechanical mode, which can amplify nonlinear optical responses of the system and reduce mechanical thermal noise and photon shot noise.
The Hamiltonian formulation of the system reads
\begin{equation}
\hat H = {{\hat H}_{mech}} + {{\hat H}_{opt}} + {{\hat H}_{OPA}} + {{\hat H}_{drive}}
,\label{H}
\end{equation}
with
\begin{equation}
\begin{aligned}
{{\hat H}_{mech}} =& \frac{{{{\hat p}^2}}}{{2m}} + \frac{1}{2}m\omega _m^2{{\hat x}^2} + \frac{{\hat p_\phi ^2}}{{2m{{\left( {R + \hat x} \right)}^2}}},\\
{{\hat H}_{opt}} =& \hbar \left( {{\omega _0} + {\Delta _s}} \right){{\hat a}^\dag }\hat a - \hbar \xi {{\hat a}^\dag }\hat a\hat x,\\
{\hat H_{OPA}} =& i\hbar G({\hat a^{\dag 2}}{e^{i\theta }}{e^{ - i{\omega _g}t}} - H.c.),\\
{{\hat H}_{drive}} =& i\hbar \sqrt {{\kappa _{ex}}} \left( {{\varepsilon _l}{{\hat a}^\dag }{e^{ - i{\omega _l}t}} + {\varepsilon _p}{{\hat a}^\dag }{e^{ - i{\omega _p}t}} - H.c.} \right)
,\label{Hfen}
\end{aligned}
\end{equation}
where ${\hat p}$, ${\hat x}$, ${\hat \phi }$, ${{\hat p}_\phi }$ describe the momentum, position, rotation angle, and angular momentum operators, with commutation relations $[ {\hat x,{\kern 1pt} \hat p} ] = [ {\hat \phi ,{\kern 1pt} {{\hat p}_\phi }} ] = i\hbar $ \cite{Davuluri11264002}. H.c. stands for the Hermitian conjugate. $\hat a\left( {{{\hat a}^\dag }} \right)$ is the annihilation (creation) operator of the cavity field with resonance frequency ${{\omega _0}}$. $\xi  = {{{\omega _0}} \mathord{\left/ {\vphantom {{{\omega _0}} R}} \right.\kern-\nulldelimiterspace} R}$ is the optomechanical coupling. ${{\hat H}_{OPA}}$ describes the coupling of the intracavity field with the OPA (pump frequency $\omega_g$). $G$ is the nonlinear gain of the OPA, which is proportional to the pump power driving amplitude. $\theta $ is the phase of the field driving the OPA \cite{Shahidani88053813}. We assume that this OPA with a second-order nonlinearity crystal is excited by a pump driving with the frequency  $\omega _g={{\omega _l} + {\omega _p}}$ \cite{Liu99033822} in Fig.~\ref{model}(c), so that the signal light and idler light in OPA have the same frequency $({\omega _l} + {\omega _p})/2$ \cite{Nation841,Leghtas347853,Adiyatullin81329,Clerk821155}. ${{\hat H}_{drive}}$ describes the interaction of the cavity field with the control field and that of the cavity field with the probe field, where ${{\kappa _{ex}}}$ is the loss caused by the resonator-fiber coupling.

In the rotating frame at the control frequency ${{\omega _l}}$, the Hamiltonian~(\ref{H}) becomes
\begin{equation}
\begin{aligned}
{{\hat H}_{eff}} =& \hbar \left( {{\Delta _0} - \xi \hat x + {\Delta _s}} \right){{\hat a}^\dag }\hat a + \frac{{{{\hat p}^2}}}{{2m}} + \frac{1}{2}m\omega _m^2{{\hat x}^2}\\
& + \frac{{\hat p_\phi ^2}}{{2m{{\left( {R + \hat x} \right)}^2}}} + i\hbar G({\hat a^{\dag 2}}{e^{ - i{\Delta _p}t}}{e^{i\theta }} - H.c.)\\
& + i\hbar \sqrt {{\kappa _{ex}}} \left[ {\left( {{\varepsilon _l} + {\varepsilon _p}{e^{ - i{\Delta _p}t}}} \right){{\hat a}^\dag } - H.c.} \right],\label{Heff}
\end{aligned}
\end{equation}
where ${\Delta _0} = {\omega _0} - {\omega _l}$ and ${\Delta _p} = {\omega _p} - {\omega _l}{\kern 1pt} $. When the control field is injected at the red-detuned sideband of the cavity resonance ($\Delta_p=\omega_m$), the transition $\left| {{n_p},{n_m} + 1} \right\rangle  \leftrightarrow \left| {{n_p} + 1,{n_m}} \right\rangle $ occurs. Moreover, $\left| {{n_p},{n_m}} \right\rangle$ couples with  $\left| {{n_p} + 1,{n_m}} \right\rangle $ through the probe field which is in resonance with the cavity mode ($\omega_p=\omega_0$). In this case, the  destructive interference of these two excitation pathways occurs, which leads to OMIT \cite{Weis3301520} in Fig.~\ref{model}(c) with pump frequency $\omega_g=\omega_l+\omega_p$ (see Sec.~II-IV, Sec.~VI) and Fig.~\ref{model}(d) with pump frequency $\omega_g=2\omega_l$  (see Sec.~V), where OPA has almost no
influence on the interference paths. With the operator expectation values defined by $a \equiv \langle {\hat a} \rangle $ , $x \equiv \langle {\hat x} \rangle $ , $\phi  \equiv \langle {\hat \phi } \rangle $, and ${p_\phi } \equiv \langle {{{\hat p}_\phi }} \rangle $, the Heisenberg-Langevin equations of the spinning optomechanical system can be derived as
\begin{align}
&\dot a =  - \left[ {\kappa  + i\left( {{\Delta _0} - \xi x + {\Delta _s}} \right)} \right]a \nonumber \\
&\qquad + \sqrt {{\kappa _{ex}}} \left( {{\varepsilon _l} + {\varepsilon _p}{e^{ - i{\Delta _p}t}}} \right)+ 2G{a^ * }{e^{i\theta }}{e^{ - i{\Delta _p}t}},\label{Heq1}\\
&m(\ddot x + {\Gamma _m}\dot x + \omega _m^2x) = \hbar \xi {a^ * }a + \frac{{p_\phi ^2}}{{m{R^3}}},\label{Heq2}\\
&\dot \phi  = \frac{{{p_\phi }}}{{m{R^2}}},\label{Heq3}\\
&{{\dot p}_\phi } = 0,\label{Heq4}
\end{align}
where $\kappa  = ({\kappa _a} + {\kappa _{ex}})/2$ and $\Gamma_m$ are the dissipations of the cavity and the damping of the mechanical mode, respectively. The derivation of Eqs.~(\ref{Heq1})-(\ref{Heq4}) can be found in Appendix. Focusing on the mean response of the system to the probe field, we write the operators for their expectation values by means of the mean-field approximation and safely ignore the quantum noise terms with strong driving conditions.

In this case, we assume the control field is much stronger than the probe field (${\varepsilon _l} \gg {\varepsilon _p}$), which induces that we can use the perturbation method to deal with Eqs.~(\ref{Heq1})-(\ref{Heq4}). The control field provides a steady-state solution of the system, while the probe field is treated as the perturbation of the steady state. We then follow the standard procedure, which decomposes the expectation value of all operators as a sum of their steady-state value and small fluctuations around the steady-state value \cite{Weis3301520,Xiong86013815}
\begin{small}
\begin{equation}
\begin{aligned}
a=&{a_s} + A_1^ + {e^{ - i{\Delta _p}t}} + A_1^ - {e^{i{\Delta _p}t}} + A_2^ + {e^{ - 2i{\Delta _p}t}} + A_2^ - {e^{2i{\Delta _p}t}},\\
x=&{x_s} + X_1^ + {e^{ - i{\Delta _p}t}} + X_1^ - {e^{i{\Delta _p}t}} + X_2^ + {e^{ - 2i{\Delta _p}t}} + X_2^ - {e^{2i{\Delta _p}t}},
\label{x}
\end{aligned}
\end{equation}
\end{small}
in which $A_2^ + $ ($A_2^ - $) is the amplitude of second-order upper (lower) sideband and corresponds to the responses at the original frequencies $2{\omega _p} - {\omega _l}$ ($3{\omega _l} - 2{\omega _p}$). We are committed to the fundamental OMIT and its second-order sideband process so that the higher-order sidebands in Eq.~(\ref{x}) are ignored. By substituting Eq.~(\ref{x}) into Eqs.~(\ref{Heq1})-(\ref{Heq4}) and comparing the coefficients of the same order, the steady-state solutions are obtained as
\begin{equation}
\begin{aligned}
{a_s}=&\frac{{\sqrt {{\kappa _{ex}}} {\varepsilon _l}}}{{\kappa + i\Delta }},\\
{x_s}=&\frac{{\hbar \xi {{\left| {{a_s}} \right|}^2}}}{{m\omega _m^2}} + R{\left( {\frac{\Omega }{{{\omega _m}}}} \right)^2}
,\label{wentaijie}
\end{aligned}
\end{equation}
where $\Delta {\rm{ = }}{\Delta _0} - \xi {x_s} + {\Delta _s}$, and $\Omega {\rm{ = }}{{d\phi } \mathord{\left/ {\vphantom {{d\theta } {dt}}} \right. \kern-\nulldelimiterspace} {dt}}$ is the angular velocity of the spinning resonator. It is clear that the revolving speed of the resonator and Sagnac-Fizeau shift ${\Delta _s}$ affect the values of both the mechanical displacement ${x_s}$ and intracavity photon number ${\left| {{a_s}} \right|^2}$.
Substituting Eq.~(\ref{x}) into Eqs.~(\ref{Heq1})-(\ref{Heq4}), we gain six algebra equations, which can be divided into two groups. The first group describes the linear response of the probe field
\begin{equation}
\begin{split}
{\sigma _1}\left( {{\Delta _p}} \right)A_1^ +  &= i\xi {a_s}X_1^ +  + 2G{e^{i\theta }}a_s^* + \sqrt {{\kappa _{ex}}} {\varepsilon _p},\\
{\sigma _2}\left( {{\Delta _p}} \right)A_1^{ - *} &=  - i\xi a_s^*X_1^ + ,\\
\chi \left( {{\Delta _p}} \right)X_1^ +  &= \hbar \xi ({a_s}A_1^{ - *} + a_s^*A_1^ + ),\label{first1}
\end{split}
\end{equation}
while the second group corresponds to the second-order sideband process
\begin{equation}
\begin{split}
{\sigma _1}(2{\Delta _p})A_2^ + &=i\xi ({a_s}X_2^ +  + A_1^ + X_1^ + ) + 2G{e^{i\theta }}A_1^{ -  * },\\
{\sigma _2}\left( {2{\Delta _p}} \right)A_2^{ - *} &=- i\xi (a_s^*X_2^ +  + A_1^{ - *}X_1^ + ),\\
\chi \left( {2{\Delta _p}} \right)X_2^ +  &=\hbar \xi (a_s^*A_2^ +  + {a_s}A_2^{ - *} + A_1^{ - *}A_1^ + ),\label{second1}
\end{split}
\end{equation}
with
\begin{eqnarray}
{\sigma _1}\left( {n{\Delta _p}} \right) &=& \kappa  + i\Delta  - in{\Delta _p},\nonumber\\
{\sigma _2}\left( {n{\Delta _p}} \right) &=& \kappa  - i\Delta  - in{\Delta _p},\nonumber\\
\chi \left( {n{\Delta _p}} \right) &=& m(\omega _m^2 - i{\Gamma _m}n{\Delta _p} -\Delta _p^2).\nonumber
\end{eqnarray}

Moreover, we can easily get the linear and second-order nonlinear responses of the system
\begin{equation}
\begin{split}
A_1^ +  &= \frac{{D + {\sigma _2}\left( {{\Delta _p}} \right)\chi \left( {{\Delta _p}} \right)}}{{{f_3}\left( {{\Delta _p}} \right)}}(\sqrt {{\kappa _{ex}}} {\varepsilon _p} + 2G{e^{i\theta }}a_s^*),\\
X_1^ +  &= \frac{{\hbar \xi a_s^*{\sigma _2}\left( {{\Delta _p}} \right)}}{{D + {\sigma _2}\left( {{\Delta _p}} \right)\chi \left( {{\Delta _p}} \right)}}A_1^ +, \\
A_1^{ - *} &= \frac{{ - i\xi a_s^*}}{{{\sigma _2}\left( {{\Delta _p}} \right)}}X_1^ +,\label{jie11}
\end{split}
\end{equation}
and
\begin{equation}
\begin{split}
A_2^ +  &= \frac{{ - D{\xi ^2}{a_s}X_1^{ + 2} + i\xi {f_1}A_1^ + X_1^ +  - 2i\xi G{e^{i\theta }}a_s^*{f_2}X_1^ + }}{{{\sigma _2}\left( {{\Delta _p}} \right){f_3}\left( {2{\Delta _p}} \right)}},\\
X_2^ +  &= \frac{{\hbar \xi [{\sigma _2}(2{\Delta _p})a_s^*A_2^ +  + {\sigma _2}(2{\Delta _p})A_1^ + A_1^{ - *} - i\xi {a_s}A_1^{ - *}X_1^ + ]}}{{{f_2}}},\\
A_2^ -  &= \frac{{i\xi }}{{{\sigma _2}{{\left( {2{\Delta _p}} \right)}^*}}}({a_s}X_2^ -  + A_1^ - X_1^ - ),\label{jie12}
\end{split}
\end{equation}
where
\begin{equation}
\begin{split}
D &= i\hbar {\xi ^2}{\left| {{a_s}} \right|^2},\nonumber\\
{f_1} &= iD{\Delta _p} + {\sigma _2}\left( {{\Delta _p}} \right){\sigma _2}\left( {2{\Delta _p}} \right)\chi \left( {2{\Delta _p}} \right),\nonumber\\
{f_2} &= D + {\sigma _2}\left( {2{\Delta _p}} \right)\chi \left( {2{\Delta _p}} \right),\nonumber\\
{f_3}\left( {n{\Delta _p}} \right) &= 2iD\Delta  + {\sigma _1}(n{\Delta _p}){\sigma _2}\left( {n{\Delta _p}} \right)\chi \left( {n{\Delta _p}} \right).\nonumber
\end{split}
\end{equation}
By using the standard input-output relations, i.e.,
\begin{equation}
{a_{out}}(t) = {a_{in}}(t) - \sqrt {{\kappa _{ex}}}  a(t),
\label{cz123}
\end{equation}
we obtain the expectation value of the output field of this system
\begin{equation}
\begin{aligned}
{a_{out}}(t) =& {C_1}{e^{ - i{\omega _l}t}} + {C_2}{e^{ - i{\omega _p}t}} - \sqrt {{\kappa _{ex}}}  A_1^ - {e^{ - i(2{\omega _l} - {\omega _p})t}}\\
& - \sqrt {{\kappa _{ex}}}  A_2^ + {e^{ - i(2{\omega _p} - {\omega _l})t}} - \sqrt {{\kappa _{ex}}}  A_2^ - {e^{ - i(3{\omega _l} - 2{\omega _p})t}},\label{aout}
\end{aligned}
\end{equation}
where ${C_1} = {\varepsilon _l} - \sqrt {{\kappa _{ex}}}  a{}_s$ and ${C_2} = {\varepsilon _p} - \sqrt {{\kappa _{ex}}}  A_1^ + $. The first term of Eq.~(\ref{aout}) denotes the output with control frequency ${{\omega _l}}$, while the second and third terms describe the anti-Stokes and Stokes fields, respectively. The terms $ - \sqrt {{\kappa _{ex}}}  A_2^ + {e^{ - i(2{\omega _p} - {\omega _l})t}}$ and $ - \sqrt {{\kappa _{ex}}}  A_2^ - {e^{ - i(3{\omega _l} - 2{\omega _p})t}}$  are concerned in the second-order upper and lower sidebands \cite{Xiong86013815}.

Subsequently, we introduce the dimensionless quantity to define the efficiency of the second-order upper and lower sidebands \cite{Xiong86013815,Teufel471204}
\begin{eqnarray}
{\eta _1} = \left| { - \frac{{\sqrt {{\kappa _{ex}}} A_2^ + }}{{{\varepsilon _p}}}} \right|,\label{eta1}\\
{\eta _2} = \left| { - \frac{{\sqrt {{\kappa _{ex}}} A_2^ - }}{{{\varepsilon _p}}}} \right|,\label{eta2}
\end{eqnarray}
where the amplitude of the probe pulse is treated as a basic scale to gauge the amplitudes of the output sidebands ${\eta _1}$ and ${\eta _2}$.
The associated group delay of the second-order upper sideband turns out to be \cite{Safavi47269,Lu1,Lu2}
\begin{equation}
{\tau _1} = {\left. {\frac{{d{\kern 1pt} \arg \left({ - \frac{{\sqrt {{\kappa _{ex}}} A_2^ + }}{{{\varepsilon _p}}}}\right)}}{{2 d{\Delta _p}}}} \right|_{{\Delta _p} = {\omega _m}}}.\label{tau1}
\end{equation}
A positive group delay (${\tau _1} > 0$) corresponds to slow light phenomenon, while a negative group delay (${\tau _1} < 0$) corresponds to fast light phenomenon \cite{Safavi47269,Milonni}.

%figure2
\begin{figure}[t]
\centerline{
\includegraphics[width=9.5cm, height=10cm, clip]{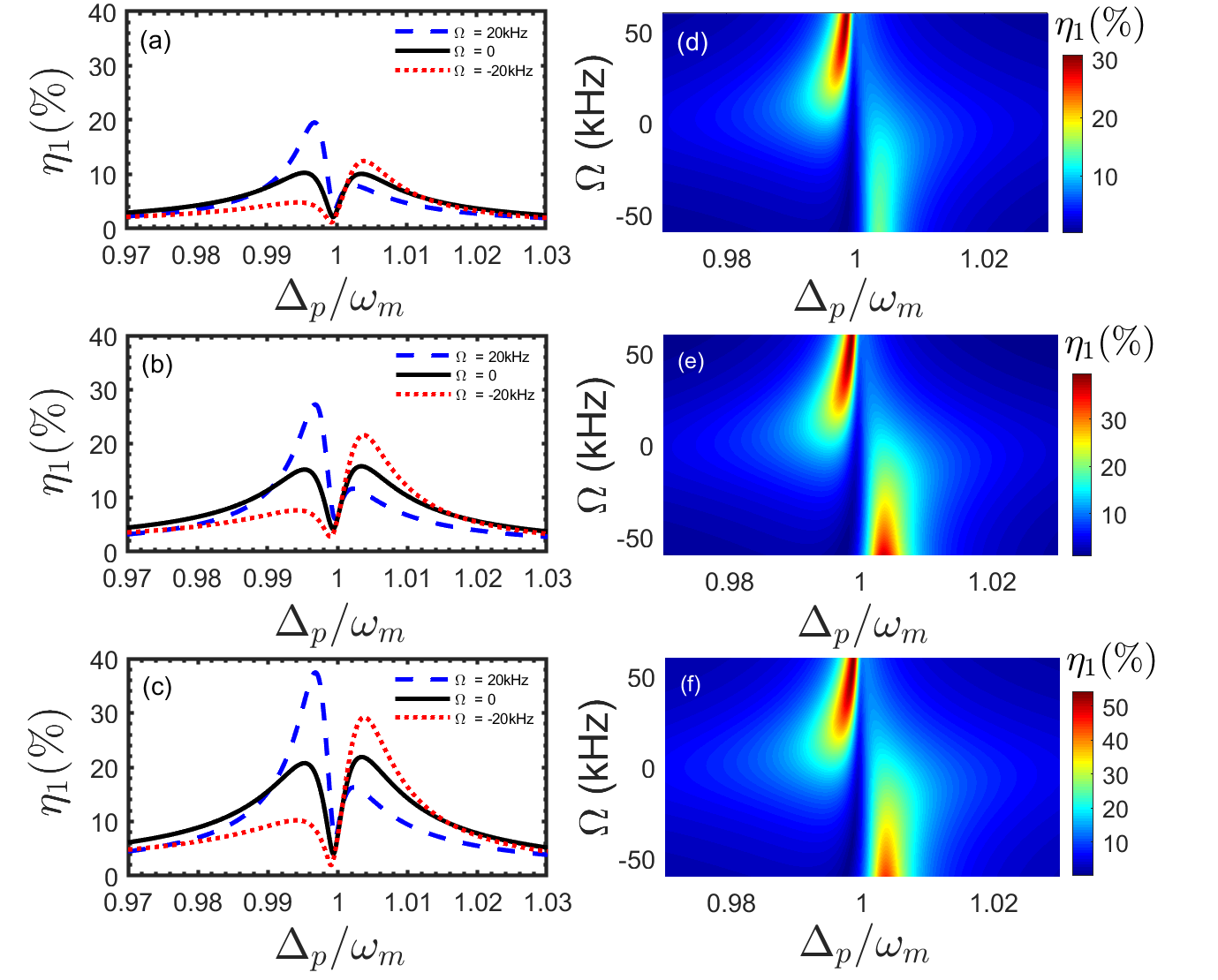}}
\caption{The efficiency ${\eta _1}$ of the second-order upper sideband generation as a function of ${\Delta _p}$ for different values of $\Omega $ and incident directions of light, where the nonlinear gain and phase of the probe field of the OPA are fixed as (a) $G = 0,\theta  = 0$; (b) $G = 0.2\kappa ,\theta  = 0$; (c) $G = 0.2\kappa ,\theta  = 3\pi /2$. ${\eta _1}$ varies with ${\Delta _p}$ and $\Omega $ under different values (d) $G = 0,\theta  = 0$; (e) $G = 0.2\kappa ,\theta = 0$; (f) $G = 0.2\kappa ,\theta  = 3\pi /2$. Other parameters are ${P_p} = 0.05{P_l}$, ${P_l} = 1$ mW, $\lambda  = 1550$ nm, $R = 0.25$ mm, $m = 25$ ng, $n = 1.44$, $Q = {\omega _0}/\kappa  = 4.5 \times {10^7}$, ${\omega _m} = 100$ MHz, ${\Gamma _m} = 0.1$ MHz, ${\kappa _{a}} = {\kappa _{ex}} = {\omega _0}/Q$, ${P_p} = 0.05{P_l}$, and ${\Delta _0} = {\omega _m}$, respectively. With the parameters, we  obtain the Sagnac-Fizeau shift $\Delta_s$=($15.082$MHz, $0$, $15.082$MHz) or $\Delta_s/\omega_m$=($0.1508$, $0$, $-0.1508$), which corresponds to the angular velocity $\Omega=$ ($20$kHz, 0, $-20$kHz) of the cavity.} \label{Eta1}
\end{figure}

\section{Results and discussions}
In our numerical simulations, to demonstrate that the observation of the second-order sidebands in a resonator assisted by OPA is within current experimental reach, we calculate Eqs.~(\ref{eta1})-(\ref{tau1}) with parameters from Refs.\cite{Maayani558569,Righini34435,Zhang1321}: $\lambda  = 1550$ nm, $R = 0.25$ mm (the resonator radius),
$m = 25$ ng, $n = 1.44$, $Q = {\omega _0}/\kappa  = 4.5 \times {10^7}$, ${\omega _m} = 100$ MHz, ${\Gamma _m} = 0.1$ MHz, ${\kappa _{a}} = {\kappa _{ex}} = {\omega _0}/Q$,
${P_p} = 0.05{P_l}$, and ${\Delta _0} = {\omega _m}$, respectively. We rotate the resonator clockwise, where $\Omega  > 0$ stands for the light coming from the left-hand side and $\Omega  < 0$ denotes the light coming from the right-hand side.

To see the influence of resonator rotation and OPA on the second-order sideband generation, the efficiency of second-order upper sideband generation is investigated as a function of frequency ${\Delta _p}/{\omega _m}$ shown in Fig.~\ref{Eta1}. In Fig.~\ref{Eta1}(a), we discuss that the efficiency ${\eta _1}$ of the second-order upper sideband varies with ${\Delta _p}$ without the participation of OPA, i.e., the nonlinear gain of the OPA $G = 0$, the phase of the field driving the OPA $\theta  = 0$. Under the resonator stationary, we find two located peaks of second-order sideband spectra and a local minimum near the resonance condition ${\Delta _p} = {\omega _m}$. By spinning the resonator, the peak position of ${\eta _1}$ has different moves when the driving fields come from different directions. By adjusting the frequency ${\Delta _p}/{\omega _m}$, we can get enhanced efficiency of the second-order sideband while driving the resonator from one direction and suppressed efficiency while driving from the opposite direction. For example, within ${\Delta _p}/{\omega _m}$ in the range from $0.99$ to $1$, ${\eta _1}$ is enhanced in the case of $\Omega  > 0$, while it is suppressed in the case of $\Omega  < 0$. Obviously, this spinning-induced direction-dependent nonreciprocal behavior can be attributed to the optical Sagnac effect induced by a spinning resonator. As shown in Fig.~\ref{Eta1}(b), the efficiency ${\eta _1}$ gets larger in the presence of OPA. To be more specific, for $G = 0.2\kappa ,\theta  = 0$ and $\Omega  = 20$ kHz, the efficiency ${\eta _1}$ can increase from ${\rm{19}}{\rm{.5\% }}$ to $27.2{\rm{\% }}$ at ${\Delta _p} = 0.997{\omega _m}$. When the system is driven from the right, i.e., $\Omega  =  - 20$ kHz, the efficiency ${\eta _1}$ also can increase from $12.4{\rm{\% }}$ to $21.6{\rm{\% }}$ at ${\Delta _p} = 1.004{\omega _m}$. Fig.~\ref{Eta1}(c) shows that the efficiency ${\eta _1}$ can also be adjusted by tuning $\theta $. What can be seen clearly is when $\theta $ changes from $\theta  = 0$ to $\theta  = 3\pi /2$, in the case of $G = 0.2\kappa $ and $\Omega  = 20$ kHz, the maximum value of ${\eta _1}$ increases to $37.4{\rm{\% }}$. In the case of $\Omega  =  - 20$ kHz, the maximum value increases to $29.2{\rm{\% }}$. We see that the efficiency of the second-order upper sideband is sensitive to the variation of the nonlinear gain of the OPA and phase of the field driving the OPA, which indicates the advantage of using a hybrid nonlinear system.  According to Eqs.~(\ref{jie11})-(\ref{jie12}), such phenomena coming from the amplitudes of second-order sidebands $A_2^ + $ and $A_2^ - $ are related directly to the Sagnac-Fizeau shift and OPA. With the purpose of seeing the influence of the OPA on the second-order sideband generation more clearly, the efficiency ${\eta _1}$ as a function of both ${\Delta _p}$ and $\Omega $ is shown in Fig.~\ref{Eta1}(d)-(f).

%figure3
\begin{figure}[t]
\centerline{
\includegraphics[width=9.2cm, height=8cm, clip]{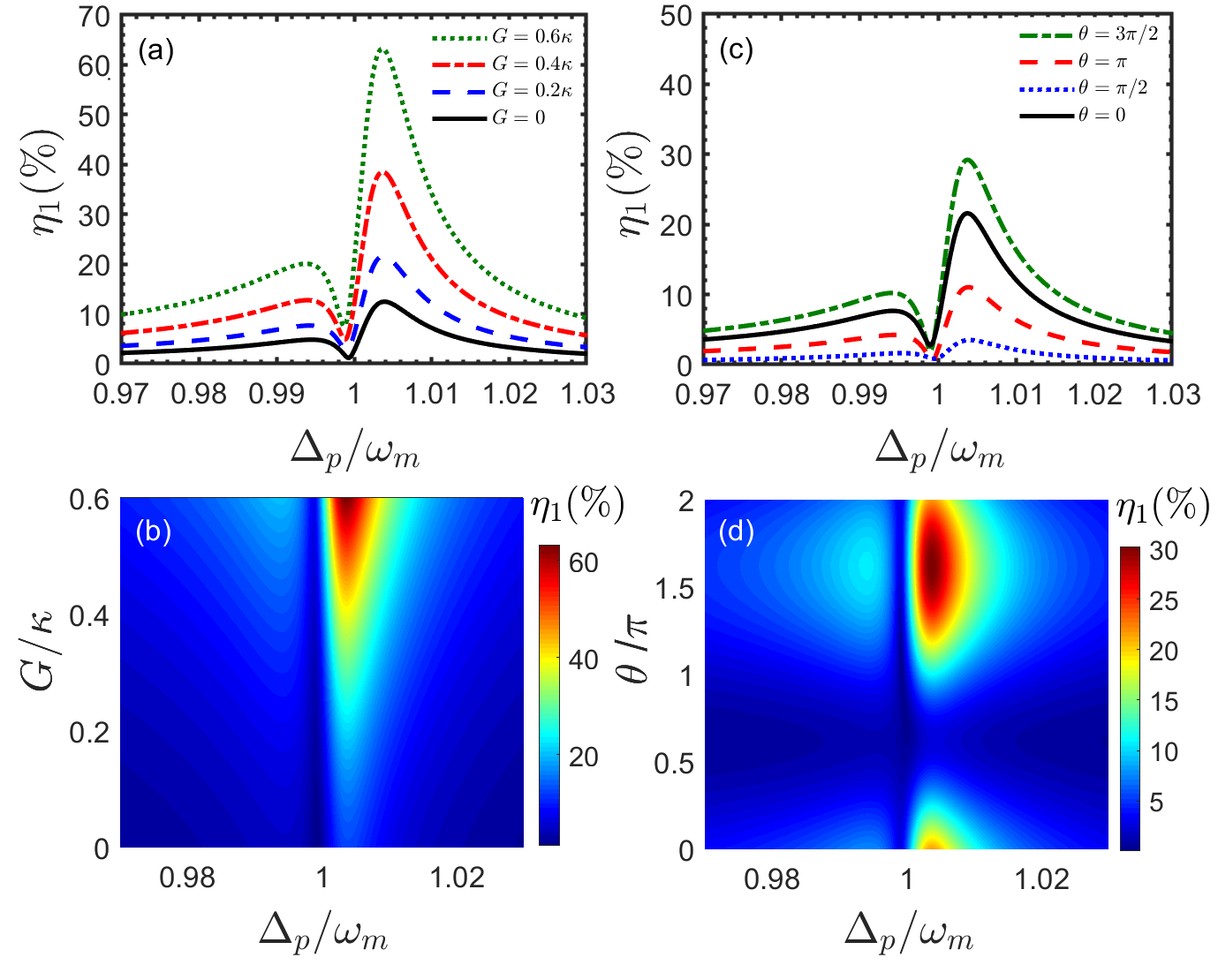}}
\caption{The efficiency ${\eta _1}$ of the second-order upper sideband generation as a function of the probe-pulsed detuning ${\Delta _p}$ for different (a) nonlinear gain $G$ of the OPA with $\theta  = 0$ and (c) phase $\theta $ with $G = 0.2\kappa $. (b) ${\eta _1}$ varies with ${\Delta _p}$ and $G$ under $\theta  = 0$. (d) ${\eta _1}$ varies with ${\Delta _p}$ and $\theta $ under $G = 0.2\kappa $. The angular velocity of the spinning resonator is fixed at $\Omega  =  - 20$ kHz. Other parameters are the same as  Fig.~\ref{Eta1}.} \label{Eta3}
\end{figure}

To explore the role of OPA in this resonator, we illustrate the efficiency ${\eta _1}$ of the second-order upper sideband versus the probe-pulsed detuning ${\Delta _p}$ with different nonlinear gain $G$ of the OPA and phase  $\theta $ of the field driving the OPA, when the system is driven from the right-hand side ($\Omega  =  - 20$ kHz) in Fig.~\ref{Eta3}. We find when the nonlinear gain $G$ of the OPA increases from $0$ to $G = 0.6\kappa $, the efficiency ${\eta _1}$ can be significantly enhanced in Fig.~\ref{Eta3}(a). The enhancement effect at the probe-pulsed detuning ${\Delta _p}/{\omega _m} < 1$ is much weaker than at ${\Delta _p}/{\omega _m} > 1$. Fig.~\ref{Eta3}(c) shows that the second-order sideband behavior of the output field can also be adjusted by tuning $\theta $. In the case of $G = 0.2\kappa $, we find that compared with $\theta  = 0$, both $\theta = \pi /2$ and $\theta  = \pi $ result in lower efficiency ${\eta _1}$ of the second-order upper sideband, but $\theta  = 3\pi /2$ leads to enhanced efficiency. In Fig.~\ref{Eta3}(d), ${\eta _1}$ as a function of detuning ${\Delta _p}$ and the phase $\theta $ of the OPA is plotted. In the range shown, the maximum value of efficiency ${\eta _1}$ is about $30.2\% $ at $\theta  = 1.6\pi$ and ${\Delta _p} = 1.004{\omega _m}$. Specifically, the efficiency is enhanced when $\theta  \in (1.6\pi ,{\kern 1pt} {\kern 1pt} 2\pi )$ and suppressed at other values. Besides, as is illustrated in Fig.~\ref{Eta3}(b) and (d), regardless of what nonlinear gain $G$ and $\theta $ is to increase, the located maximums of the efficiency ${\eta _1}$ are still located at the same position of the probe-pulsed detuning. This phenomenon can be explained by Refs.\cite{Weis3301520,Xiong86013815}, which shows there are some connections between OMIT and the second-order sideband process. When OMIT occurs, the second-order sideband process is subdued. The linewidth of the OMIT window is related to the intracavity photon number
\begin{equation}\label{width}
{\Gamma _{OMIT}} \approx {\Gamma _{\rm{m}}} + \frac{{{\xi ^2}x_{zpf}^2}}{\kappa }{\left| {{a_s}} \right|^2},
\end{equation}
where ${x_{zpf}} = \sqrt {\hbar /2m{\omega _{\rm{m}}}} $. By perturbation theory, we can get the intracavity photon number ${\left| {{a_s}} \right|^2}$ in Eq.~(\ref{wentaijie}), which is independent of other perturbation terms such as probe pulse and nonlinear gain of the OPA. That is to say, the positions of these local maximums of sideband spectra only depend on the intrinsic structural parameters of an optomechanical system and intensity of the control field. As a result, the OPA not only improves the sideband efficiency of the second-order sideband but also keeps the locality of maximum values of the sideband efficiency.

%figure4
\begin{figure}[t]
\centerline{
\includegraphics[width=9.2cm, height=10cm, clip]{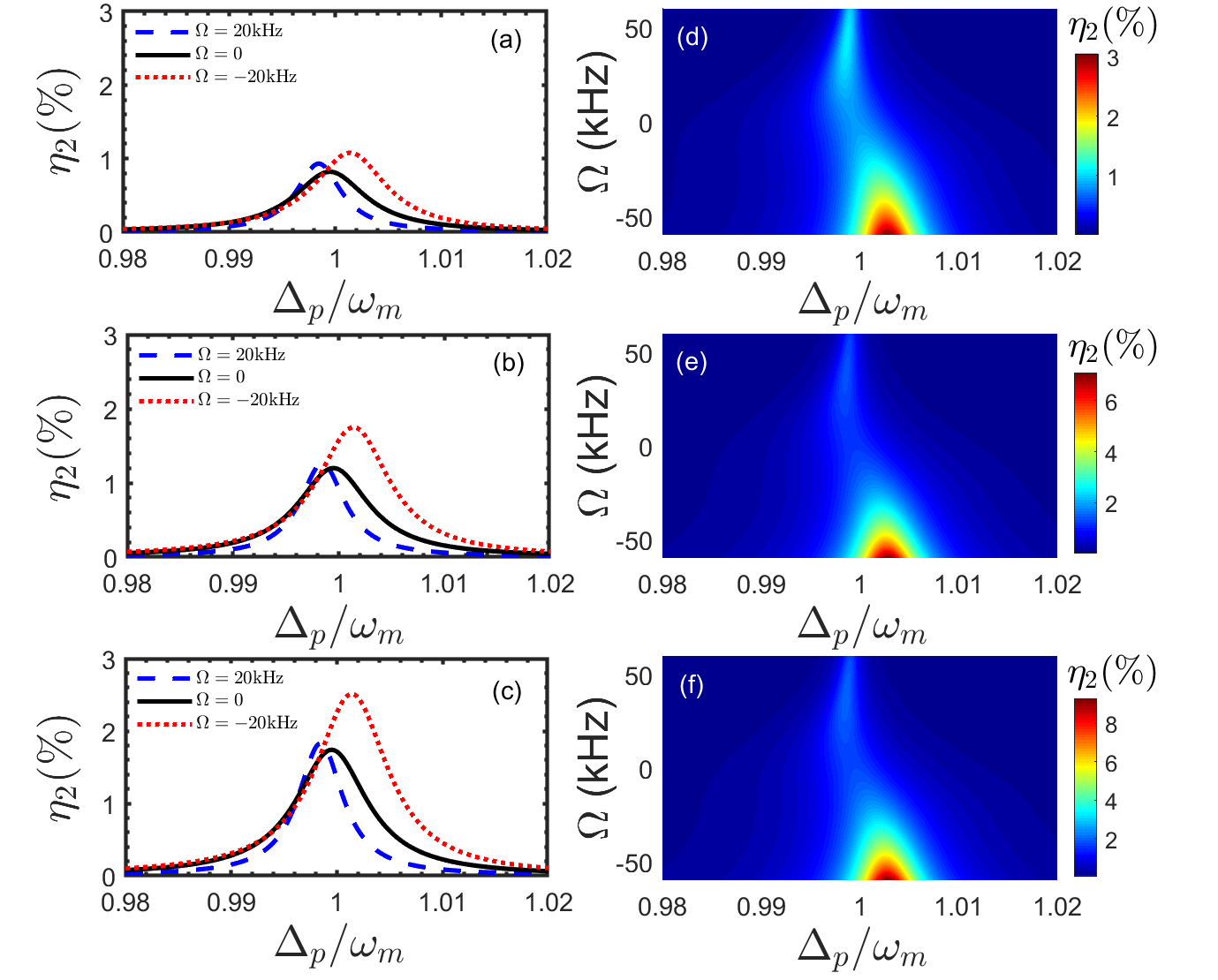}}
\caption{The efficiency ${\eta _2}$ of the second-order lower sideband generation as a function of ${\Delta _p}$ for different values of $\Omega $ and incident directions of light, where the nonlinear gain and phase of the probe field of the OPA are fixed as (a) $G = 0,\theta  = 0$; (b) $G = 0.2\kappa ,\theta  = 0$; (c) $G = 0.2\kappa ,\theta  = 3\pi /2$. ${\eta _2}$ varies with ${\Delta _p}$ and $\Omega $ under different values (d) $G = 0,\theta  = 0$; (e) $G = 0.2\kappa ,\theta = 0$; (f) $G = 0.2\kappa ,\theta  = 3\pi /2$. Other parameters are the same as  Fig.~\ref{Eta1}.} \label{Eta2}
\end{figure}

In Fig.~\ref{Eta2}, we discuss the influence of resonator rotation and OPA on the second-order lower sideband generation. As shown in Fig.~\ref{Eta2}(a), unlike the second-order upper sideband, the second-order lower sideband has no local minimum but only one peak. The efficiency is much smaller than the second-order upper sideband. In detail, with neither resonator rotation nor the OPA drive, ($G = 0$, $\Omega  = 0$), both peaks of ${\eta _1}$ are about $19.6\% $ and the peak of ${\eta _2}$ is only $0.82\% $. Furthermore, the second-order lower sideband exhibits non-reciprocal characteristics due to the rotation of the resonator, which is more pronounced at ${\Delta _p}/{\omega _m} > 1$. In detail, compared with the stationary resonator (i.e., no spinning with $\Omega  = 0$), the spinning resonator increases for $\Omega  =-20$ kHz, while it decreases for $\Omega  =20$ kHz at ${\Delta _p}/{\omega _m} > 1$ in Fig.~\ref{Eta2}(a). In Fig.~\ref{Eta2}(d), we find that for the same resonator speed, the enhancement effect is more pronounced when the device is driven from the right side ($\Omega  < 0$) than from the left side ($\Omega  > 0$). For example, for $\Omega  =  - 60$ kHz, the maximum value of ${\eta _2}$ is $3.04\% $ at ${\Delta _p}/{\omega _m} = 1.003$. For $\Omega  =  60$ kHz, the maximum value of ${\eta _2}$ is $0.97\% $ at ${\Delta _p}/{\omega _m} = 0.999$. In Fig.~\ref{Eta2}(b) and (c), as with the second-order upper sideband, the presence of OPA significantly improves the efficiency of the second-order lower sideband, which also keeps the locality of maximum values of the sideband efficiency. In detail, for $\Omega  =  - 20$ kHz, when the nonlinear gain $G$ of OPA increases from 0 to $0.2\kappa $, the maximum value of ${\eta _2}$ increases from $1.08\% $ to $1.75\% $ at ${\Delta _p}/{\omega _m} = 1.001$. Besides, when the phase $\theta $ of the OPA increases from 0 to $3\pi /2$, the maximum value of ${\eta _2}$ can be increased to $2.51\% $, which is more than twice the value without OPA.

We show that the presence of OPA only causes a change in the peak of ${\eta _1}$ and has almost no influence on asymmetry (see black-solid line in Fig.~\ref{Eta1}(a)-(c) for $\Omega=0$, and black-solid line in Fig.~\ref{Eta2}(a)-(c) for $\Omega=0$).

Without the OPA ($G=0$), the asymmetric line shape of ${\eta _1}$ with regard to ${\Delta _p} = {\omega _m}$ and the ${\eta _2}$ peak being not exactly at ${\Delta _p} = {\omega _m}$ come from the spinning of the resonator. In this case, the mean mechanical displacement ${x_s}$ in Eq.~(\ref{wentaijie}) is made up of two terms: the first term is proportional to the intracavity photon number $|{a_s}|^2$, which is closely related to the Sagnac-Fizeau shift ${\Delta _s}=\frac{{nR\Omega {\omega _0}}}{c}( {1 - \frac{1}{{{n^2}}} - \frac{\lambda }{n}\frac{{dn}}{{d\lambda }}} )$ in Eq.~(\ref{deltas})  or, equivalently, very sensitive to the angular velocity $\Omega$ of resonator and incident  direction of input fields, thus giving rise to the nonreciprocal behavior. The second term $R{(\Omega /{\omega _m})^2}$ of ${x_s}$ makes the mechanical displacement larger due to the rotation. The existence of these two terms together affects the second-order upper and lower sidebands in Eqs.~(\ref{eta1}) and (\ref{eta2}), which lead to the asymmetry of ${\eta _1}$ with regard to ${\Delta _p} = {\omega _m}$ and the ${\eta _2}$ peak being not exactly at ${\Delta _p} = {\omega _m}$ shown in Fig.~\ref{Eta1} to Fig.~\ref{Eta2}.

%To be specific, the second term $R{( {\frac{\Omega }{{{\omega _m}}}} )^2}$ of ${x_s}$ in Eq.~(11) originates from the system Hamiltonian in Eq.~(4). Due to the rotation (assuming $G=0$), there is an extra rotational kinetic energy  term $\hat p_\phi ^2/[2m{( {R + \hat x} )^2}]$ in the Hamiltonian of our model,

In this case, $R{(\Omega /{\omega _m})^2}$ of ${x_s}$ in Eq.~(\ref{wentaijie}) originates from an extra term in the Hamiltonian of our model due to the rotation, i.e., the rotational kinetic energy  term $\hat p_\phi ^2/[2m{( {R + \hat x} )^2}]$ in Eq.~(\ref{Hfen}), which is different from the usual situation in Ref.\cite{Xiong86013815}. Since $x/R \ll 1$ ($x = \langle {\hat x} \rangle $ denotes the expectation value of $\hat x$), the term $\hat p_\phi ^2/[2m{( {R + \hat x} )^2}]$ is approximately equal to $- \hat p_\phi ^2\hat x/(m{R^3}) + \hat p_\phi ^2/(2m{R^2})$ (neglecting second and higher order small quantities about $x/R$). This means that there is an extra force $- \hat p_\phi ^2\hat x/(m{R^3})$ exerted on the mechanical mode making it deviate from its original equilibrium position.

%figure5
\begin{figure}[t]
\centerline{
\includegraphics[width=9cm, height=6.8cm, clip]{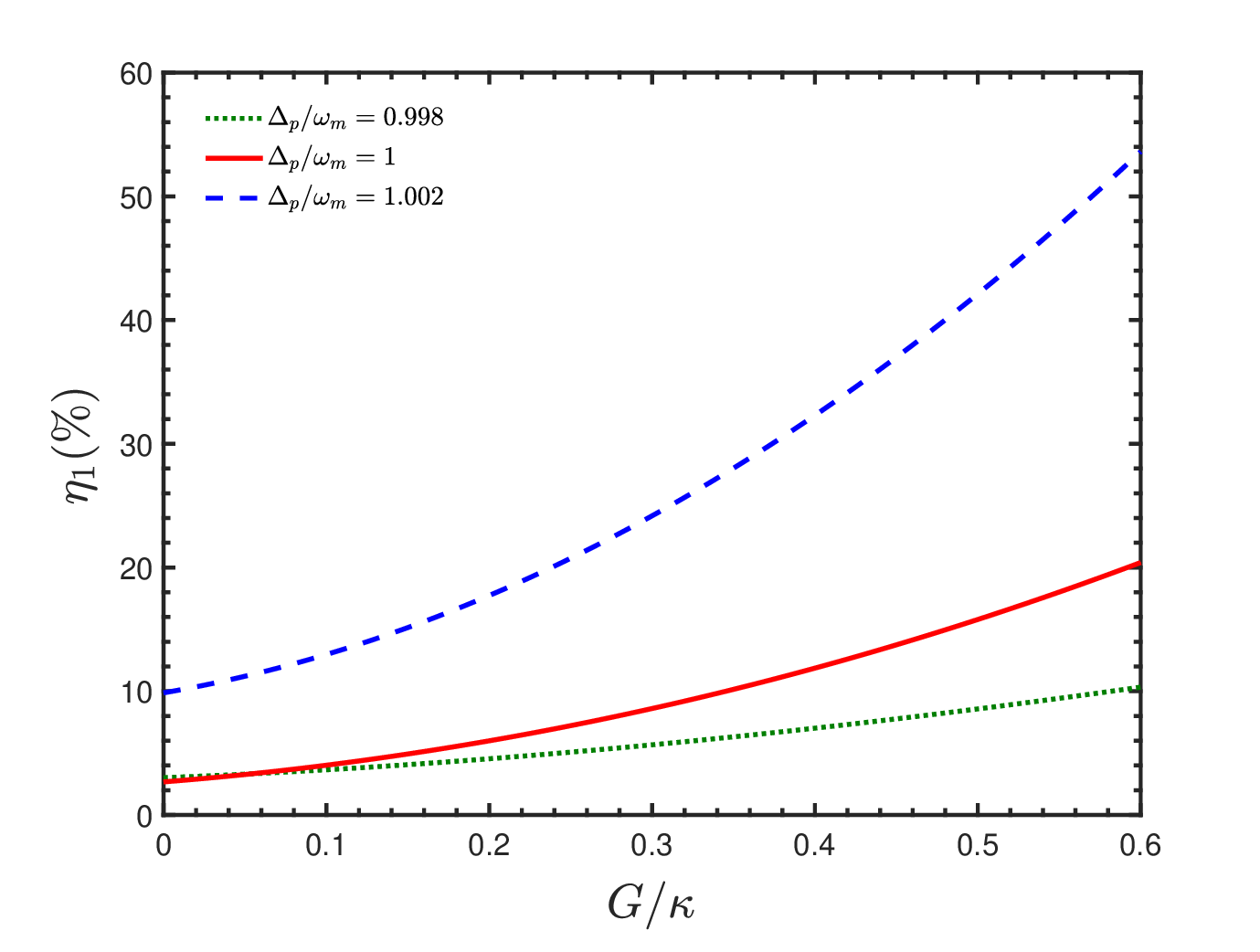}}

\caption{The efficiency ${\eta _1}$ of the second-order upper sideband generation as a function of the nonlinear gain $G$ of OPA for different probe-pulsed detuning ${\Delta _p}$, where $\theta=0$ and $\Omega=0$. Other parameters are the same as  Fig.~\ref{Eta1}.}\label{BianG}
\end{figure}

To clearly see the influence of the nonlinear gain $G$ of OPA on the second-order sideband generation, the efficiency ${\eta _1}$ is investigated as a function of the nonlinear gain $G$ for different probe-pulsed detuning ${\Delta _p}$, as shown in Fig.~\ref{BianG}. In detail, when $G$ increases from $0$ to $0.6\kappa $ in the case of ${\Delta _p}/{\omega _m} = 1.002$, the system provides an enhancement of more than five times for the sideband efficiency ${\eta _1}$. In general, with the nonlinear gain $G$ increasing, the efficiency ${\eta _1}$ of the second-order upper sideband generation increases obviously. The reason is that when the OPA is pumped at $\omega_g={\omega _l} + {\omega _p}$, i.e., twice the frequency of the anti-Stokes field, the parametric frequency conversion between this anti-Stokes field and phonon mode can provide another way to generate an optical second-order sideband, leading to the enhancement of a second-order sideband.

\section{Tunable slow and fast light}
%figure6
\begin{figure}[h]
\centerline{
\includegraphics[width=9cm, height=8cm, clip]{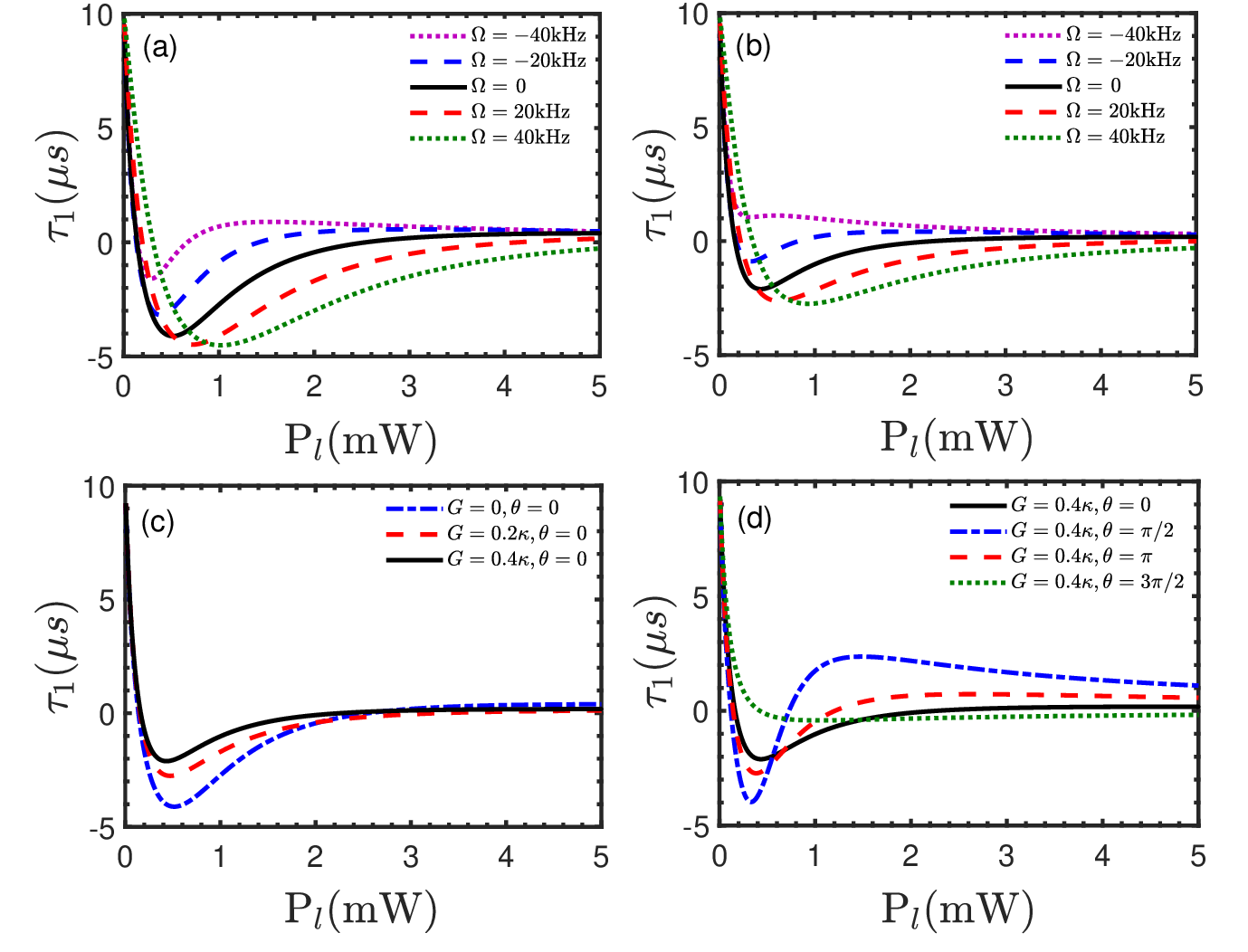}}
\caption{Optical group delay of the second-order upper sideband ${\tau _1}$ is plotted as a function of ${P_l}$ with different values of $\Omega $ and incident directions of light (a) without OPA and (b) in the presence of OPA effect at $G = 0.4\kappa $ and $\theta=0$. ${\tau _1}$ is plotted as a function of ${P_l}$ with different (c) nonlinear gain $G$ and (d) the phase $\theta $ of the field driving the OPA, where $\Omega=0$. Other parameters are the same as  Fig.~\ref{Eta1}.}\label{Tau2}
\end{figure}

We know the slow light effect is an important result of OMIT, which can be described by the optical group delay \cite{Safavi47269,He351649,Li635090,Mirza2725515,Liao11698}. It is similar to that of electromagnetically induced transparency, in the region of the narrow transparency window the rapid phase dispersion can cause the group delay
%expressed as
%${\tau _1} = {\left. {\frac{{d{\kern 1pt} \arg \left({ - \frac{{\sqrt {{\kappa _{ex}}} A_2^ + }}{{{\varepsilon _p}}}}\right)}}{{2 d{\Delta _p}}}} \right|_{{\Delta _p} = {\omega _m}}}$
given by Eq.~(\ref{tau1}). A positive group delay (${\tau _1} > 0$) corresponds to slow light propagation and a negative group delay (${\tau _1} < 0$) denotes fast light propagation.

In the previous work \cite{Safavi47269,Milonni}, it has been demonstrated that the delay of the transmitted light is only relevant to the pump power in a conventional optomechanical system. In our model, we see clearly from Fig.~\ref{Tau2} that the delay time of the second-order upper sideband can be adjusted not only by tuning the speed and direction of rotation of the resonator but also by adjusting the nonlinear gain of the OPA and phase of the field driving the OPA. In Fig.~\ref{Tau2}(a) and (b), we investigate the group delay of the second-order upper sideband ${\tau _1}$ as a function of control laser power ${P_l}$ for different $\Omega $. We find that when the resonator is stationary ($\Omega  = 0$), with the power increasing, ${\tau _1}$ tends to advance and even switches into fast light. However, in the presence of resonator rotation, the delay time of the second-order upper sideband will be prolonged at high control powers, which is useful for storage. In detail, as shown in Fig.~\ref{Tau2}(a), for a resonator speed of $20$ kHz, the group delay time ${\tau _1}$ increases when the resonator is driven from the right side ($\Omega  = - 20$ kHz) and decreases when the resonator is driven from the left side ($\Omega  = 20$ kHz). The group delay can still reach the conversion from fast light to slow light at this point. Increasing the resonator speed to $40$ kHz, at high control power, when the resonator is driven from the right side ($\Omega  =  - 40$ kHz), the group delay of the second-order sideband is always positive, i.e., slow light is obtained. When the resonator is driven from the left side ($\Omega  = 40$ kHz), the group delay is always negative and fast light can be obtained. At this point, the switching between fast and slow light disappears. In Fig.~\ref{Tau2}(b), we show the results of group delay ${\tau _1}$ versus control laser power $P_l$ in the presence of OPA. In the low power range, the addition of OPA increases the value of ${\tau _1}$. More interestingly, at $\Omega  =  - 40$ kHz, the fast and slow light conversion behavior of the group delay disappears, where only slow light effect is obtained.

Now we discuss the influence of the presence of OPA on the delay time of the second-order sideband. In Fig.~\ref{Tau2}(c) and (d), we display the group delay ${\tau _1}$ as a function of the control power ${P_l}$ for different parameters of nonlinear gain $G$ and phase $\theta $ of the field driving the OPA, where the resonator is stationary. When the OPA is considered in the optomechanical system, as is expected, the delay time of the second-order upper sideband generation obviously increases with the increasing power. With the nonlinear gain $G$ increasing from $0$ to $0.4\kappa $, the group delay ${\tau _1}$ accordingly increases, while the trend of switching between fast and slow light effects remains unchanged. In Fig.~\ref{Tau2}(d), we see that the ${\tau _1}$ is sensitive to the variation of the phase of the OPA. When $\theta  = \pi /2$, ${\tau _1}$ exhibits a significant transition from fast to slow light, in other words the delay time significantly reduces at low power and increases at high power. Interestingly, for $\theta  = 3\pi /2$, the valley of the ${\tau _1}$ disappears in the low power range, where the group delay exhibits a fast light effect (${\tau _1} < 0$) in the high power range. Physically, from Eq.~(\ref{jie12}), when the OPA is added inside the optomechanical coupled system, the quantum interference effect between the probe field and second-order sideband process is related directly to the phase of the OPA, so that the optical-response properties for the probe field become phase-sensitive.

%figure7
\begin{figure}[h]
\centerline{
\includegraphics[width=9cm, height=6.8cm, clip]{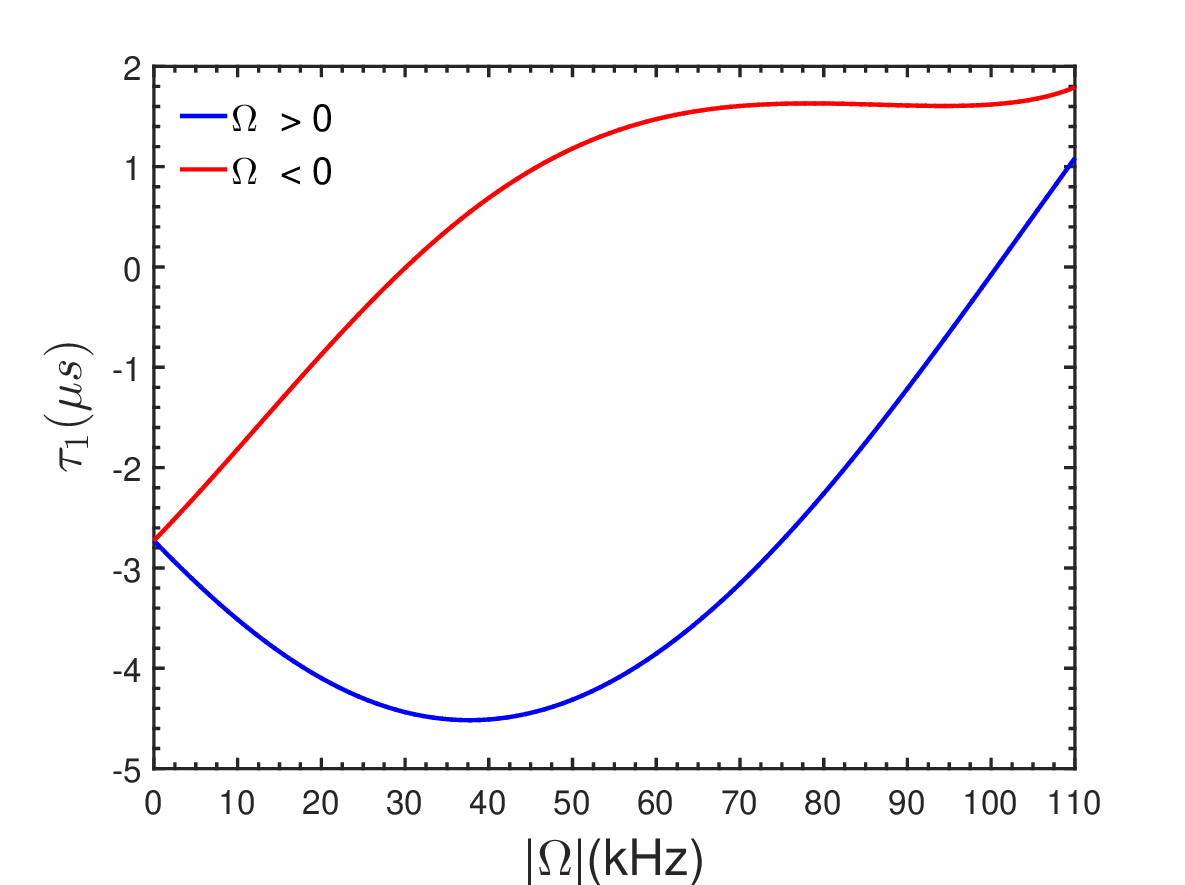}}
\caption{The group delay of the second-order upper sideband ${\tau _1}$ varies with the spinning angular velocity $\left| \Omega  \right|$ at $\Omega  > 0$ and $\Omega  < 0$, where $G = 0.4\kappa $ and $\theta=0$. The power of the control field ${P_l}$ is $1$ mW. Other parameters are the same as  Fig.~\ref{Tau2}.}\label{Tau_omega}
\end{figure}

As shown in Fig.~\ref{Tau_omega}, the group delay ${\tau _1}$ varies with the rotation speed of the resonator $\left| \Omega  \right|$ at a fixed control power, where the red sideband ${{\Delta _p} = {\omega _m}}$ is also presented. We find the group delay can achieve the transition from fast to slow light regardless of the direction of incidence of the input fields but with very significant differences. If $\Omega  > 0$ (the driving fields come from the left-hand side of the fiber), when the rotation speed reaches $101$ kHz, the group delay ${\tau _1}$ experiences the conversion from ${\tau _1} < 0$ to ${\tau _1} > 0$. However, if $\Omega  < 0$ (driving from the right-hand side of the fiber), when the rotation speed reaches $30$ kHz, ${\tau _1}$ experiences the conversion from ${\tau _1} < 0$ to ${\tau _1} > 0$. Therefore, we realize the conversion between the fast light and slow light by controlling the incident direction of the input fields in the spinning system.

%figure8
\begin{figure}[t]
\centering
\begin{minipage}{0.5\textwidth}
\includegraphics[width=1\textwidth]{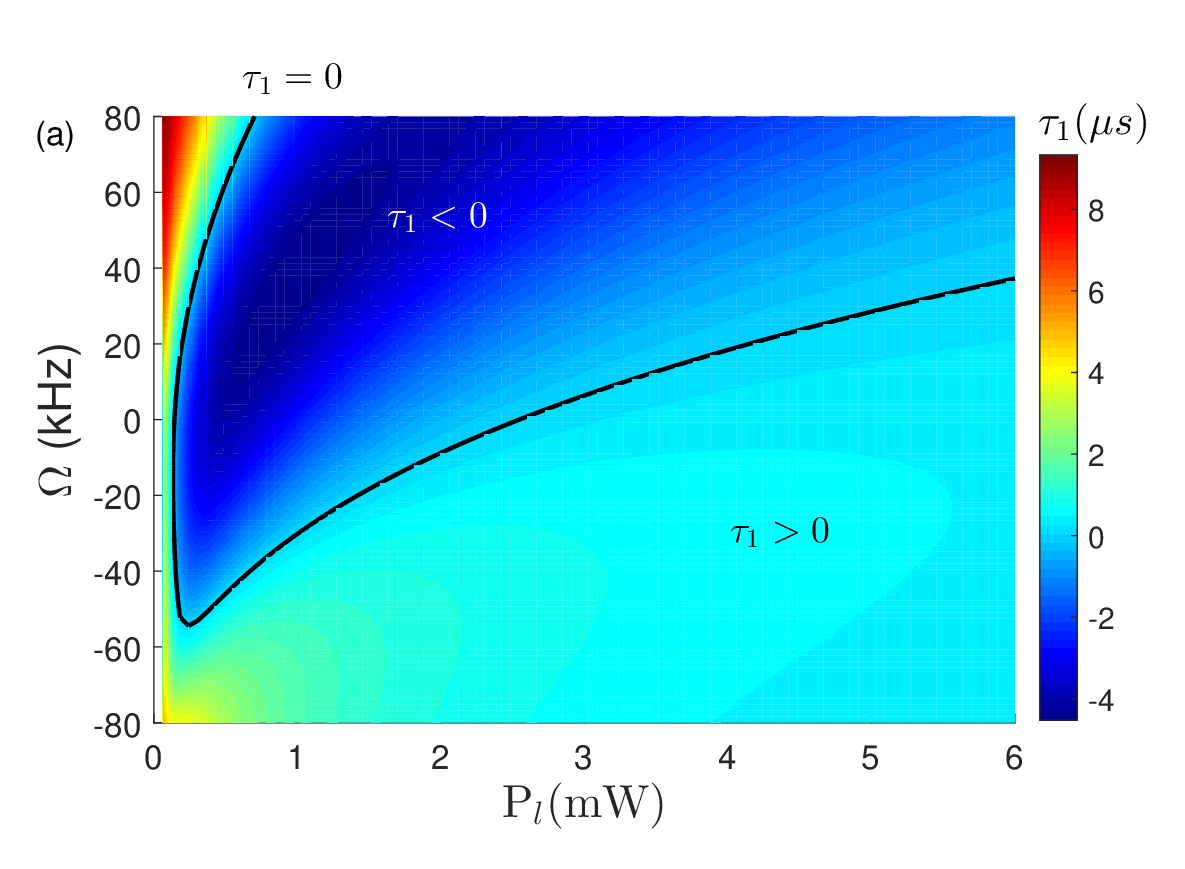}
\end{minipage}

\begin{minipage}{0.5\textwidth}
\includegraphics[width=1\textwidth]{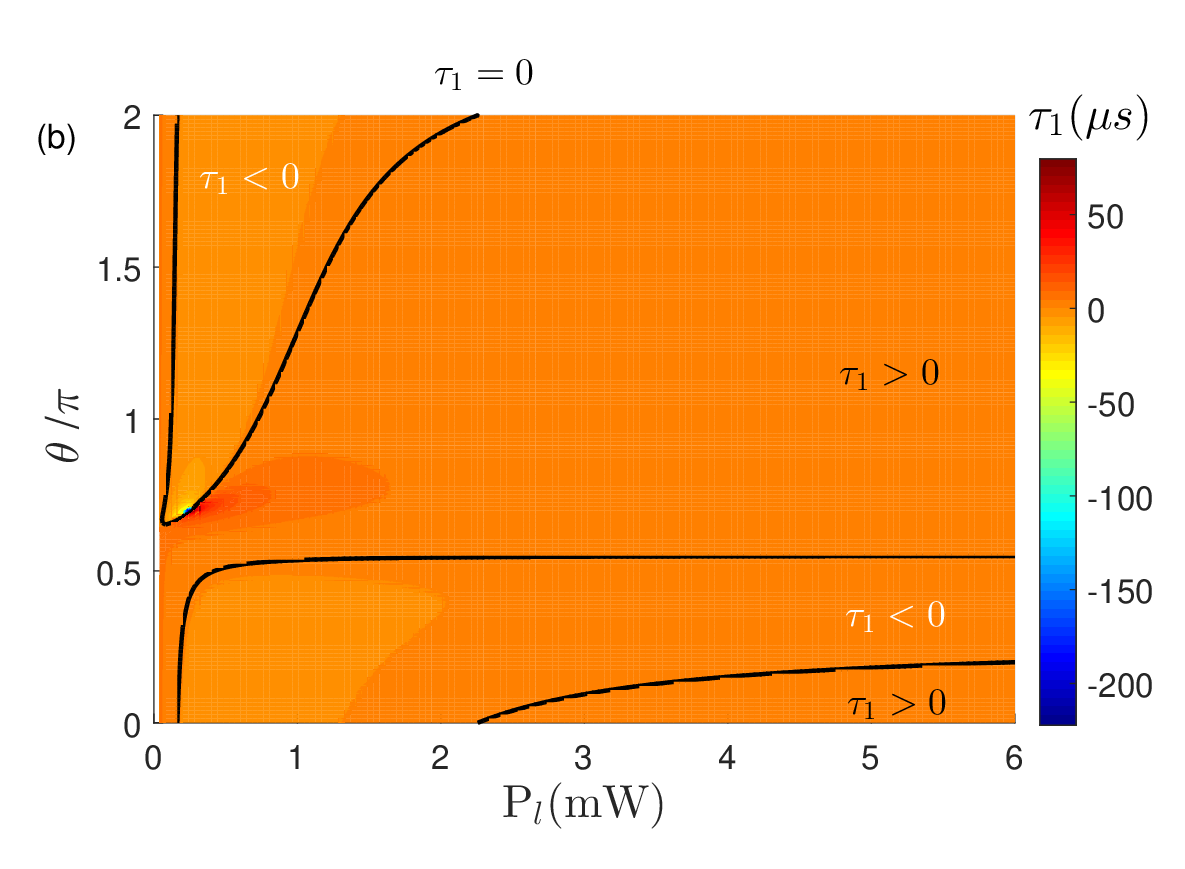}
\end{minipage}

\caption{(a) The group delay of the second-order upper sideband ${\tau _1}$ varies with ${P_l}$ and $\Omega $, where $G = 0$ and $\theta=0$. (b) ${\tau _1}$ varies with ${P_l}$ and $\theta $ at $G = 0.4\kappa $ and $\Omega=0$. The black curves correspond to ${\tau _1} = 0$. Other parameters are the same as  Fig.~\ref{Tau2}.}\label{qyc3weipltheta}
\end{figure}
In the above discussion, we see that the group delay of the second-order upper sideband is sensitive to the variation of the rotation speed of the resonator, the direction of incidence of the input fields, and the phase of the field driving the OPA. In Fig.~\ref{qyc3weipltheta}(a), the group delay  ${\tau _1}$ of the second-order upper sideband is plotted as the function of control power ${P_l}$ and the rotation speed of the resonator $\Omega $. In Fig.~\ref{qyc3weipltheta}(b),  ${\tau _1}$ is plotted as the function of control power ${P_l}$ and the phase $\theta $ of the field driving the OPA. The black curves correspond to ${\tau _1} = 0$. In this case, we can obtain the slow light effect or fast light effect by properly selecting the values of ${P_l}$, $\Omega $, and $\theta $. Moreover, a tunable switch from fast to slow light can be realized by adjusting their values.

\section{Influence of changing the driving frequency of OPA on the efficiency}
The optical degenerate parametric amplifier (OPA), a second-order optical crystal in nature, can generate pairs of down-converted photons and show nearly perfect single or dual squeezing \cite{Gerry,Li100023838,Clerk821155,Nation841,Leghtas347853,Shen100023814}. As we all know, placing an OPA pumped by an external laser in the optomechanical cavity can modulate the optomechanical coupling, which can lead to optical amplification directly  \cite{Adamyan92053818}. We can discuss the influence of different pump frequencies of the driving OPA on the sidebands and compare the amplification of the second-order sidebands in both cases. Now, we vary the frequency of the laser field driving the OPA, so that the OPA is excited by a pump drive with the frequency $\omega _g=2{\omega _l}$ \cite{Li100023838} in Fig.~\ref{model}(d). The pump photon with frequency $\omega_g=2{\omega _l}$ is down-converted into
an identical pair of photons with frequency ${\omega _l}$ after passing through the second-order nonlinearity crystal. ${{\hat H}_{OPA}}$ reads
\begin{equation}\label{Hopa}
{\hat H_{OPA}} = i\hbar G({\hat a^{\dag 2}}{e^{i\theta }}{e^{ - 2i{\omega _l}t}} - H.c.).
\end{equation}
The total Hamiltonian of the system in the rotating frame at the laser frequency ${{\omega _l}}$ is given by
\begin{equation}\label{Heff1}
\begin{aligned}
{{\hat H}_{eff}} =&\hbar \left( {{\Delta _0} - \xi \hat x + {\Delta _s}} \right){{\hat a}^\dag }\hat a + \frac{{{{\hat p}^2}}}{{2m}} + \frac{1}{2}m\omega _m^2{{\hat x}^2}\\
 &+\frac{{\hat p_\phi ^2}}{{2m{{\left( {R + \hat x} \right)}^2}}} + i\hbar G({\hat a^{\dag 2}}{e^{i\theta }} - H.c.)\\
 &+ i\hbar \sqrt {{\kappa _{ex}}} \left[ {\left( {{\varepsilon _l} + {\varepsilon _p}{e^{ - i{\Delta _p}t}}} \right){{\hat a}^\dag } - H.c.} \right].
\end{aligned}
\end{equation}
We can get the equations of motion
\begin{align}
&\dot a =  - \left[ {\kappa  + i\left( {{\Delta _0} - \xi x + {\Delta _s}} \right)} \right]a \nonumber \\
&\qquad + \sqrt {{\kappa _{ex}}} \left( {{\varepsilon _l} + {\varepsilon _p}{e^{ - i{\Delta _p}t}}} \right)+2G{e^{i\theta }}{a^*},\label{Heq12}\\
&m(\ddot x + {\Gamma _m}\dot x + \omega _m^2x) = \hbar \xi {a^ * }a + \frac{{p_\phi ^2}}{{m{R^3}}},\label{Heq22}\\
&\dot \phi  = \frac{{{p_\phi }}}{{m{R^2}}},\label{Heq32}\\
&{{\dot p}_\phi } = 0,\label{Heq42}
\end{align}
where we write the operators for their expectation values by the mean-field approximation. The steady-state solutions of the system are obtained as
\begin{equation}
\begin{split}
  {\tilde a_s}  &= \frac{{2G{e^{i\theta }} + \kappa  - i\tilde \Delta  }}{{{\kappa ^2} + {{\tilde \Delta}  ^2} - 4{G^2}}}, \\
  {{\tilde x}_s}  &= \frac{{\hbar \xi {{\left| {{\tilde a_s} } \right|}^2}}}{{m\omega _m^2}} + R{\left( {\frac{\Omega }{{{\omega _m}}}} \right)^2},\label{wentaijie1}
\end{split}
\end{equation}
where $\tilde \Delta {\rm{ = }}{\Delta _0} - \xi {{\tilde x}_s} + {\Delta _s}$. It is worth noting that here, unlike Eq.~(\ref{wentaijie}), the intracavity photon number ${\left| {{{\tilde a}_s}} \right|^2}$ and displacement of mechanical oscillator ${{\tilde x}_s}$ strongly depend on the magnitude of nonlinear gain $G$ and phase $\theta $ of the OPA.
Eqs.~(\ref{Heq12})-(\ref{Heq42}) can be solved analytically with the linearized ansatz
\begin{small}
\begin{equation}
\begin{aligned}
a=&{\tilde a_s} + \tilde A_1^ + {e^{ - i{\Delta _p}t}} + \tilde A_1^ - {e^{i{\Delta _p}t}} + \tilde A_2^ + {e^{ - 2i{\Delta _p}t}} + \tilde A_2^ - {e^{2i{\Delta _p}t}},\\
x=&{{\tilde x}_s} + \tilde X_1^ + {e^{ - i{\Delta _p}t}} + \tilde X_1^ - {e^{i{\Delta _p}t}} + \tilde X_2^ + {e^{ - 2i{\Delta _p}t}} + \tilde X_2^ - {e^{2i{\Delta _p}t}}.
\nonumber
\end{aligned}
\end{equation}
\end{small}
After the ansatz, we obtain six algebra equations, which can be divided into two groups
\begin{equation}
\begin{split}
{{\tilde \sigma }_1}\left( {{\Delta _p}} \right)\tilde A_1^ +  &= i\xi {{\tilde a}_s}\tilde X_1^ +  + 2G{e^{i\theta }}\tilde A_1^{ - *} + \sqrt {{\kappa _{ex}}} {\varepsilon _p},\\
{{\tilde \sigma }_2}\left( {{\Delta _p}} \right)\tilde A_1^{ - *} &=  - i\xi \tilde a_s^*\tilde X_1^ +  + 2G{e^{-i\theta }}\tilde A_1^ + , \\
\chi \left( {{\Delta _p}} \right)\tilde X_1^ +  &= \hbar \xi ({{\tilde a}_s}\tilde A_1^{ - *} + \tilde a_s^*\tilde A_1^ + ),\label{first2}
\end{split}
\end{equation}
and
\begin{equation}
\begin{split}
{{\tilde \sigma }_1}\left( {2{\Delta _p}} \right)\tilde A_2^ +  &= i\xi ({{\tilde a}_s}\tilde X_2^ +  + \tilde A_1^ + \tilde X_1^ + ) + 2G{e^{i\theta }}\tilde A_2^{ - *},\\
{{\tilde \sigma }_2}\left( {2{\Delta _p}} \right)\tilde A_2^{ - *} &=  - i\xi (\tilde a_s^*\tilde X_2^ +  + \tilde A_1^{ - *}\tilde X_1^ + ) + 2G{e^{-i\theta }}\tilde A_2^ + , \\
\chi \left( {2{\Delta _p}} \right)\tilde X_2^ +  &= \hbar \xi ({{\tilde a}_s}\tilde A_2^{ - *} + \tilde a_s^*\tilde A_2^ +  + \tilde A_1^ + \tilde A_1^{ - *}),\label{second2}
\end{split}
\end{equation}
with
\begin{eqnarray}
{{\tilde \sigma }_1}\left( {n{\Delta _p}} \right) &=& \kappa  + i{\tilde \Delta }  - in{\Delta _p},\nonumber\\
{{\tilde \sigma }_2}\left( {n{\Delta _p}} \right) &=& \kappa  - i{\tilde \Delta }  - in{\Delta _p},\nonumber\\
\chi \left( {n{\Delta _p}} \right) &=& m(\omega _m^2 - i{\Gamma _m}n{\Delta _p} -\Delta _p^2).\nonumber
\end{eqnarray}

We get the linear and second-order nonlinear responses of the system
\begin{equation}
\begin{split}
\tilde A_1^ +  &= \frac{{\tilde D + {{\tilde \sigma }_2}({\Delta _p})\chi ({\Delta _p})}}{{{{\tilde f}_4}({\Delta _p}) + {{\tilde f}_3}({\Delta _p})}}\sqrt {{\kappa _{ex}}} {\varepsilon _p},\\
\tilde X_1^ +  &= \frac{{\hbar \xi [2G{e^{-i\theta }}{{\tilde a}_s} + \tilde a_s^*{{\tilde \sigma }_2}({\Delta _p})]}}{{\tilde D + {{\tilde \sigma }_2}({\Delta _p})\chi ({\Delta _p})}}\tilde A_1^ +   ,\\
\tilde A_1^{ - *} &=\frac{{ - i\xi \tilde a_s^*}}{{{{\tilde \sigma }_2}({\Delta _p})}}\tilde X_1^ +  + \frac{{2G{e^{-i\theta }}}}{{{{\tilde \sigma }_2}({\Delta _p})}}\tilde A_1^ + ,\label{jie21}
\end{split}
\end{equation}
and
\begin{equation}
\begin{split}
\tilde A_2^ +  &= \frac{{i\hbar {\xi ^2}{{\tilde f}_6}\tilde A_1^ + \tilde A_1^{ - *} + {{\tilde f}_7}\tilde A_1^{ - *}\tilde X_1^ +  + i\xi {{\tilde f}_2}\tilde A_1^ + \tilde X_1^ + }}{{{{\tilde f}_4}\left( {2{\Delta _p}} \right) + {{\tilde f}_3}\left( {2{\Delta _p}} \right)}},\\
\tilde X_2^ +  &= \frac{{\hbar \xi [ {{{\tilde f}_5}\tilde A_2^ +  - i\xi {{\tilde a}_s}\tilde A_1^{ -  * }\tilde X_1^ +  + {{\tilde \sigma }_2}\left( {2{\Delta _p}} \right)\tilde A_1^ + \tilde A_1^{ - *}} ]}}{{{{\tilde f}_2}}},\\
\tilde A_2^ -  &= \frac{{i\xi }}{{{{\tilde \sigma }_2}{{\left( {2{\Delta _p}} \right)}^*}}}({{\tilde a}_s}\tilde X_2^ -  + \tilde A_1^ - \tilde X_1^ - ) + \frac{{2G{e^{ i\theta }}}}{{{{\tilde \sigma }_2}{{\left( {2{\Delta _p}} \right)}^*}}}\tilde A_2^{ +  * } ,\label{jie22}\\
\end{split}
\end{equation}
where
\begin{equation}
\begin{split}
\tilde D &= i\hbar {\xi ^2}{\left| {{{\tilde a}_s}} \right|^2},\nonumber\\
{{\tilde f}_2} &= \tilde D + {{\tilde \sigma }_2}\left( {2{\Delta _p}} \right)\chi \left( {2{\Delta _p}} \right),\nonumber\\
{{\tilde f}_3}\left( {n{\Delta _p}} \right) &= 2i\tilde D\Delta  + {{\tilde \sigma }_1}(n{\Delta _p}){{\tilde \sigma }_2}\left( {n{\Delta _p}} \right)\chi \left( {n{\Delta _p}} \right),\nonumber\\
{{\tilde f}_4}\left( {n{\Delta _p}} \right) &= 2i\hbar {\xi ^2}G\left( {\tilde a_s^{*2}{e^{i\theta }} - \tilde a_s^2{e^{ - i\theta }}} \right) - 4{G^2}\chi \left( {n{\Delta _p}} \right),\nonumber\\
{{\tilde f}_5} &= 2G{e^{ - i\theta }}{{\tilde a}_s} + \tilde a_s^*{{\tilde \sigma }_2}\left( {2{\Delta _p}} \right),\nonumber\\
{{\tilde f}_6} &=  - 2G{e^{i\theta }}\tilde a_s^* + \tilde a_s{{\tilde \sigma }_2}\left( {2{\Delta _p}} \right),\nonumber\\
{{\tilde f}_7} &= \hbar {\xi ^3}\tilde a_s^2 - 2i\xi G{e^{i\theta }}\chi \left( {2{\Delta _p}} \right).\nonumber\\
\end{split}
\end{equation}
We obtain the amplitude of the sidebands, which are substituted into the efficiency of the second-order upper sideband ${{\tilde \eta }_1} = | - {{\sqrt {{\kappa _{ex}}} \tilde A_2^ + } \mathord{/ {\vphantom {{\sqrt {{\kappa _{ex}}} \tilde A_2^ + } {{\varepsilon _p}}}}\kern-\nulldelimiterspace} {{\varepsilon _p}}}|$ and second-order lower sideband ${{\tilde \eta }_2} = | - \sqrt {{\kappa _{ex}}} \tilde A_2^ - /{\varepsilon _p}|$.

%figure9Here, $\theta  = 0$ and $G = 0.2\kappa $ are fixed in (a) and (b), respectively.
\begin{figure}[t]
\centerline{
\includegraphics[width=9.2cm, height=8cm, clip]{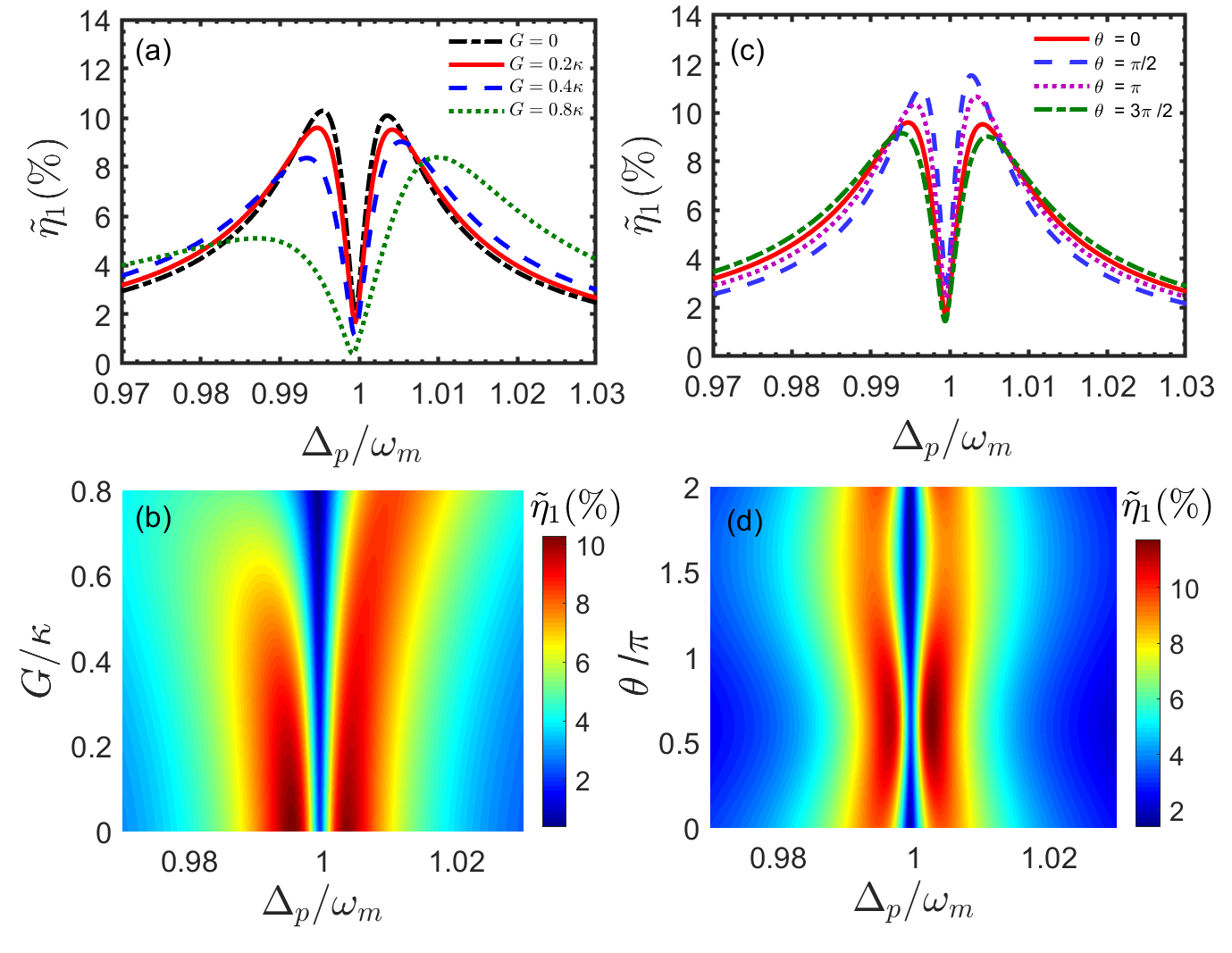}}
\caption{The efficiency ${{\tilde \eta }_1}$ of the second-order upper sideband generation as a function of the probe-pulsed detuning ${\Delta _p}$ for different (a) nonlinear gain $G$ under $\theta  = 0$ and (c) phase $\theta $ of the field driving the OPA under $G = 0.2\kappa $. (b) ${{\tilde \eta }_1}$ varies with ${\Delta _p}$ and $G$ under value $\theta  = 0$. (d) ${{\tilde \eta }_1}$ varies with ${\Delta _p}$ and $\theta $ under $G = 0.2\kappa $. The resonator is stationary ($\Omega  =  0$). Other parameters are the same as  Fig.~\ref{Eta1}.} \label{2wl_Gtheta}
\end{figure}
To illustrate the different influences on the second-order sidebands of the OPA excited by a pump drive of frequency $\omega_g=2{\omega _l}$, the efficiency of the second-order upper sideband generation with the resonator stationary is investigated as a function of frequency ${\Delta _p}/{\omega _m}$ in Fig.~\ref{2wl_Gtheta}. As shown in Fig.~\ref{2wl_Gtheta}(a), in the absence of the OPA, the efficiency ${{\tilde \eta }_1}$ possesses two near-symmetrical peaks and a local minimum near the resonance condition ${\Delta _p}/{\omega _m} = 1$. When $G \ne 0$, with the nonlinear gain $G$ of OPA increasing, the peak of efficiency ${{\tilde \eta }_1}$ decreases gradually. But in the driven frequency ${\Delta _p}$ range away from the resonance condition ${\Delta _p} = {\omega _m}$, such as ${\Delta _p} > 1.01{\omega _m}$, the efficiency ${{\tilde \eta }_1}$ is enhanced. Moreover, it is noted that the larger the nonlinear gain $G$ of OPA is, the wider the linewidth of the suppressive windows of the efficiency ${{\tilde \eta }_1}$ is. Due to the presence of OPA, the suppressive window will be asymmetric. The result can be applied to determining the excitation number of atoms and plays important roles in nonlinear media in the optical properties of the output field.
Interestingly, when $G$ increases to $G = 0.8\kappa $, a clear asymmetric linear pattern of the efficiency ${{\tilde \eta }_1}$ emerges, with a much larger peak at ${\Delta _p} = 1.01{\omega _m}$ than at ${\Delta _p} = 0.987{\omega _m}$. In Fig.~\ref{2wl_Gtheta}(c), we discuss the efficiency ${{\tilde \eta }_1}$ under different phase $\theta $ of the field driving the OPA. We find that the phase $\theta $ amplifies the efficiency of the second-order sideband generation, so that the peak of ${{\tilde \eta }_1}$ increases from $9.52\% $ to $11.53\% $ for $\theta  = \pi /2$. This is due to the fact that the degenerate parametric amplifier is a phase-sensitive amplifier, where the phase relationship between the control laser and signal laser driving the degenerate parametric amplifier determines the direction of the energy flow, i.e., whether the signal light is effectively amplified or not.
In Fig.~\ref{2wl_Gtheta}(b) and (d), ${{\tilde \eta }_1}$ as a function of the detuning ${\Delta _p}$ and phase $\theta $ of the field driving the OPA is shown. We can see that the efficiency of the second-order sideband generation is sensitive to both the nonlinear gain $G$ and phase $\theta $ changes of the OPA. When ${\Delta _p} \in ( {{\omega _m},1.02{\omega _m}} )$, the influence of the $G$ and $\theta $ on the efficiency ${{\tilde \eta }_1}$ becomes more obvious. Specially, as shown in Fig.~\ref{2wl_Gtheta}(d), it can be found that at $\theta  \in ( {0,1.28\pi } )$, the efficiency ${{\tilde \eta }_1}$ is amplified. When $\theta  = 0.64\pi $ and ${\Delta _p} = 1.003{\omega _m}$, ${{\tilde \eta }_1}$ obtains the maximum value $11.73\% $.

%figure10 Here, $\theta  = 0$ and $G = 0.2\kappa $ are fixed in (a) and (b), respectively.
\begin{figure}[t]
\centerline{
\includegraphics[width=9.2cm, height=8cm, clip]{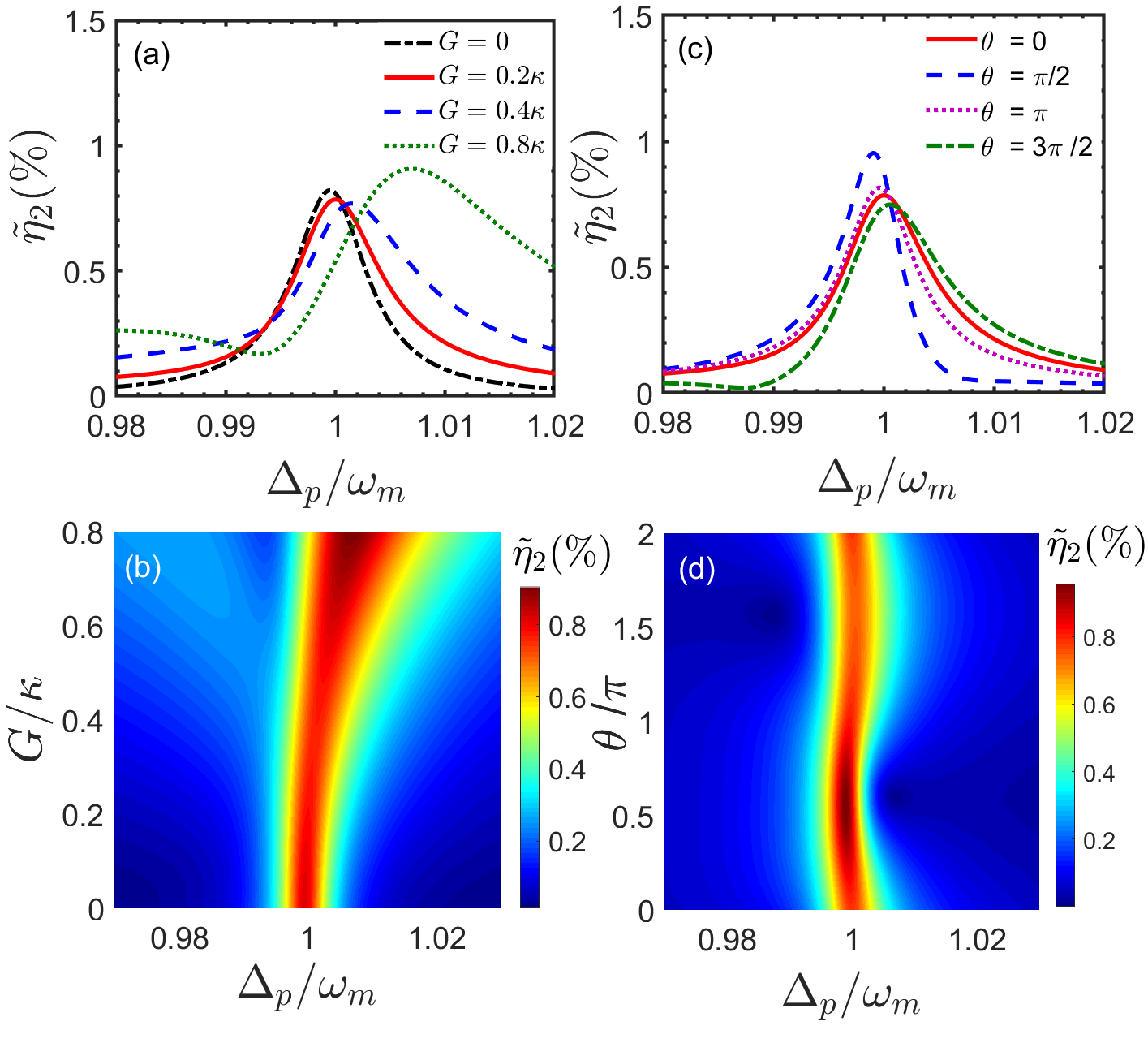}}
\caption{The efficiency ${{\tilde \eta }_2}$ of the second-order lower sideband generation as a function of the probe-pulsed detuning ${\Delta _p}$ for different (a) nonlinear gain $G$ under $\theta  = 0$ and (c) phase $\theta $ of the field driving the OPA under $G = 0.2\kappa $. (b) ${{\tilde \eta }_2}$ varies with ${\Delta _p}$ and $G$ under value $\theta  = 0$. (d) ${{\tilde \eta }_2}$ varies with ${\Delta _p}$ and $\theta $ under $G = 0.2\kappa $. The resonator is stationary ($\Omega  =  0$). Other parameters are the same as  Fig.~\ref{Eta1}.} \label{2wleta2_Gtheta}
\end{figure}
Next, we discuss the influence of the OPA on the second-order lower sideband efficiency ${{\tilde \eta }_2}$. In Fig.~\ref{2wleta2_Gtheta}(a) and (c), we can see that both $G$ and $\theta $ change the peak of ${{\tilde \eta }_2}$ (The detailed results refer to Fig.~\ref{2wleta2_Gtheta}(b) and (d)). As $G$ increases, the position of the peak shifts to the right, i.e., a larger value of ${\Delta _p}$ is needed to bring ${{\tilde \eta }_2}$ to its maximum. In particular, when $G = 0.8\kappa $, ${{\tilde \eta }_2}$ appears as a local minimum at ${\Delta _p} = 0.993{\omega _m}$. As shown in Fig.~\ref{2wleta2_Gtheta}(d), ${{\tilde \eta }_2}$ is amplified when $\theta  \in \left( {0,1.14\pi } \right)$, which obtains the maximum value of $0.95\% $.
In general, when the pump laser frequency driving the OPA is $\omega_g=2{\omega _l}$, the nonlinear gain $G$ of the OPA is not significant for the amplification of the second-order upper and lower sidebands. Compared with the case, where the pump laser frequency driving OPA is $\omega_g={{\omega _l} + {\omega _p}}$, $G$ can change the linewidth of the suppressive window of ${{\tilde \eta }_2}$ and localization of the sideband efficiency maximum.

%figure11  In this case, $G = 0.2\kappa $ is fixed in (c) and (d).
\begin{figure}[t]
\centerline{
\includegraphics[width=9.2cm, height=8cm, clip]{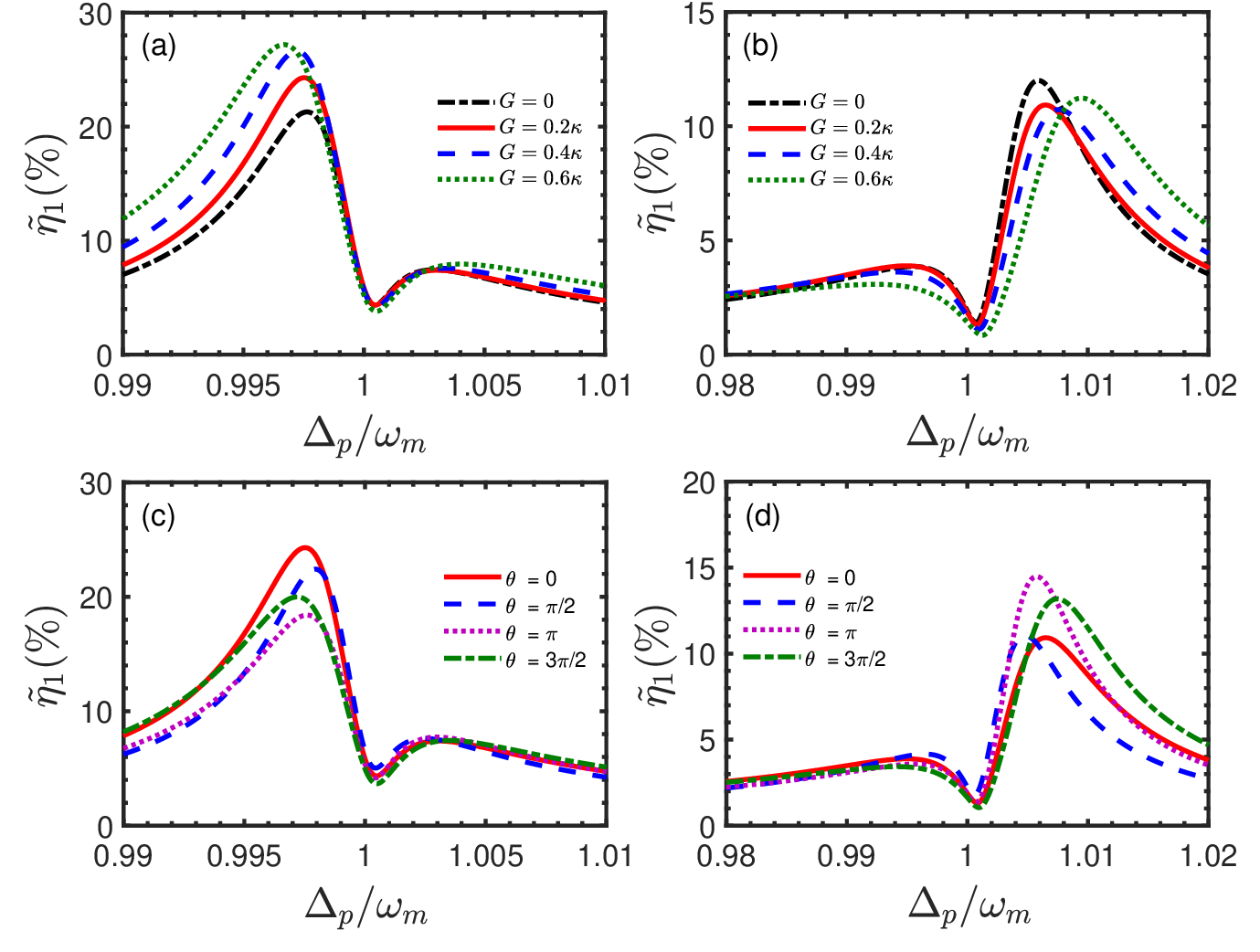}}
\caption{The efficiency ${{\tilde \eta }_1}$ of the second-order upper sideband generation as a function of the probe-pulsed detuning ${\Delta _p}$ for different $G$ with $\theta  = 0$ at (a) $\Omega = 20$ kHz and (b) $\Omega = -20$ kHz. The efficiency ${{\tilde \eta }_1}$ as a function of the probe-pulsed detuning ${\Delta _p}$ for different $\theta $ with $G = 0.2\kappa $ at (c) $\Omega = 20$ kHz and (d) $\Omega = -20$ kHz. Other parameters are the same as  Fig.~\ref{Eta1}.} \label{2wl_omega}
\end{figure}
As shown in Fig.~\ref{2wl_omega}, we discuss the influence of the OPA on the second-order upper sideband generation when the resonator is rotating. In Fig.~\ref{2wl_omega}(a), it can be seen that when the system is driven from the left-hand side ($\Omega  =  20$ kHz), the increase of the nonlinear gain $G$ of the OPA enhances the second-order sideband peak. However, the effect of the OPA in the transmission window (near ${\Delta _p}/{\omega _m} = 1$) is small, while at ${\Delta _p}/{\omega _m} < 0.996$ and ${\Delta _p}/{\omega _m} > 1.004$, the OPA has a significant enhancement effect. In Fig.~\ref{2wl_omega}(b), we find that when the system is driven from the right-hand side ($\Omega = -20$ kHz), changing the nonlinear gain $G$ cannot enhance the second-order sideband peak. But the increase in the nonlinear gain $G$ of the OPA still makes the linewidth of the efficiency ${{\tilde \eta }_1}$ broaden. In Fig.~\ref{2wl_omega}(c) and (d), ${{\tilde \eta }_1}$ as a function of detuning ${\Delta _p}$ for the different $\theta $ at $G = 0.2\kappa $ is plotted. In this case, $\Omega  = 20$ kHz and $\Omega  = - 20$ kHz are fixed in Fig.~\ref{2wl_omega}(c) and (d), respectively. In detail, the second-order sideband peak is significantly enhanced when $\theta  = \pi $ at $\Omega  = - 20$ kHz, but decreased at $\Omega = 20$ kHz.

%figure12
\begin{figure}[t]
\centerline{
\includegraphics[width=9.5cm, height=9.5cm, clip]{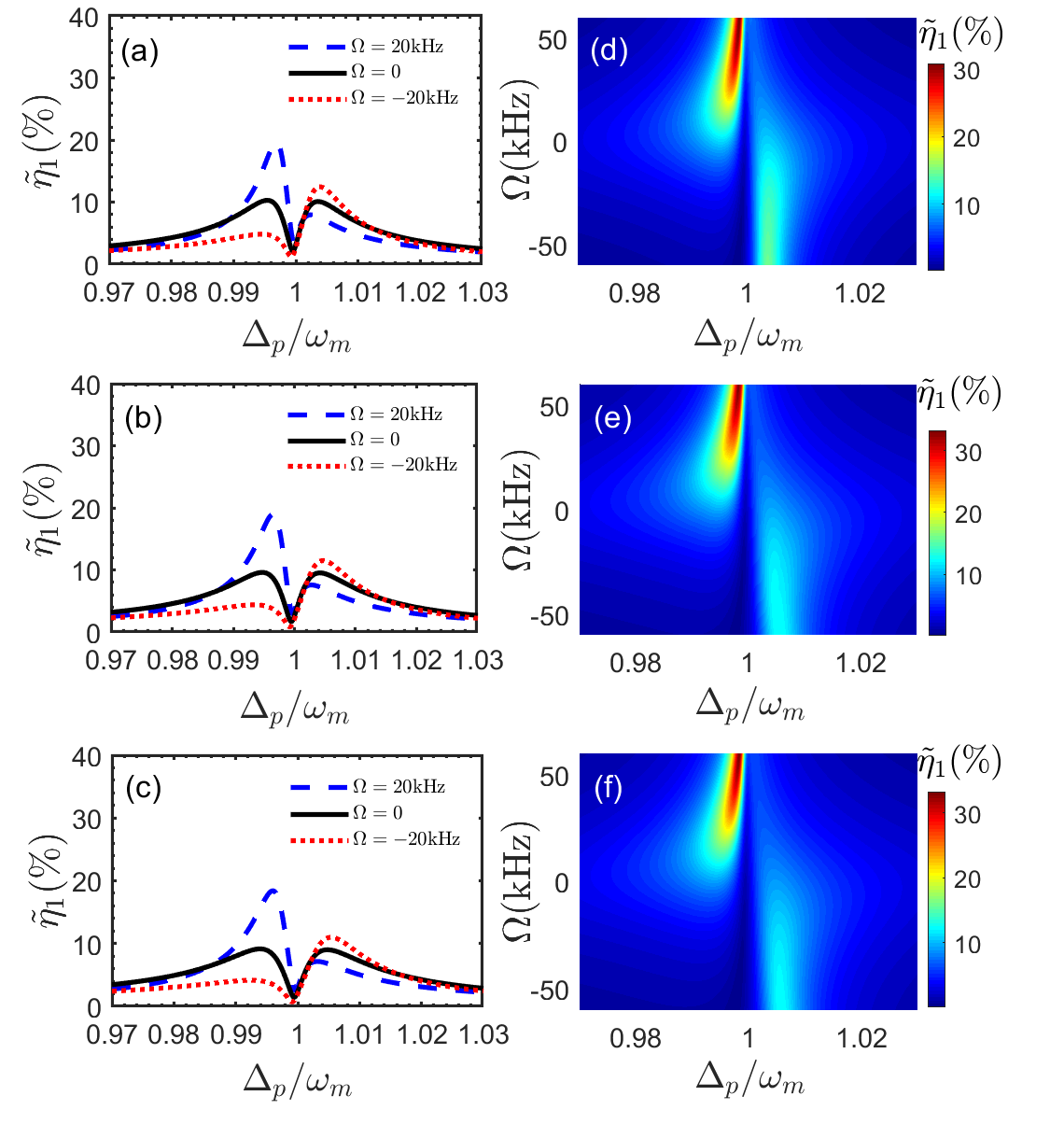}}
\caption{The efficiency ${{\tilde \eta }_1}$ of the second-order upper sideband generation as a function of ${\Delta _p}$ under different values of $\Omega $ and incident directions of light, where the nonlinear gain and phase of the probe field of the OPA are fixed as (a) $G = 0,\theta  = 0$; (b) $G = 0.2\kappa ,\theta  = 0$; (c) $G = 0.2\kappa ,\theta  = 3\pi /2$. ${\eta _1}$ varies with ${\Delta _p}$ and $\Omega $ under different values (d) $G = 0,\theta  = 0$; (e) $G = 0.2\kappa ,\theta = 0$; (f) $G = 0.2\kappa ,\theta  = 3\pi /2$. These parameters are the same as  Fig.~\ref{Eta1}.} \label{2wl_3w_G}
\end{figure}

%figure13
\begin{figure}[h]
\centerline{
\includegraphics[width=9.2cm, height=8cm, clip]{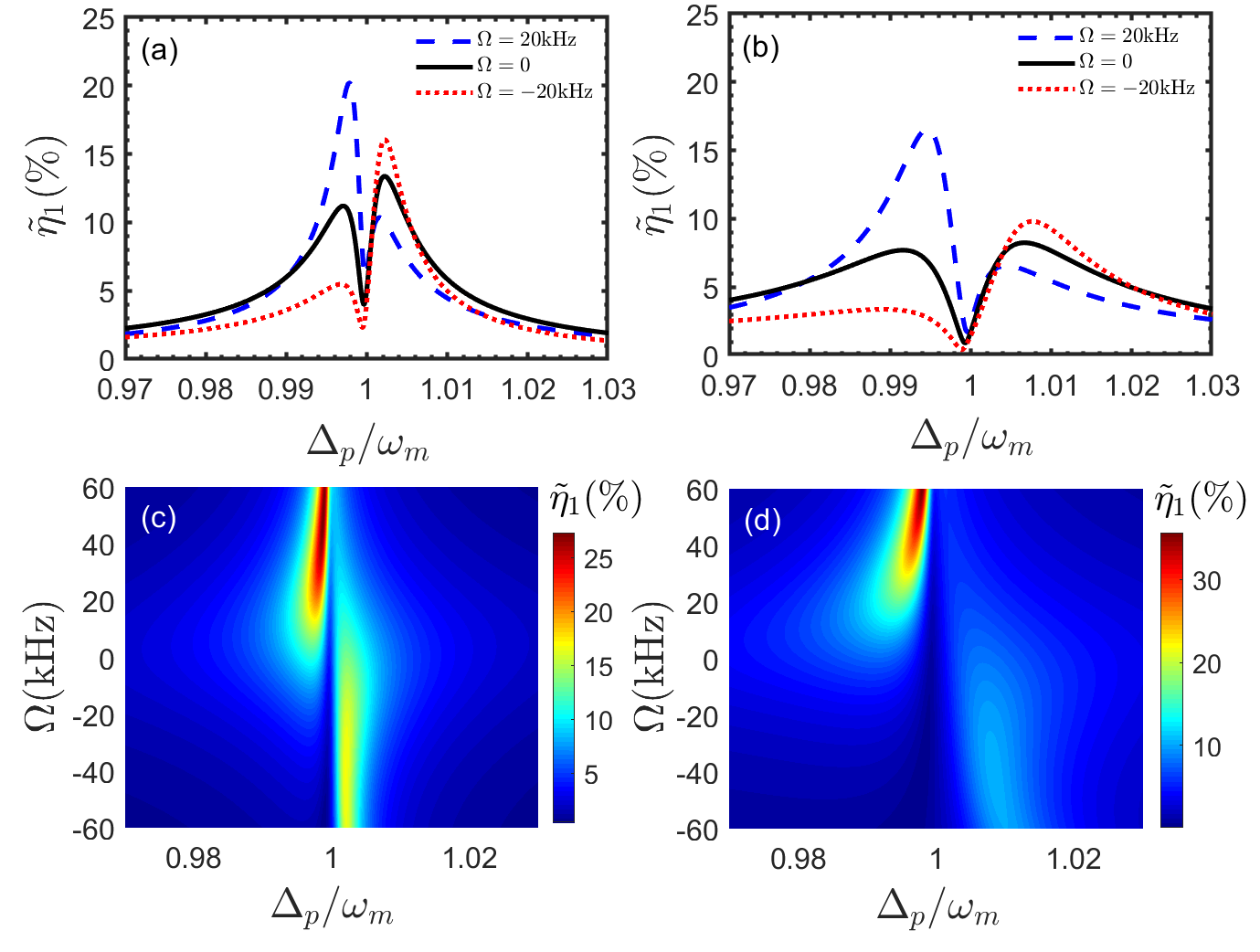}}
\caption{(a)(b) The efficiency ${{\tilde \eta }_1}$ of the second-order upper sideband generation as a function of ${\Delta _p}$ under different values of $\Omega $ and incident directions of light. (c)(d) ${{\tilde \eta }_1}$ varies with ${\Delta _p}$ and $\Omega $. The parameters chosen are (a)(c) $G = 0.4\kappa$, $\theta  = \pi /2$ and (b)(d) $G = 0.4\kappa $, $\theta  = 3\pi /2$. Other parameters are the same as Fig.~\ref{Eta1}.} \label{2wl_3w_Theta}
\end{figure}

In the above discussions, we note that when the frequency $\omega_g$ of the laser field driving the OPA is changed from ${{\omega _l} + {\omega _p}}$ to $2{\omega _l}$, the influence of the resonator speed, the direction of incidence of the input fields, the nonlinear gain of the OPA and phase of the field driving the OPA on the second-order sideband efficiency has a significant difference in the system. In Figs.~\ref{2wl_3w_G} and~\ref{2wl_3w_Theta}, we find in such a hybrid nonlinear system containing the OPA, the spinning-induced direction-dependent nonreciprocal behavior remains. We fix the clockwise speed of the resonator at $20$ kHz and vary the nonlinear gain $G$ and phase $\theta $ of the field driving the OPA, plotting ${{\tilde \eta }_1}$ as a function of ${\Delta _p}$ and $\Omega $ when the spinning system is driven from the left-hand side and right-hand side, respectively. In Fig.~\ref{2wl_3w_G}, we choose the same OPA gain as in Fig.~\ref{Eta1} to compare two different OPA cases ($\omega_g={{\omega _l} + {\omega _p}}$ and $\omega_g=2{\omega _l}$). When the control laser frequency driving the OPA is $\omega_g=2{\omega _l}$, changing the nonlinear gain $G$ can not enhance the second-order sideband peak. The efficiency of the second-order upper sideband is not sensitive to the variation of the nonlinear gain of the OPA and phase of the field driving the OPA. while it is interesting that we can see with the resonator speed increasing, the second-order sideband peak shifts to the right regardless of the direction from which the system is driven as shown in Fig.~\ref{2wl_3w_G}(e) and (f). Furthermore, there are also similarities between the two different OPA cases, such as compared with the case where the system is driven from the right side ($\Omega  < 0$), the influence of resonator rotation on the second-order sideband enhancement is much more significant when the system is driven from the left side ($\Omega  > 0$).

\section{Nonreciprocal second-order sidebands in non-Markovian systems}
When the system interacts with the environment, the dynamics of the system affected by the environment behaves the dissipation or the backflow oscillation
of the photon from the environment, where the former corresponds to the Markovian approximation, while the latter exhibits non-Markovian effects \cite{breuer2002,breuer1032104012009,breuer880210022016,Vega015001}. In Sec.II-Sec.V, we have studied the optomechanical second-order sidebands under the Markovian approximation. In this section, we investigate the influences of non-Markovian effects on the efficiency of second-order sidebands. For this purpose, we consider that the cavity interacts with the non-Markovian environment consisting of a series of boson modes (eigenfrequency ${\omega _k}$) \cite{Xiong2019100,Cialdi2019100,Tang201297,Groblacher20156,Liu20117,Hoeppe2012108,Xu201082,Madsen2011106,Guo2021126,Khurana201999,Uriri2020101,Liu2020102,Anderson199347,Li2022129,breuer880210022016,Vega015001}, \textbf{where the non-Markovian environment couples to an external reservoir.} In a rotating frame defined by ${{\hat U}_S}(t) = \exp [ - i{\omega _l}t({{\hat a}^\dag }\hat a + \sum\nolimits_k {\hat b_k^\dag } {{\hat b}_k}+\sum_j {\hat c_j^\dag {{\hat c}_j}} )]$, the total Hamiltonian (\ref{Heff}) is changed to
\begin{equation}
\begin{aligned}
{{\hat H}_{eff}} =& \hbar \left( {{\Delta _0} - \xi \hat x + {\Delta _s}} \right){{\hat a}^\dag }\hat a + \frac{{{{\hat p}^2}}}{{2m}} + \frac{1}{2}m\omega _m^2{{\hat x}^2}\\
& + \frac{{\hat p_\phi ^2}}{{2m{{\left( {R + \hat x} \right)}^2}}} + i\hbar G({\hat a^{\dag 2}}{e^{ - i{\Delta _p}t}}{e^{i\theta }} - H.c.)\\
& + i\hbar \sqrt {{\kappa _{ex}}} \left[ {\left( {{\varepsilon _l} + {\varepsilon _p}{e^{ - i{\Delta _p}t}}} \right){{\hat a}^\dag } - H.c.}\right]\\
& + \hbar \sum\limits_k {{\Delta _k}\hat b_k^\dag {{\hat b}_k}}  + i\hbar \sum\limits_k {({g_k}\hat a} \hat b_k^\dag  - H.c.)\\
& + \hbar \sum\limits_{j} {({{\tilde \omega }_{j}} - {\omega _{l}})\hat c_{j}^\dag {{\hat c}_{j}}}  + i\hbar \sum\limits_{jk} {({v_{jk}}{{\hat c}_{j}}} \hat b_k^\dag  - H.c.) ,\label{Heff1}
\end{aligned}
\end{equation}
where ${\Delta _k} = {\omega _k} - {\omega _l}$ defines the detuning of $k$th mode (eigenfrequency ${\omega _k}$) of the non-Markovian environment from the driving field. ${{\hat b}_k}(\hat b_k^\dag )$ is the annihilation (creation) operator. ${{g_k}}$ is the coupling coefficient between cavity and environment. \textbf{${{v_{jk}}}$ denotes the coupling strength between the $k$th mode  of the non-Markovian environment and $j$th mode of the external reservoir with frequency ${{{\tilde \omega }_{j}}}$. ${{{\hat c}_{j}}}$ and $\hat c_{j}^\dag$ represent annihilation and creation operators of the external reservoir, respectively.} The dynamics of the system can be derived as
\begin{small}
\begin{align}
&\frac{d}{{dt}}\hat a(t) =  - [ {\frac{\kappa }{2} + i( {{\Delta _0} - \xi \hat x(t) + {\Delta _s}} )} ]\hat a(t)- \sum\limits_k {g_k^*{{\hat b}_k}(t)} \nonumber \\
&\qquad \qquad + \sqrt {{\kappa _{ex}}} ( {{\varepsilon _l} + {\varepsilon _p}{e^{ - i{\Delta _p}t}}} )+ 2G{{\hat a}^\dag }(t){e^{i\theta }}{e^{ - i{\Delta _p}t}},\label{Heq13}\\
&\frac{d}{{dt}}{{\hat b}_k}(t) =  - i{\Delta _k}{{\hat b}_k}(t) + {g_k}\hat a(t) + \sum\limits_{j} {{v_{jk}}{{\hat c}_{j}}}(t), \label{Heq23}\\
&\frac{d}{{dt}}{{\hat c}_{j}}(t) =  - i({{\tilde \omega }_{j}} - {\omega _l}){{\hat c}_{j}}(t) - \sum\limits_{{k_1}} {{v_{j{k_1}}^*}{{\hat b}_{{k_1}}}}(t) ,\label{Heq23c}\\
&\frac{{{d^2}}}{{d{t^2}}}\hat x(t) + {\Gamma _m}\frac{d}{{dt}}\hat x(t) + \omega _m^2\hat x(t) = \frac{{\hbar \xi }}{m}{{\hat a}^\dag }(t)\hat a(t) + \frac{{\hat p_\phi ^2(t)}}{{{m^2}{R^3}}},\label{Heq33}\\
&\frac{d}{{dt}}\hat \phi (t) = \frac{{{{\hat p}_\phi }(t)}}{{m{R^2}}},\label{Heq43}\\
&\frac{d}{{dt}}{{\hat p}_\phi }(t) = 0,\label{Heq53}
\end{align}
\end{small}
where the intrinsic loss rate ${\kappa _a} = \kappa /2$ is phenomenologically added in above equations. Eq.~(\ref{Heq23c}) gives
\begin{equation}
\begin{aligned}
{{\hat c}_{j}}(t) =& {e^{ - i({{\tilde \omega }_{j}} - {\omega _l})t}}{{\hat c}_{j}}(0) \\
&- \sum\limits_{{k_1}} {{v_{j{k_1}}^*}} \int_0^t {{e^{ - i({{\tilde \omega }_{j}} - {\omega _l})(t - \tau )}}{{\hat b}_{{k_1}}}(\tau )} d\tau .\label{afsm}
\end{aligned}
\end{equation}
Substituting Eq.~(\ref{afsm}) into Eq.~(\ref{Heq23}), we get
\begin{equation}
\begin{aligned}
\frac{d}{{dt}}{{\hat b}_k}(t) =&  - i{\Delta _k}{{\hat b}_k}(t) + {g_k}\hat a(t) + \sqrt {2\pi } {c_{k,in}}\\
 & - \sum\limits_{{k_1}} {\int_0^t {{D_{k{k_1}}}(t - \tau )} {{\hat b}_{{k_1}}}(\tau )d\tau } ,\label{afsssdsm}
\end{aligned}
\end{equation}
\textbf{where the input-field operator of the reservoir ${\hat c_{k,in}}(t) = \frac{1}{{\sqrt {2\pi } }}\sum_{j} {{v_{jk}}{e^{ - i({{\tilde \omega }_{j}} - {\omega _l})t}}} {{\hat c}_{j}}(0)$, the correlation function ${D_{k{k_1}}}(t - \tau ) = \sum_j {{v_{jk}}v_{j{k_1}}^*{e^{ - i({{\tilde \omega }_j} - {\omega _l})(t - \tau )}} = \int {{{\tilde J}_{k{k_1}}}(\omega )} } {e^{ - i(\omega  - {\omega _l})(t - \tau )}}d\omega $, and the spectral density of the reservoir  ${{\tilde J}_{k{k_1}}}(\omega ) = \sum_j {{v_{jk}}v_{j{k_1}}^*\delta (\omega  - {{\tilde \omega }_j})} $ with $\delta (\omega)$ being the Dirac delta function. Taking ${{\tilde J}_{k{k_1}}}(\omega ) = \frac{{{\mu _k}}}{\pi }{\delta _{k{k_1}}}$ ($\delta _{k{k_1}}$ represents the Kronecker delta symbol, i.e., $\delta _{k{k_1}}=1$ for $k=k_1$, otherwise $\delta _{k{k_1}}=0$), and then ${D_{k{k_1}}}(t- \tau) = 2{\mu _k}\delta (t- \tau){\delta _{k{k_1}}}$  \cite{breuer2002,Gardiner1711022027}, we obtain}
\begin{equation}
\begin{aligned}
\frac{d}{{dt}}{{\hat b}_k}(t) =  - i{{\tilde \Delta }_k}{{\hat b}_k}(t) + {g_k}\hat a(t) + \sqrt {2\pi } {\hat c_{k,in}},\label{af23545m}
\end{aligned}
\end{equation}
\textbf{with ${{\tilde \Delta }_k} = {\Delta _k} - i\mu _k$. To simplify the calculation, we assume $\mu _k \equiv \mu$ below, where $\mu$ denotes the decay from the non-Markovian environment coupling to an external reservoir.} The solution of Eq.~(\ref{af23545m}) is
\begin{equation}
\begin{aligned}
{{\hat b}_k}(t) =& {{\hat b}_k}(0){e^{ - i{{\tilde \Delta }_k}t}} + {g_k}\int_0^t \hat a(\tau ){e^{ - i{{\tilde \Delta }_k}(t - \tau )}}{d\tau } \\
&+ \sqrt {2\pi } \int_0^t {{\hat c}_{k,in}}(\tau ){e^{ - i{{\tilde \Delta }_k}(t - \tau )}}{d\tau }.\label{fmbk}
\end{aligned}
\end{equation}
The first term on the right-hand side of Eq.~(\ref{fmbk}) represents the freely propagating parts of the environmental fields and the second term describes the influence of non-Markovian environment on the cavity. The third term on the right-hand side of Eq.~(\ref{fmbk}) denotes the influence of the input-field operator of the reservoir on the non-Markovian environment. Substituting Eq.~(\ref{fmbk}) into Eq.~(\ref{Heq13}), we obtain an integro-differential equation
\begin{equation}
\begin{aligned}
\frac{d}{{dt}}\hat a(t) =&  - [ {\frac{\kappa }{2} + i( {{\Delta _0} - \xi \hat x(t) + {\Delta _s}} )} ]\hat a(t)\\
& + \sqrt {{\kappa _{ex}}} ( {{\varepsilon _l} + {\varepsilon _p}{e^{ - i{\Delta _p}t}}} ) + 2G{{\hat a}^\dag }(t){e^{i\theta }}{e^{ - i{\Delta _p}t}}\\
&   + {\hat {K} }(t) + {{\hat {L}} }(t) - \int_0^t \hat a(\tau )f(t - \tau ){d\tau },\label{afm}
\end{aligned}
\end{equation}
\textbf{where ${\hat {K} }(t) =  - \sum_k {g_k^*{{\hat  b}_k}(0){e^{ - i{\tilde \Delta _k}t}}}  =   \int_{ - \infty }^\infty  { {h ^*}} (t - \tau ){\hat {a}_{in}}(\tau )d\tau$, $\hat {L}(t) =  - \sqrt {2\pi } \sum_k {g_k^*\int_0^t {{\hat c}_{k,in}}(\tau ){e^{ - i{{\tilde \Delta }_k}(t - \tau )}}} {d\tau }$, the input-field operator ${\hat {a}_{in}}(t) = \frac{1}{{\sqrt {2\pi } }}\sum_k {{e^{ - i{\tilde \Delta _k}t}}{{\hat b}_k}(0)} $, the impulse response function $h(t) = \frac{{ - 1}}{{\sqrt {2\pi } }}\sum_k {{e^{i{{\tilde \Delta }_k}t}}{g_k}}  \equiv \frac{{ - 1}}{{\sqrt {2\pi } }}\int {{e^{i(\omega  - {\omega _l})t + \mu t}}g(\omega )} d\omega$ (we have made the replacement ${g_k} \to g(\omega )$ in the continuum limit), and the correlation function $f(t) = \sum_k {{{\left| {{g_k}} \right|}^2}{e^{ - i{{\tilde \Delta }_k}t}}}  = \int {J(\omega ){e^{ - i(\omega  - {\omega _l})t - \mu t}}d\omega } $ with the spectral density of the non-Markovian environment $J(\omega ) = \sum\nolimits_k {|{g_k}{|^2}} \delta (\omega  - {\omega _k})$. Both ${\hat {a}}_{in}(t)$ and ${\hat c_{k,in}}$ are the input fields with zero expectation value ${ a}_{in}(t) = \langle {{\hat{ a}_{in}}(t)} \rangle  = 0$ and ${{\rm{c}}_{k,in}}(t) = \langle {{{\hat{c}}_{k,in}}(t)} \rangle  = 0$ for the environment and reservoir initialization in the vacuum states, which lead to ${{K} }(t)= \langle {\hat{K} }(t) \rangle  = 0$ and ${{L} }(t)= \langle {\hat{L} }(t) \rangle  = 0$.} We define the spectral response function as
\begin{eqnarray}
g(\omega ) = \sqrt {\frac{{{\kappa _{ex}}}}{{2\pi }}} \frac{{{\lambda _1}}}{{{\lambda _1} - i(\omega  - {\omega _l})}},\label{zzzz1}
\end{eqnarray}
where ${\lambda _1}$ is the environmental spectrum width and ${\kappa _{ex}} = \kappa $ is the cavity dissipation through the input and output ports. The spectral density of the environment is \cite{shen880338352013,zhang870321172013,diosi850341012012,xiong860321072012,shen950121562017}
\begin{eqnarray}
\begin{aligned}
J(\omega ) = \frac{{{\kappa _{ex}}}}{{2\pi }}\frac{{\lambda _1^2}}{{\lambda _1^2 + {(\omega - \omega_l) ^2}}},
\label{spectral_density}
\end{aligned}
\end{eqnarray}
which corresponds to the Lorentzian spectral density. With Eqs.~(\ref{zzzz1}) and (\ref{spectral_density}), we get $h (\tau  - t) =  - \sqrt {{\kappa _{ex}}} {\lambda _1}{e^{ - {({\lambda _1} + i\mu )}(t - \tau )}}\theta (t - \tau )$ and $f(t - \tau ) = \frac{1}{2}{\kappa _{ex}}{\lambda _1}{e^{ - {({\lambda _1} + i\mu )}\left| {t - \tau } \right|}}$, where $\theta (t - t')$ is the unit step function, $\theta (t - t') = 1$ for $t \ge t'$, which represents a Gaussian Ornstein-Uhlenbeck process \cite{uhlenbeck368231930,gillespie5420841996,jing1052404032010}.

%figure14
\begin{figure*}[t]
\centerline{
\includegraphics[width=18.1cm, height=7.6cm, clip]{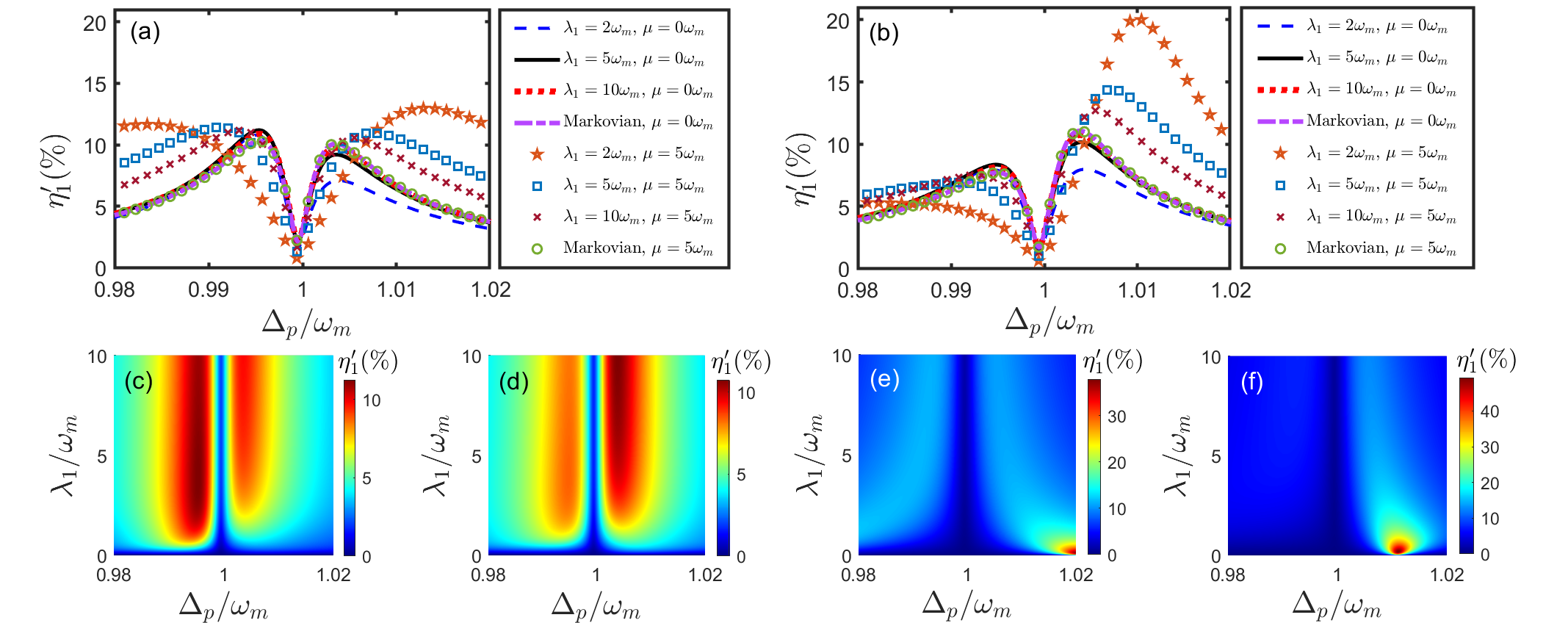}}
\caption{(a)(b) The efficiency ${\eta^{\prime}_1}$ of the second-order upper sideband generation as a function of ${\Delta _p}$, which corresponds to the Markovian and non-Markovian environments with the different environmental spectrum width ${\lambda _1}$ without the OPA involvement $(G=0)$. (c)(d)(e)(f) ${\eta^{\prime}_1}$ varies with ${\Delta _p}$ and ${\lambda _1}$. The rotation speed is set as (a)(c)(e) $\Omega  = 0$ and (b)(d)(f) $\Omega  = 7.7$ kHz. \textbf{The parameter $\mu$ denotes the decay from the non-Markovian environment coupling to an external reservoir, where $\mu=0$ for (c) and (d), while $\mu=5\omega_m$ for (e) and (f).} Other parameters are the same as  Fig.~\ref{Eta1}.} \label{lambda_bian1_fm}
\end{figure*}
For convenience, we take the expectation values of the operator equations by defining $a \equiv \langle {\hat a} \rangle $ , $x \equiv \langle {\hat x} \rangle $ , $\phi  \equiv \langle {\hat \phi } \rangle $ and ${p_\phi } \equiv \langle {{{\hat p}_\phi }} \rangle $. The steady-state solution of the non-Markovian system can be obtained from Eq.~(\ref{afm}) as
\begin{equation}
\begin{aligned}
{a^{\prime}_s}=&\frac{{\sqrt {{\kappa _{ex}}} {\varepsilon _l}}}{{\kappa  + i{\Delta '} }},\\
{x^{\prime}_s}=&\frac{{\hbar \xi {{| {a^{\prime}_s} |}^2}}}{{m\omega _m^2}} + R{\left( {\frac{\Omega }{{{\omega _m}}}} \right)^2}
,\label{wentaijie3}
\end{aligned}
\end{equation}
where $\Delta '{\rm{ = }}{\Delta _0} - \xi {x^{\prime}_s} + {\Delta _s}$.
We make the ansatz
\begin{small}
\begin{equation}
\begin{aligned}
a=& {a'_s} + {{A^{\prime}}_1^ +} {e^{ - i{\Delta _p}t}} +{{A^{\prime}}_1^ -} {e^{i{\Delta _p}t}} + {{A^{\prime}}_2^ +} {e^{ - 2i{\Delta _p}t}} + {{A^{\prime}}_2^ -}{e^{2i{\Delta _p}t}},\\
x=&{x'_s} + {{X^{\prime}}_1^ +} {e^{ - i{\Delta _p}t}} +{{X^{\prime}}_1^ -} {e^{i{\Delta _p}t}} + {{X^{\prime}}_2^ +} {e^{ - 2i{\Delta _p}t}} + {{X^{\prime}}_2^ -}{e^{2i{\Delta _p}t}},
\nonumber
\end{aligned}
\end{equation}
\end{small}
We get the linear response of the probe field
\begin{small}
\begin{equation}
\begin{aligned}
{\sigma '_1}( {{\Delta _p}} ){{A^{\prime}}_1^ +}  &= \Lambda ( {{\Delta _p}} )[i\xi {a'_s}{{X^{\prime}}_1^ +}  + 2G{e^{i\theta }}{{a^{\prime}}_s^ *} + \sqrt {{\kappa _{ex}}} {\varepsilon _p}],\\
{\sigma '_2}( {{\Delta _p}} ){{A^{\prime}}_1^ {-*}} &=  - i\xi \Lambda ( {{\Delta _p}} ){{a^{\prime}}_s^ *}{{A^{\prime}}_1^ +} , \\
\chi ( {{\Delta _p}} ){{X^{\prime}}_1^ +}  &= \hbar \xi ({a'_s}{{A^{\prime}}_1^ {-*}} + {{a^{\prime}}_s^ *}{{A^{\prime}}_1^ +}),\label{first3}
\end{aligned}
\end{equation}
\end{small}
and second-order sideband process
\begin{small}
\begin{equation}
\begin{aligned}
{\sigma '_1}(2{\Delta _p}){{A^{\prime}}_2^ +}  &= \Lambda ( {2{\Delta _p}} )[i\xi ({a'_s}{{X^{\prime}}_2^ +} + {{A^{\prime}}_1^ +} {{X^{\prime}}_1^ +} ) + 2G{e^{i\theta }}{{A^{\prime}}_1^ {-*}}],\\
{\sigma '_2}( {2{\Delta _p}} ){{A^{\prime}}_2^ {-*}} &=  - i\xi \Lambda ( {2{\Delta _p}} )({{a^{\prime}}_s^ *}{{X^{\prime}}_2^ +}  +{{A^{\prime}}_1^ {-*}}{{X^{\prime}}_1^ +} ) , \\
\chi ( {2{\Delta _p}} ){{X^{\prime}}_2^ +}  &= \hbar \xi ({{a^{\prime}}_s^ *}{{A^{\prime}}_2^ +}  + {a'_s}{{A^{\prime}}_2^ {-*}} + {{A^{\prime}}_1^ {-*}}{{A^{\prime}}_1^ +} ),\label{second3}
\end{aligned}
\end{equation}
\end{small}
with
\begin{eqnarray}
\Lambda \left( {n{\Delta _p}} \right) &=& {{\lambda _1} + i\mu } - in{\Delta _p},\nonumber\\
{\sigma '_1}\left( {n{\Delta _p}} \right) &=& \kappa {\lambda _1} - \frac{{i\kappa n{\Delta _p}}}{2} + \Lambda ( {n{\Delta _p}} )(i\Delta  - in{\Delta _p}),\nonumber\\
{\sigma '_2}\left( {n{\Delta _p}} \right) &=& \kappa {\lambda _1} - \frac{{i\kappa n{\Delta _p}}}{2} - \Lambda ( {n{\Delta _p}} )(i\Delta + in{\Delta _p}),\nonumber\\
\chi \left( {n{\Delta _p}} \right) &=& m(\omega _m^2 - i{\Gamma _m}n{\Delta _p} -\Delta _p^2).\nonumber\\
\label{second3xxc}
\end{eqnarray}
Through the derived non-Markovian input-output relation by Eq.~(\ref{afm}), we obtain the expected value of the output field
\begin{equation}
{a_{out}}(t)  =  {a_{in}}(t)+\int_0^t { h (\tau-t )} a(\tau )d\tau. \label{shuru}
\end{equation}

Thus in the non-Markovian case, the efficiency of second-order upper sideband is defined as
\begin{equation}\label{xiaolv}
{\eta^{\prime}_1} = \left| { - \frac{{\sqrt {{\kappa _{ex}}} {\lambda _1}{{A^{\prime}}_2^ +} \frac{1}{{{{\lambda _1} + i\mu } - 2i{\Delta _p}}}}}{{{\varepsilon _p}}}} \right|.
\end{equation}
\textbf{With Eq.~(\ref{xiaolv}), we consider two cases (i) and (ii) separately.}

\textbf{(i) In the first case, we take the decay $\mu=0$ in Eq.~(\ref{xiaolv}).}
In Fig.~\ref{lambda_bian1_fm}(a) with the decay $\mu=0$, resonator stationary but without the participation of the OPA, we show the efficiency of second-order upper sideband generation as a function of ${\Delta _p}$ with the different spectral width of environment ${\lambda _1}$. For a given spectral width of environment, decreasing from ${\lambda _1} = 10{\omega _m} \sim 2{\omega _m}$, we find from the figure that the second-order upper sideband ${\eta^{\prime}_1}$ gradually decreases, whose two located peaks become increasingly asymmetric in the non-Markovian environment. Interestingly, from Fig.~\ref{lambda_bian1_fm}(b) with the decay $\mu=0$, when the light comes from the right side and $\Omega  = 7.7$ kHz, ${\eta^{\prime}_1}$ becomes symmetric in the non-Markovian environment at ${\lambda _1} = 2{\omega _m}$. That is, by controlling the rotation speed of the resonator and incident direction of the input fields, the symmetry of the second-order sideband is restored, but with a change in height compared with the Markovian environment. With the purpose of seeing the influence of the environmental spectrum width on the second-order sideband generation more clearly, the efficiency ${\eta^{\prime}_1}$ as a function of both ${\Delta _p}$ and ${\lambda _1}$ is shown in Fig.~\ref{lambda_bian1_fm}(c) and (d) with the decay $\mu=0$.
%figure15
\begin{figure}[t]
\centerline{
\includegraphics[width=8.0cm, height=5.6cm, clip]{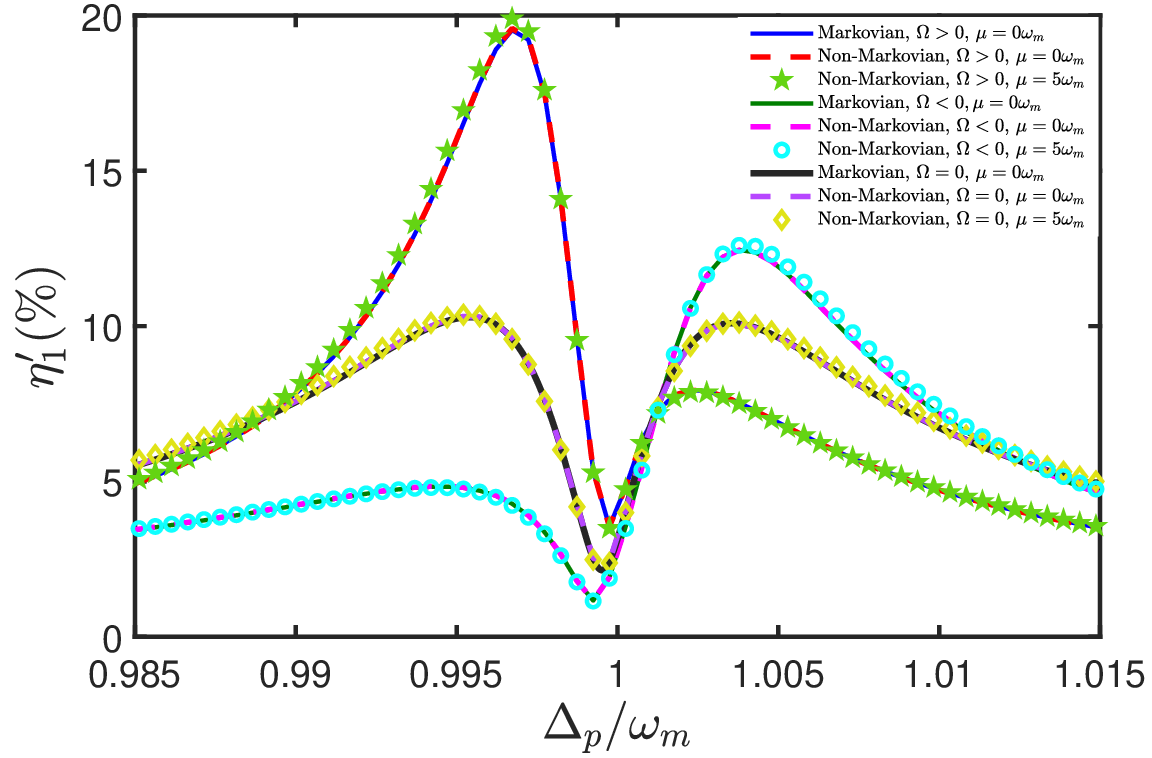}}
\caption{The efficiency ${\eta^{\prime}_1}$ of the second-order upper sideband generation as a function of ${\Delta _p}$ under different values of $\Omega $ and incident directions of light, where we take $G=0$. This figure shows the consistency of nonreciprocal second-order sidebands between non-Markovian limit with ${\lambda _1} = 200{\omega _m}$ and Markovian approximation. Other parameters are the same as  Fig.~\ref{Eta1}.} \label{m_fm_duibi2}
\end{figure}

As the spectrum width of the environment is further increased, the efficiency of second-order upper sideband generation increases. For the sake of clarity, we separately draw the non-Markovin case and the Markovian limit case where the environmental spectrum width ${\lambda _1} = 200{\omega _m}$ for the condition that the resonator is stationary and no OPA is involved in Fig.~\ref{m_fm_duibi2} with the decay $\mu=0$. This figure shows the consistency of nonreciprocal second-order upper sideband between non-Markovian limit with ${\lambda _1} = 200{\omega _m}$ and Markovian approximation, regardless of the incident direction of the input fields. This originates from the fact that the correlation function $f(t)$ and impulse response function $h (t)$ tend to $\kappa_{ex} \delta (t)$ and $-\sqrt {\kappa_{ex}}  \delta (t)$ in the wideband limit (i.e., the spectrum width $\lambda_1$ approaches infinity), respectively, which leads to Eqs.~(\ref{afm}) and (\ref{shuru}) in the non-Markovian regime returning back to Eqs.~(\ref{Heq1}) and (\ref{cz123}) under the Markovian approximation.

%figure16
\begin{figure*}[t]
\centerline{
\includegraphics[width=18.1cm, height=8.1cm, clip]{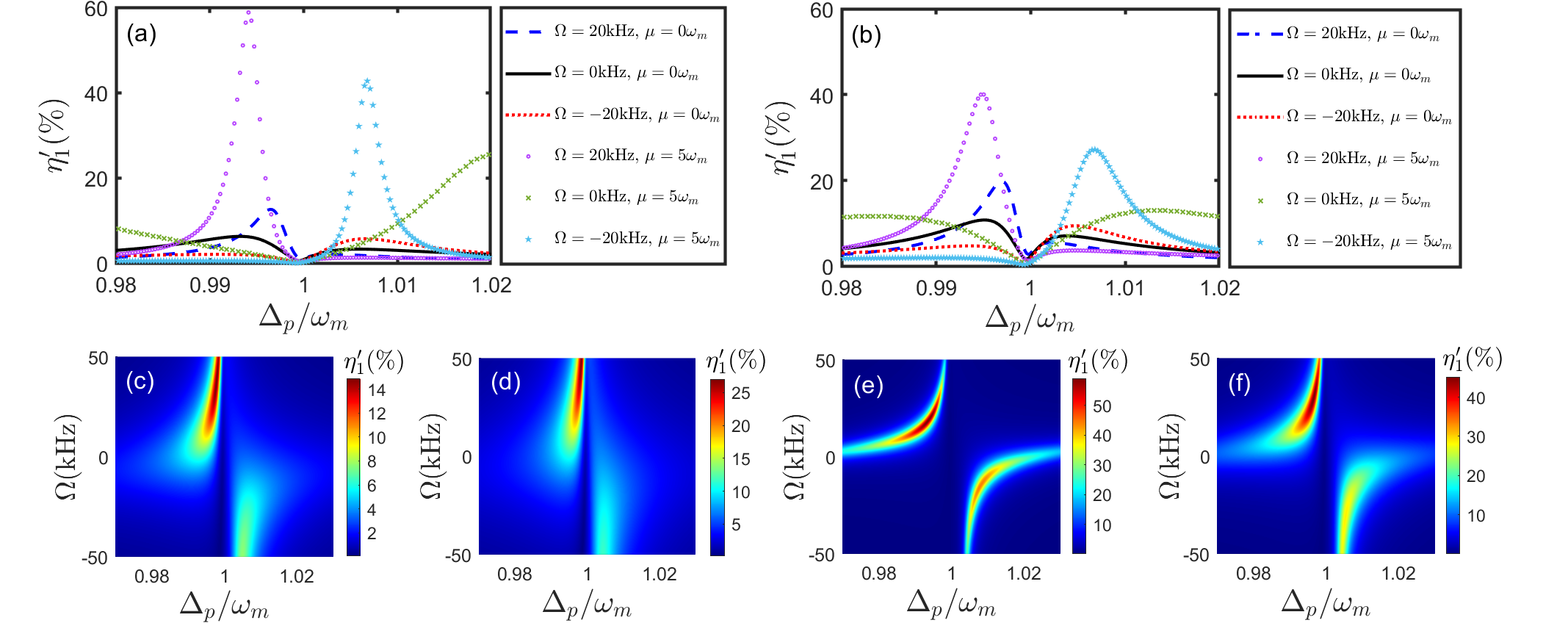}}
\caption{(a)(b) The efficiency ${\eta^{\prime}_1}$ of the second-order upper sideband generation as a function of ${\Delta _p}$ under different values of $\Omega $ and incident directions of light in the non-Markovian environment and without the participation of the OPA $(G=0)$. (c)(d)(e)(f) ${\eta^{\prime}_1}$ varies with ${\Delta _p}$ and $\Omega $. The environmental spectrum widths are (a)(c)(e) ${\lambda _1} = 0.5{\omega _m}$ and (b)(d)(f) ${\lambda _1} = 2{\omega _m}$, respectively. \textbf{(c) and (d) take  the decay $\mu=0$, while the decay $\mu=5\omega_m$ corresponds to (e) and (f).} Other parameters are the same as  Fig.~\ref{Eta1}.} \label{fhy_fm}
\end{figure*}
Fig.~\ref{fhy_fm}(a)-(d) with the decay $\mu=0$ shows the spinning-induced direction-dependent nonreciprocal behavior of second-order upper sideband in the non-Markovian environment but without the participation of the OPA. We note that on the one hand, the efficiency of second-order sideband ${\eta^{\prime}_1}$ is very sensitive to the environmental spectrum width. On the other hand, the operating bandwidth for observing an obvious nonreciprocal enhancement of second-order sideband changes in the non-Markovian environment. Compared with the Markovian environment in Fig.~\ref{m_fm_duibi2} with the decay $\mu=0$, the operating bandwidth becomes significantly wider at frequency ${\Delta _p} > {\omega _m}$ and narrower at ${\Delta _p} < {\omega _m}$.

%figure17
\begin{figure}[t]
\centerline{
\includegraphics[width=9.0cm, height=7.6cm, clip]{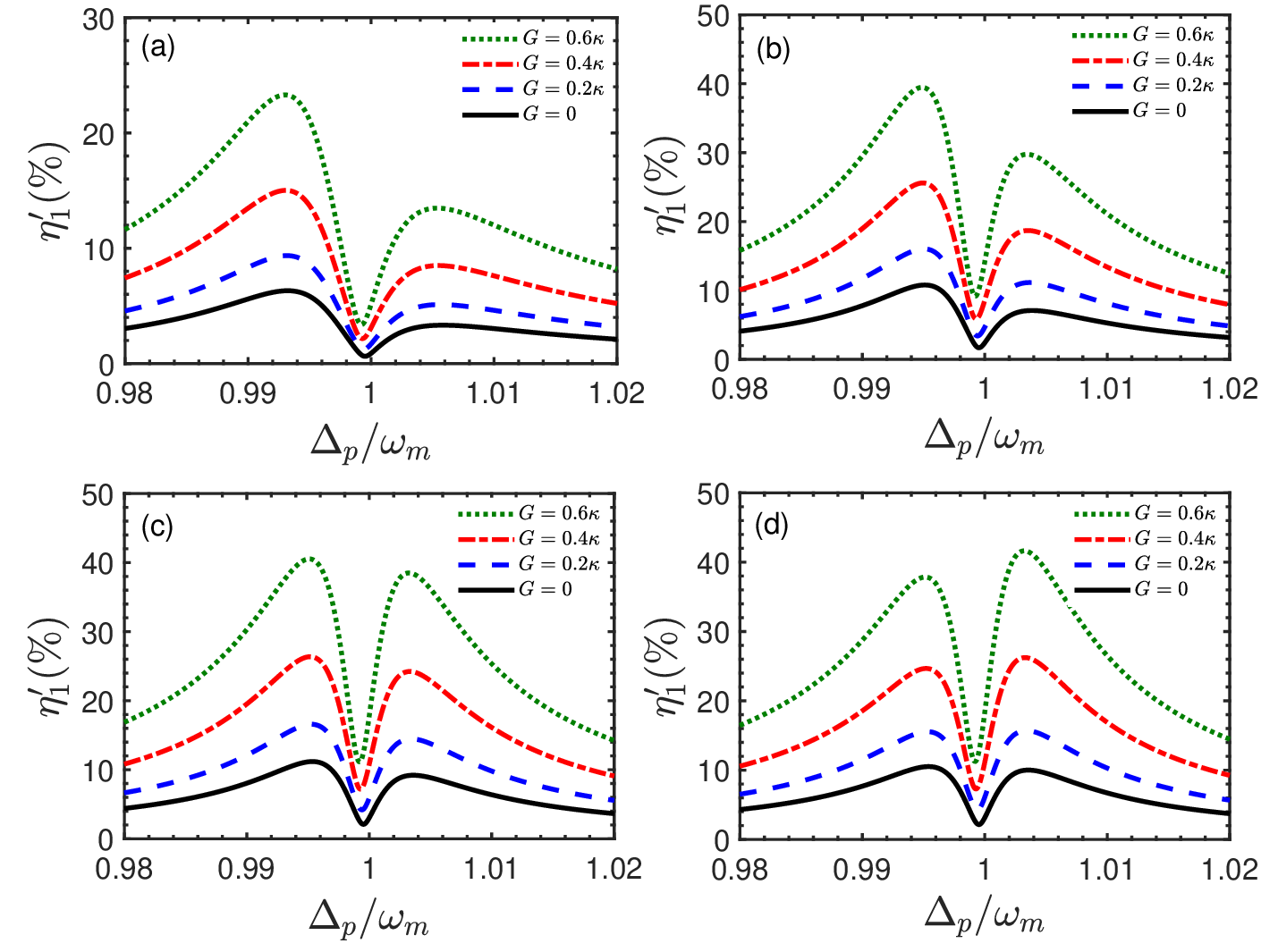}}
\caption{The efficiency ${\eta^{\prime}_1}$ of the second-order upper sideband generation as a function of ${\Delta _p}$ for different nonlinear gain $G$ of the OPA, where $\theta=0$, $\Omega=0$, and the decay $\mu=0$. The environmental spectrum widths are (a) ${\lambda _1} = 0.5{\omega _m}$, (b) ${\lambda _1} = 2{\omega _m}$, (c) ${\lambda _1} = 5{\omega _m}$, and (d) ${\lambda _1} = 30{\omega _m}$, respectively. Other parameters are the same as  Fig.~\ref{Eta1}.} \label{bianlambdaG}
\end{figure}

Figs.~\ref{lambda_bian1_fm}, \ref{m_fm_duibi2} and \ref{fhy_fm} with the decay  $\mu=0$ present the influence of pure non-Markovian effect on the second-order sideband without the participation of the OPA ($G=0$). In Fig.~\ref{bianlambdaG} with the decay $\mu=0$, we show the variation of second-order upper sideband efficiency in the presence of both non-Markovian effect and OPA. As expected, when the nonlinear gain $G$ of the OPA increases from $0$ to $0.6\kappa $, the efficiency ${\eta^{\prime}_1}$ is significantly enhanced. Moreover, the non-Markovian effect is more pronounced for ${\eta^{\prime}_1}$ when the environmental spectrum width is small (i.e., ${\lambda _1} < 2{\omega _m}$). As shown in Fig.~\ref{bianlambdaG}(d) with the decay $\mu=0$ at ${\lambda _1} = 30{\omega _m}$, the enhancement effect of the OPA for second-order sideband is almost identical to the case of Markovian limit.

\textbf{(ii) In the second case, we take the decay $\mu=5\omega_m$ in Eq.~(\ref{xiaolv}). The influences of the decay from the non-Markovian environment coupling to an external reservoir on the efficiency of second-order upper sidebands are shown in Figs.~\ref{lambda_bian1_fm}, \ref{m_fm_duibi2} and \ref{fhy_fm} with $\mu=5\omega_m$. We find that the decay $\mu$ has large influences on the efficiency of second-order upper sidebands in non-Markovian regimes, while it has almost no influence on the efficiency of second-order upper sidebands under the Markovian approximation. This is because the decay $\mu$ is comparable to the spectral width $\lambda_1 $ of the non-Markovian environment revealed from Eqs.~(\ref{second3xxc}) and (\ref{xiaolv}) (see Fig.~\ref{lambda_bian1_fm}(a)(b)(e)(f) and Fig.~\ref{fhy_fm}(a)(b)(e)(f)) since the spectral width $\lambda_1 $ takes finite values in non-Markovian regimes. However, the spectral width $\lambda_1$ tends to infinity (i.e., $\lambda_ 1 \to \infty$) under the Markovian approximation, which leads to that the decay $\mu$ is negligible compared with the spectral width $\lambda_1 $ due to $\mu \ll \infty $ in Eqs.~(\ref{second3xxc}) and (\ref{xiaolv}) (see Fig.~\ref{lambda_bian1_fm}(a)(b) and  Fig.~\ref{m_fm_duibi2}).}

\section{Conclusion}

In summary, we have theoretically studied the second-order OMIT sidebands and group delays in a spinning resonator containing an optical parametric amplifier. We discuss the influence of the OPA driven by different pumping frequencies on the second-order sideband generation. The results show that the second-order sidebands in the rotating resonator can be greatly enhanced in the presence of the OPA and still remain the nonreciprocal behavior due to the optical Sagnac effect. The second-order sidebands can be adjusted simultaneously by the pumping frequency and phase of the field driving the OPA, the gain coefficient of the OPA, the rotation speed of the resonator, and the incident direction of the input fields. When the OPA is excited by a pump driving with the frequency $\omega_g={{\omega _l} + {\omega _p}}$, the higher nonlinear gain of the OPA is, the stronger the second-order sidebands are. At this point, the OPA can only enhance the second-order sidebands but cannot change the position of the peaks and the non-reciprocal nature due to resonator rotation, which maintains the localization of the maximum value of the sideband efficiency. When the OPA is excited by a pump driving with the frequency $\omega_g=2{\omega _l}$, the nonlinear gain of the OPA cannot enhance the second-order sidebands, which can only be achieved by adjusting the phase of the field driving the OPA. The OPA can also change the linewidth of the suppressive window of the second-order sidebands, which can be applied to determining the excitation number of atoms and plays important roles in nonlinear media in the optical properties of the output field. Combining the Sagnac transformation and the presence of the OPA, we demonstrate that the group delay of the second-order upper sideband can be tuned by adjusting the nonlinear gain and phase of the field driving the OPA, the rotation speed of the resonator and incident direction of the input fields, which allows us to realize a tunable switch from slow light to fast light in the spinning optomechanical system. Moreover, we extend the study of second-order sidebands from the Markovian to the non-Markovian bath, which consists of a collection of infinite oscillators (bosonic photonic modes). We find the second-order OMIT sidebands in a spinning resonator exhibit a transition from the non-Markovian to Markovian regime by controlling environmental spectral width. Finally, we investigate the influences of the decay from the non-Markovian environment coupling to an external reservoir on the efficiency of second-order upper sidebands.

These results indicate the advantage of using a hybrid nonlinear system and contribute to a better understanding of light propagation in nonlinear optomechanical devices, which provides potential applications for precision measurement, optical communications, and quantum sensing. Expansions of the above non-Markovian nonreciprocal second-order sidebands to various general nonlinear physical models, e.g., (1) ${\chi ^{( 2 )}}$ nonlinear
materials ${{\hat a}^2}{{\hat b}^\dag } + \hat b{{\hat a}^{\dag 2}}$ \cite{zz1,zz2}, (2) Kerr nonlinear mediums ${{\hat a}^{\dag 2}}{{\hat a}^2}$ \cite{zz3,zz4}, and (3)  quadratic optomechanical systems ${{\hat a}^\dag }\hat a(\hat b + {{\hat b}^\dag })^2$ \cite{Aspelmeyer861391,Thompson45272,jack630438032001zs,jack630438032001zzz,jack630438032001z1s,jack630438032001zs3}, deserve future investigations.

\section{ACKNOWLEDGMENTS}
This work was supported by National Natural Science Foundation of China under Grants No. 12274064, Scientific Research Project for Department of Education of Jilin Province under Grant No. JJKH20190262KJ, and Natural Science Foundation of Jilin Province (subject arrangement project) under Grant No. 20210101406JC.

\section*{}
\appendix*
\section{Derivation of Eqs.~(\ref{Heq1})-(\ref{Heq4})}
In order to give the origin of $\Gamma_m$ in Eq.~(\ref{Heq2}), we add the coupling Hamiltonian $\hat H_{CL}$  \cite{Weiss1999,Caldeira1983347,Caldeira5871983,Grabert1151988,Spiechowicz052107,
Einsiedler0222282020,Sinha051111,Sun5118451995} between the mechanical mode and a Bosonic bath  consisting of a set of harmonic oscillators with mass $m_l$ and frequency $\Omega_l$ to Eq.~(\ref{Heff}) as follows
\begin{equation}
\begin{aligned}
\hat H_{CL} =\sum\limits_l {\left[ {\frac{{\hat P_l^2}}{{2{M_l}}} + \frac{{{M_l}\Omega _l^2}}{2}{{\left( {{{\hat C}_l} - \frac{{{v_l}}}{{{M_l}\Omega _l^2}}\hat x} \right)}^2}} \right]} ,
\label{H0HI}
\end{aligned}
\end{equation}
where $\hat C_l$ and $\hat P_l$ are the coordinate and momentum of the harmonic oscillators, respectively, while $v_l$ denotes coupling strength between mechanical mode and bath. The counterterm proportional to $\hat x^2$ is typically introduced in the Hamiltonian, which accounts for a renormalization of the central oscillator frequency due to the interaction with the bath \cite{Weiss1999,Caldeira1983347,Caldeira5871983,Grabert1151988,Spiechowicz052107,
Einsiedler0222282020,Sinha051111,Sun5118451995}.
The Heisenberg equations read
\begin{eqnarray}
\frac{d}{{dt}}\hat a &=&  - \left[ {\kappa  + i\left( {{\Delta _0} - \xi x + {\Delta _s}} \right)} \right]\hat a\nonumber\\
&& + \sqrt {{\kappa _{ex}}} ({\varepsilon _l} + {\varepsilon _p}{e^{ - i{\Delta _p}t}}) + 2G{{\hat a}^\dag }{e^{i\theta }}{e^{ - i{\Delta _p}t}},\nonumber\\
\label{Heisenbergz123}\\
\frac{d}{{dt}}\hat x &=& \frac{{\hat p}}{m},
\label{Heisenbergz1}\\
\frac{d}{{dt}}\hat p &=&
 - m\omega _m^2\hat x + \sum\limits_l {{v_l}{{\hat C}_l}}  - \sum\limits_l {\frac{{v_l^2}}{{{M_l}\Omega _l^2}}} \hat x\nonumber\\
 &&+ \hbar \xi {{\hat a}^\dag }\hat a + \frac{{\hat p_\phi ^2}}{{m{R^3}}},
\label{Heisenbergz2}\\
\frac{d}{{dt}}{{\hat C}_l} &=& \frac{{{{\hat P}_l}}}{{{M_l}}},
\label{Heisenbergz3}\\
\frac{d}{{dt}}{{\hat P}_l} &= & - {M_l}\Omega _l^2{{\hat C}_l}+{v_l}\hat x,
\label{Heisenbergz4}\\
\frac{d}{{dt}}\hat \phi  &=& \frac{{{\hat p_\phi }}}{{m{R^2}}},\label{Heq3c}\\
\frac{d}{{dt}}\hat {{p} }_\phi &=& 0,\label{Heq4c}
\end{eqnarray}
where the Heisenberg operator ${\hat x}(t)$ is abbreviated as $\hat x \equiv \hat x(t) = {e^{i{{\hat H}_T}t/\hbar }}\hat x(0){e^{i{{\hat H}_T}t/\hbar }}$ with ${{{\hat H}_T}}={{\hat H}_{eff}}+\hat H_{CL}$ (${\hat H}_{eff}$ is given by Eq.~(\ref{Heff})), and the other operators also have similar expressions. Eqs.~(\ref{Heq1})(\ref{Heq3})(\ref{Heq4}) are consistent with Eqs.~(\ref{Heisenbergz123})(\ref{Heq3c})(\ref{Heq4c}), respectively.  Differentiating Eqs.~(\ref{Heisenbergz1}) and~(\ref{Heisenbergz3}), together with Eqs.~(\ref{Heisenbergz2}) and~(\ref{Heisenbergz4}), we have
\begin{eqnarray}
&&m\left[ {\frac{{{d^2}}}{{d{t^2}}}\hat x + \omega _m^2\hat x} \right] =   \sum\limits_l {{v_l}{{\hat C}_l}} - \sum\limits_l {\frac{{v_l^2}}{{{M_l}\Omega _l^2}}} \hat x \nonumber \\
&&\ \ \ \ \ \ \ \ \ \ \ \ \ \ \ \ \ \ \ \ \ \ \ \ \ \ + \hbar \xi {{\hat a}^\dag }\hat a + \frac{{\hat p_\phi ^2}}{{m{R^3}}},\label{Heisenbergz1a}\\
&&\frac{{{d^2}}}{{d{t^2}}}{{\hat C}_l} + \Omega _l^2{{\hat C}_l} =   \frac{{{v_l}}}{{{M_l}}}\hat x.
\label{Heisenbergz2a}
\end{eqnarray}
The solution of Eq.~(\ref{Heisenbergz2a}) is
\begin{equation}
\begin{aligned}
{{\hat C}_l} =& {{\hat C}_l}(0)\cos {\Omega _l}t + \frac{{{{\hat P}_l}(0)}}{{{M_l}{\Omega _l}}}\sin {\Omega _l}t\\
& + {v_l}\int_0^t {\frac{{\sin {\Omega _l}(t - \tau )}}{{{M_l}{\Omega _l}}}} \hat x(\tau )d\tau . \label{omegalttau}
\end{aligned}
\end{equation}
Substituting Eq.~(\ref{omegalttau}) into Eq.~(\ref{Heisenbergz1a}) gives
\begin{equation}
\begin{aligned}
&m\left[ {\frac{{{d^2}}}{{d{t^2}}}\hat x + \omega _m^2\hat x + \int_0^t {\eta (t - \tau )} \hat x(\tau )d\tau } \right] +\sum\limits_l {\frac{{v_l^2}}{{{M_l}\Omega _l^2}}}  \hat x\\
&= \hat F(t) + \hbar \xi {{\hat a}^\dag }\hat a + \frac{{\hat p_\phi ^2}}{{m{R^3}}}, \label{omegalz1za}
\end{aligned}
\end{equation}
with $\hat F(t) =   \sum\nolimits_l {{v_l}} [{{\hat C}_l}(0)\cos {\Omega _l}t + ({{\hat P}_l}(0)/{M_l}{\Omega _l})\sin {\Omega _l}t]$. The kernel $\eta(t)$ equals $\frac{{d{\alpha}(t)}}{{dt}}$, where the correlation function ${\alpha}(t) = \sum\nolimits_l {v_l^2\cos {\Omega _l}t/(m{M_l}\Omega _l^2)}\equiv \int {I(\omega )} \cos (\omega )d\omega$ with the spectral density $I(\omega ) = \sum\nolimits_l {\frac{{v_l^2}}{{m{M_l}\Omega _l^2}}\delta (\omega  - {\Omega _l})}$. Taking expectation values (The states of each part for the system are initially prepared in their respective vacuum states) to Eq.~(\ref{omegalz1za}) leads to
\begin{equation}
\begin{aligned}
&m\left[ {\frac{{{d^2}}}{{d{t^2}}}x + \omega _m^2x + \int_0^t {\eta (t - \tau )} x(\tau )d\tau } \right] +\sum\limits_l {\frac{{v_l^2}}{{{M_l}\Omega _l^2}}}  x\\
&=  \hbar \xi {a^ * }a + \frac{{p_\phi ^2}}{{m{R^3}}}, \label{omegalz1za11}
\end{aligned}
\end{equation}
where we have used the expectation value ${{F(t)}} = \langle {{{\hat F(t)}}} \rangle $ of $\hat F(t)$ equalling zero. With the partial integration and $x(0)= 0$ (the expectation value of $\hat x(0)$ is ${{x(0)}} = \langle {{{\hat x(0)}}} \rangle $), Eq.~(\ref{omegalz1za11}) is reduced as
\begin{equation}
\begin{aligned}
m[ \ddot x + \int_0^t {\alpha (t - \tau )} \dot x(\tau )d\tau  + \omega _m^2x]  = \hbar \xi {a^ * }a + \frac{{p_\phi ^2}}{{m{R^3}}}. \label{omegalzxz1z}
\end{aligned}
\end{equation}
With the Lorentzian spectral density $I(\omega ) = {\Gamma _m}{\Lambda ^2}/[\pi ({\omega ^2} + {\Lambda ^2})]$ \cite{breuer2002,breuer1032104012009,breuer880210022016,Vega015001}, we obtain $\alpha (t) = {\Gamma _m}\Lambda {e^{ -\Lambda|t|}}$, where the parameter $\Lambda$ defines the spectral width of the bath, which is connected to the bath correlation time $T_B$ by the relation $T_B = \Lambda^{-1}$, while the time scale for the state of the system changing is given by $T_S = \Gamma _m^{-1}$. Under the Markovian approximation ($\Lambda\to \infty $), we get
\begin{equation}
\begin{aligned}
\alpha (t)\to2\Gamma _m\delta(t).
\label{omezxw1}
\end{aligned}
\end{equation}
Eq.~(\ref{Heq2}) can be obtained by substituting Eq.~(\ref{omezxw1}) into  Eq.~(\ref{omegalzxz1z}), where we have used the identity $\int_0^t {\delta (t - \tau )} \dot x(\tau )d\tau  = \frac{1}{2}\dot x(t)$ \cite{Gardiner1711022027}.


\begin{references}

\bibitem{Aspelmeyer861391} M. Aspelmeyer, T. J. Kippenberg, and F. Marquardt, Cavity optomechanics, Rev. Mod. Phys. \textbf{86}, 1391 (2014).

\bibitem{Aspelmeyer6529} M. Aspelmeyer, P. Meystre, and K. Schwab, Quantum optomechanics, Phys. Today \textbf{65}, 29 (2012).

\bibitem{Kippenberg3211172} T. J. Kippenberg and K. J. Vahala, Cavity optomechanics: Back-action at the mesoscale, Science \textbf{321}, 1172 (2008).

\bibitem{Marquardt240} F. Marquardt and S. M. Girvin, Optomechanics, Physics \textbf{2}, 40 (2009).

\bibitem{Sainadh92033824} U. S. Sainadh and M. A. Kumar, Effects of linear and quadratic dispersive couplings on optical squeezing in an optomechanical system, Phys. Rev. A \textbf{92}, 033824 (2015).

\bibitem{Metzger4321002} C. H. Metzger and K. Karrai, Cavity cooling of a microlever, Nature (London) \textbf{432}, 1002 (2004).

\bibitem{Gigan44467} S. Gigan, H. R. B\"{o}hm, M. Paternostro, F. Blaser, G. Langer, J. B. Hertzberg, K. C. Schwab, D. B\"{a}uerle, M. Aspelmeyer, and A. Zeilinger, Self-cooling of a micromirror by radiation pressure, Nature (London) \textbf{444}, 67 (2006).

\bibitem{Schliesser5509} A. Schliesser, O. Arcizet, R. Rivi{\`e}re, G. Anetsberger, and T. J. Kippenberg, Resolved-sideband cooling and position measurement of a micromechanical oscillator close to the Heisenberg uncertainty limit, Nat. Phys. \textbf{5}, 509 (2009).

\bibitem{Arcizet44471} O. Arcizet, P. F. Cohadon, T. Briant, M. Pinard, and A. Heidmann, Radiation-pressure cooling and optomechanical instability of a micromirror, Nature (London) \textbf{444}, 71 (2006).

\bibitem{Meystre525215} P. Meystre, A short walk through quantum optomechanics, Ann. Phys. (Berlin) \textbf{525}, 215 (2013).



\bibitem{Abramovici256325} A. Abramovici, W. E. Althouse, R. W. P. Drever, Y. G\"{u}rsel, S. Kawamura, F. J. Raab, D. Shoemaker, L. Sievers, R. E. Spero, K. S. Thorne, R. E. Vogt, R. Weiss, S. E. Whitcomb, and M. E. Zucker, LIGO: The laser interferometer gravitational-wave observatory, Science \textbf{256}, 325 (1992).

\bibitem{Caves4575} C. M. Caves, Quantum-mechanical radiation-pressure fluctuations in an interferometer, Phys. Rev. Lett. \textbf{45}, 75 (1980).

\bibitem{Braginsky293228} V. B. Braginsky and S. P. Vyatchanin, Low quantum noise tranquilizer for Fabry-Perot interferometer, Phys. Lett. A \textbf{293}, 228 (2002).

\bibitem{Nejad562816} A. A. Nejad, H. R. Askari, and H. R. Baghshahi, Optical bistability in coupled optomechanical cavities in the presence of Kerr effect, Appl. Opt. \textbf{56}, 2816 (2017); J. Y. Sun and H. Z. Shen, Photon blockade in non-Hermitian optomechanical systems with nonreciprocal couplings, Phys. Rev. A \textbf{107}, 043715 (2023).

\bibitem{Sarma331335} B. Sarma and A. K. Sarma, Controllable optical bistability in a hybrid optomechanical system, J. Opt. Soc. Am. B \textbf{33}, 1335 (2016).


\bibitem{Shahidani311087} S. Shahidani, M. H. Naderi, M. Soltanolkotabi, and S. Barzanjeh, Steady-state entanglement, cooling, and tristability in a nonlinear optomechanical cavity, J. Opt. Soc. Am. B \textbf{31}, 1087 (2014).

\bibitem{Jiang2213773} C. Jiang, Y. S. Cui, and K. D. Zhu, Ultrasensitive nanomechanical mass sensor using hybrid opto-electromechanical systems, Opt. Express \textbf{22}, 13773 (2014).

\bibitem{Thompson45272} J. D. Thompson, B. M. Zwickl, A. M. Jayich, F. Marquardt, S. M. Girvin, and J. G. E. Harris, Strong dispersive coupling of a high-finesse cavity to a micromechanical membrane, Nature (London) \textbf{452}, 72 (2008).

\bibitem{Bagci50781} T. Bagci, A. Simonsen, S. Schmid, L. G. Villanueva, E. Zeuthen, J. Appel, J. M. Taylor, A. S{\o}rensen, K. Usami, A. Schliesser, and E. S. Polzik, Optical detection of radio waves through a nanomechanical transducer, Nature (London) \textbf{507}, 81 (2014).

\bibitem{Andrews10321} R. W. Andrews, R. W. Peterson, T. P. Purdy, K. Cicak, R. W. Simmonds, C. A. Regal, and K. W. Lehnert, Bidirectional and efficient conversion between microwave and optical light, Nat. Phys. \textbf{10}, 321 (2014).

\bibitem{Nejad97053839} A. A. Nejad, H. R. Askari, and H. R. Baghshahi, Optomechanical detection of weak microwave signals with the assistance of a plasmonic wave, Phys. Rev. A \textbf{97}, 053839 (2018).



\bibitem{Weis3301520} S. Weis, R. Rivi\`{e}re, S. Del\'{e}glise, E. Gavartin, O. Arcizet, A. Schliesser, and T. J. Kippenberg, Optomechanically induced transparency, Science \textbf{330}, 1520 (2010).

\bibitem{Safavi47269} A. H. Safavi-Naeini, T. P. M. Alegre, J. Chan, M. Eichenfield, M. Winger, Q. Lin, J. T. Hill, D. E. Chang, and O. Painter, Electromagnetically induced transparency and slow light with optomechanics, Nature (London) \textbf{472}, 69 (2011).

\bibitem{Jia91043843} W. Z. Jia, L. F. Wei, Y. Li, and Y. X. Liu, Phase-dependent optical response properties in an optomechanical system by coherently driving the mechanical resonator, Phys. Rev. A \textbf{91}, 043843 (2015).

\bibitem{Jing59663} H. Jing, {\c{S}}. K. \"{O}zdemir, Z. Geng, J. Zhang, X.Y. L\"{u}, B. Peng, L. Yang, and F. Nori, Optomechanically-induced transparency in parity-time-symmetric microresonators, Sci. Rep. \textbf{5}, 9663 (2015).

\bibitem{Wang90023817} H. Wang, X. Gu, Y. X. Liu, A. Miranowicz, and F. Nori, Optomechanical analog of two-color electromagnetically induced transparency: Photon transmission through an optomechanical device with a two-level system, Phys. Rev. A \textbf{90}, 023817 (2014).

\bibitem{Karuza88013804} M. Karuza, C. Biancofiore, M. Bawaj, C. Molinelli, M. Galassi, R. Natali, P. Tombesi, G. Di Giuseppe, and D. Vitali, Optomechanically induced transparency in a membrane-in-the-middle setup at room temperature, Phys. Rev. A \textbf{88}, 013804 (2013).

\bibitem{Fleischhauer77633} M. Fleischhauer, A. Imamoglu, and J. P. Marangos, Electromagnetically induced transparency: Optics in coherent media, Rev. Mod. Phys. \textbf{77}, 633 (2005).

\bibitem{Agarwal81041803} G. S. Agarwal and S. M. Huang, Electromagnetically induced transparency in mechanical effects of light, Phys. Rev. A \textbf{81}, 041803(R) (2010).

\bibitem{Chang13023003} D. E. Chang, A. H. Safavi-Naeini, M. Hafezi, and O. Painter, Slowing and stopping light using an optomechanical crystal array, New J. Phys. \textbf{13}, 023003 (2011).

\bibitem{Fiore107133601} V. Fiore, Y. Yang, M. C. Kuzyk, R. Barbour, L. Tian, and H. L. Wang, Storing optical information as a mechanical excitation in a silica optomechanical resonator, Phys. Rev. Lett. \textbf{107}, 133601 (2011).

\bibitem{Zhou9179} X. Zhou, F. Hocke, A. Schliesser, A. Marx, H. Huebl, R. Gross, and T. J. Kippenberg, Slowing, advancing and switching of microwave signals using circuit nano-electromechanics, Nat. Phys. \textbf{9}, 179 (2013).

\bibitem{Hill31196} J. T. Hill, A. H. Safavi-Naeini, J. Chan, and O. Painter, Coherent optical wavelength conversion via cavity optomechanics, Nat. Commun. \textbf{3}, 1196 (2012).

\bibitem{Huang83043826} S. M. Huang and G. S. Agarwal, Electromagnetically induced transparency with quantized fields in optocavity mechanics, Phys. Rev. A \textbf{83}, 043826 (2011).

\bibitem{Xiong58050302} H. Xiong, L. G. Si, X. Y. L\"{u}, X. X. Yang, and Y. Wu, Review of cavity optomechanics in the weak-coupling regime: from linearization to intrinsic nonlinear interactions, Sci. China Phys. Mech. Astron. \textbf{58}, 1 (2015).

\bibitem{Bartolo94033841} N. Bartolo, F. Minganti, W. Casteels, and C. Ciuti, Exact steady state of a Kerr resonator with one- and two-photon driving and dissipation: Controllable Wigner-function multimodality and dissipative phase transitions, Phys. Rev. A \textbf{94}, 033841 (2016).

\bibitem{Zhou421289} Y. H. Zhou, S. S. Zhang, H. Z. Shen, and X. X. Yi, Second-order nonlinearity induced transparency, Opt. Lett. \textbf{42}, 1289 (2017).

\bibitem{Ilchenko92043903}V. S. Ilchenko, A. A. Savchenkov, A. B. Matsko, and L. Maleki, Nonlinear optics and crystalline whispering gallery mode cavities, Phys. Rev. Lett. \textbf{92}, 043903 (2004).

\bibitem{Rabl107063601} P. Rabl, Photon blockade effect in optomechanical systems, Phys. Rev. Lett. \textbf{107}, 063601 (2011).

\bibitem{Flayac96053810} H. Flayac and V. Savona, Unconventional photon blockade, Phys. Rev. A \textbf{96}, 053810 (2017).

\bibitem{Lemonde90063824} M. A. Lemonde, N. Didier, and A. A. Clerk, Antibunching and unconventional photon blockade with Gaussian squeezed states, Phys. Rev. A \textbf{90}, 063824 (2014).

\bibitem{Marino87052906} F. Marino and F. Marin, Coexisting attractors and chaotic canard explosions in a slow-fast optomechanical system, Phys. Rev. E \textbf{87}, 052906 (2013).

\bibitem{Xiong86013815} H. Xiong, L. G. Si, A. S. Zheng, X. X. Yang, and Y. Wu, Higher-order sidebands in optomechanically induced transparency, Phys. Rev. A \textbf{86}, 013815 (2012).
%%%%%
\bibitem{Bi5758} L. Bi, J. J. Hu, P. Jiang, D. H. Kim, G. F. Dionne, L. C. Kimerling, and C. A. Ross, On-chip optical isolation in monolithically integrated non-reciprocal optical resonators, Nature Photon. \textbf{5}, 758 (2011).

\bibitem{Aleahmad72129} P. Aleahmad, M. Khajavikhan, D. Christodoulides, and P. LiKamWa, Integrated multi-port circulators for unidirectional optical information transport, Sci. Rep. \textbf{7}, 2129 (2017).

\bibitem{AbdelMalek528560} F. AbdelMalek, W. Aroua, S. Haxha, and I. Flint, Light-switching-light optical transistor based on metallic nanoparticle cross-chains geometry incorporating Kerr nonlinearity, Ann. Phys. (Berlin) \textbf{528}, 560 (2016).

\bibitem{Bernier8604} N. R. Bernier, L. D. T\'{o}th, A. Koottandavida, M. A. Ioannou, D. Malz, A. Nunnenkamp, A. K. Feofanov, and T. J. Kippenberg, Nonreciprocal reconfigurable microwave optomechanical circuit, Nat. Commun. \textbf{8}, 604 (2017).

%\bibitem{Kippenberg95033901} T. J. Kippenberg, H. Rokhsari, T. Carmon, A. Scherer, and K. J. Vahala, Analysis of radiation-pressure induced mechanical oscillation of an optical microcavity, Phys. Rev. Lett. \textbf{95}, 033901 (2005).

\bibitem{H5367} H. L\"{u}, Y. J. Jiang, Y. Z. Wang, and H. Jing, Optomechanically induced transparency in a spinning resonator, Photonics Res. \textbf{5}, 367 (2017).

\bibitem{Schliesser10095015} L. Jin, J. X. Peng, Q. Z. Yuan, and X. L. Fen, Macroscopic quantum coherence in a spinning optomechanical system, Opt. Express \textbf{29}, 41191 (2021).

\bibitem{Jiang3371} B. J. Li, R. Huang, X. W. Xu, A. Miranowicz, and H. Jing,
Nonreciprocal unconventional photon blockade in a spinning optomechanical system, Photon. Res. \textbf{7}, 630 (2019).

\bibitem{Maayani558569} S. Maayani, R. Dahan, Y. Kligerman, E. Moses, A. U. Hassan, H. Jing, F. Nori, D. N. Christodoulides, and T. Carmon, Flying couplers above spinning resonators generate irreversible refraction, Nature (London) \textbf{558}, 569 (2018).

\bibitem{Shen10657} Z. Shen, Y. L. Zhang, Y. Chen, C. L. Zou, Y. F. Xiao, X. B. Zou, F. W. Sun, G. C. Guo, and C. H. Dong, Experimental realization of optomechanically induced non-reciprocity, Nat. Photonics \textbf{10}, 657 (2016).

\bibitem{Ruesink713662} F. Ruesink, M. A. Miri, A. Al\`{u}, and E. Verhagen, Nonreciprocity and magnetic-free isolation based on optomechanical interactions, Nat. Commun. \textbf{7}, 13662 (2016).

\bibitem{Fang7465} K. J. Fang, J. Luo, A. Metelmann, M. H. Matheny, F. Marquardt, A. A. Clerk, and O. Painter, Generalized non-reciprocity in an optomechanical circuit via synthetic magnetism and reservoir engineering, Nat. Phys. \textbf{13}, 465 (2017).

\bibitem{Cao118033901} Q. T. Cao, H. M. Wang, C. H. Dong, H. Jing, R. S. Liu, X. Chen, L. Ge, Q. H. Gong, and Y. F. Xiao, Experimental demonstration of spontaneous chirality in a nonlinear microresonator, Phys. Rev. Lett. \textbf{118}, 033901 (2017).

\bibitem{Li102033526} W. A. Li, G. Y. Huang, J. P. Chen, and Y. Chen, Nonreciprocal enhancement of optomechanical second-order sidebands in a spinning resonator, Phys. Rev. A \textbf{102}, 033526 (2020).

\bibitem{Jing51424} H. Jing, H. L\"{u}, {\c{S}}. K. \"{O}zdemir, T. Carmon, and F. Nori, Nanoparticle sensing with a spinning resonator, Optica \textbf{5}, 1424 (2018).

\bibitem{Chen14082005} H. J. Chen, High-resolution biomolecules mass sensing based on a spinning optomechanical system with phonon pump, Appl. Phys. Express \textbf{14}, 082005 (2021).

\bibitem{Huang121153601} R. Huang, A. Miranowicz, J. Q. Liao, F. Nori, and H. Jing, Nonreciprocal photon blockade, Phys. Rev. Lett. \textbf{121}, 153601 (2018).

\bibitem{Li7630} H. Xie, L. W. He, X. Shang, G. W. Lin, X. M. Lin, Nonreciprocal photon blockade in cavity optomagnonics, Phys. Rev. A \textbf{106}, 053707 (2022).

\bibitem{Jiang10064037} Y. Jiang, S. Maayani, T. Carmon, F. Nori, and H. Jing, Nonreciprocal phonon laser, Phys. Rev. Applied \textbf{10}, 064037 (2018).

\bibitem{Peng5332000405} R. Peng, C. S. Zhao, Z. Yang, B. Xiong, and L. Zhou, Nonreciprocal amplification in coupled-rotating cavities around exceptional points, Ann. Phys. (Berlin) \textbf{533}, 2000405 (2021).

\bibitem{Zhang207594} H. L. Zhang, R. Huang, S. D. Zhang, Y. Li, C. W. Qiu, F. Nori, and H. Jing, Breaking anti-PT symmetry by spinning a resonator, Nano Lett. \textbf{20}, 7594 (2020).

\bibitem{Li103053522} B. J. Li, \c{S}. K. \"{O}zdemir, X. W. Xu, L. Zhang, L. M. Kuang, and H. Jing, Nonreciprocal optical solitons in a spinning Kerr resonator, Phys. Rev. A \textbf{103}, 053522 (2021).
%%%%%%%%%%%%%%%%%%%opa

\bibitem{Otey93033835} S. Pina-Otey, F. Jim\'{e}nez, P. Degenfeld-Schonburg, and C. Navarrete-Benlloch, Classical and quantum-linearized descriptions of degenerate optomechanical parametric oscillators, Phys. Rev. A \textbf{93}, 033835 (2016).

\bibitem{Coillet9828} G. P. Lin, A. Coillet, and Y. K. Chembo, Nonlinear photonics with high-Q whispering-gallery-mode resonators, Adv. Opt. Photon. \textbf{9}, 828 (2017).

\bibitem{Hu100043824} C. S. Hu, L. T. Shen, Z. B. Yang, H. Z. Wu, Y. Li, and S. B. Zheng, Manifestation of classical nonlinear dynamics in optomechanical entanglement with a parametric amplifier, Phys. Rev. A \textbf{100}, 043824 (2019).

\bibitem{Adamyan92053818} H. H. Adamyan, J. A. Bergou, N. T. Gevorgyan, and G. Y. Kryuchkyan, Strong squeezing in periodically modulated optical parametric oscillators, Phys. Rev. A \textbf{92}, 053818 (2015).

\bibitem{Lu114093602} X. Y. L\"{u}, Y. Wu, J. R. Johansson, H. Jing, J. Zhang, and F. Nori, Squeezed optomechanics with phase-matched amplification and dissipation, Phys. Rev. Lett. \textbf{114}, 093602 (2015).

\bibitem{Mi67115} X. W. Mi, J. X. Bai and S. Ke-hui, Robust entanglement between a movable mirror and a cavity field system with an optical parametric amplifier, Eur. Phys. J. D \textbf{67}, 115 (2013).

\bibitem{Hu7124} C. S. Hu, X. R. Huang, L. T. Shen, Z. B. Yang, and H. Z. Wu, Enhancement of entanglement in distant micromechanical mirrors using parametric interactions, Eur. Phys. J. D \textbf{71}, 24 (2017).

\bibitem{Xuereb86013809} A. Xuereb, M. Barbieri, and M. Paternostro, Multipartite optomechanical entanglement from competing nonlinearities, Phys. Rev. A \textbf{86}, 013809 (2012).
%
\bibitem{Agarwal93043844} G. S. Agarwal and S. M. Huang, Strong mechanical squeezing and its detection, Phys. Rev. A \textbf{93}, 043844 (2016).

\bibitem{Huang79013821} S. M. Huang and G. S. Agarwal, Enhancement of cavity cooling of a micromechanical mirror using parametric interactions, Phys. Rev. A \textbf{79}, 013821 (2009).

\bibitem{Huang80033807} S. M. Huang and G. S. Agarwal, Normal-mode splitting in a coupled system of a nanomechanical oscillator and a parametric amplifier cavity, Phys. Rev. A \textbf{80}, 033807 (2009).

\bibitem{Sarma96053827} B. Sarma and A. K. Sarma, Quantum-interference-assisted photon blockade in a cavity via parametric interactions, Phys. Rev. A \textbf{96}, 053827 (2017).

\bibitem{Shen98023856} H. Z. Shen, C. Shang, Y. H. Zhou, and X. X. Yi, Unconventional single-photon blockade in non-Markovian systems, Phys. Rev. A \textbf{98}, 023856 (2018).

\bibitem{Shen101013826} H. Z. Shen, Q. Wang, J. Wang, and X. X. Yi, Nonreciprocal unconventional photon blockade in a driven dissipative cavity with parametric amplification, Phys. Rev. A \textbf{101}, 013826 (2020).


\bibitem{Qin120093601} W. Qin, A. Miranowicz, P. B. Li, X. Y. L\"{u}, J. Q. You, and F. Nori, Exponentially enhanced light-matter interaction, cooperativities, and steady-state entanglement using parametric amplification, Phys. Rev. Lett. \textbf{120}, 093601 (2018).

\bibitem{Lemonde111053602} M. A. Lemonde, N. Didier, and A. A. Clerk, Nonlinear interaction effects in a strongly driven optomechanical cavity, Phys. Rev. Lett. \textbf{111}, 053602 (2013).

\bibitem{Liu111083601} Y. C. Liu, Y. F. Xiao, Y. L. Chen, X. C. Yu, and Q. H. Gong, Parametric down-conversion and polariton pair generation in optomechanical systems, Phys. Rev. Lett. \textbf{111}, 083601 (2013).

\bibitem{Mikkelsen96043832} M. Mikkelsen, T. Fogarty, J. Twamley, and T. Busch, Optomechanics with a position-modulated Kerr-type nonlinear coupling, Phys. Rev. A \textbf{96}, 043832 (2017).

\bibitem{Ferretti85033303} T. S. Yin, X. Y. L\"{u}, L. L. Zheng, M. Wang, S. Li, and Y. Wu, Nonlinear effects in modulated quantum optomechanics, Phys. Rev. A \textbf{95}, 053861 (2017).

\bibitem{Suzuki92033823} H. Suzuki, E. Brown, and R. Sterling, Nonlinear dynamics of an optomechanical system with a coherent mechanical pump: Second-order sideband generation, Phys. Rev. A \textbf{92}, 033823 (2015).

\bibitem{Kronwald111133601} A. Kronwald and F. Marquardt, Optomechanically induced transparency in the nonlinear quantum regime, Phys. Rev. Lett. \textbf{111}, 133601 (2013).

\bibitem{Jiao18083034} Y. Jiao, H. L\"{u}, J. Qian, Y. Li, and H. Jing, Nonlinear optomechanics with gain and loss: amplifying higher-order sideband and group delay, New J. Phys. \textbf{18}, 083034 (2016).


\bibitem{Liu717637} S. P. Liu, W. X. Yang, T. Shui, Z. H. Zhu, and A. X. Chen, Tunable two-phonon higher-order sideband amplification in a quadratically coupled optomechanical system, Sci. Rep. \textbf{7}, 17637 (2017).

\bibitem{Fan65850} L. R. Fan, K. Y. Fong, M. Poot, and H. X. Tang, Cascaded optical transparency in multimode-cavity optomechanical systems, Nat. Commun. \textbf{6}, 5850 (2015).

\bibitem{Xiong423630} H. Xiong, Z. X. Liu, and Y. Wu, Highly sensitive optical sensor for precision measurement of electrical charges based on optomechanically induced difference-sideband generation, Opt. Lett. \textbf{42}, 3630 (2017).

\bibitem{Kong95033820} C. Kong, H. Xiong, and Y. Wu, Coulomb-interaction-dependent effect of high-order sideband generation in an optomechanical system, Phys. Rev. A \textbf{95}, 033820 (2017).

\bibitem{Cohen520522} J. D. Cohen, S. M. Meenehan, G. S. MacCabe, S. Gr\"{o}blacher, A. H. Safavi-Naeini, F. Marsili, M. D. Shaw, and O. Painter, Phonon counting and intensity interferometry of a nanomechanical resonator, Nature (London) \textbf{520}, 522 (2015).

\bibitem{Nunnenkamp111053603} K. B{\o}rkje, A. Nunnenkamp, J. D. Teufel, and S. M. Girvin, Signatures of nonlinear cavity optomechanics in the weak coupling regime, Phys. Rev. Lett. \textbf{111}, 053603 (2013).

\bibitem{Zhao63224211} W. Zhao, S. D. Zhang, A. Miranowicz, and H. Jing, Weak-force sensing with squeezed optomechanics, Sci. China Phys. Mech. Astron. \textbf{63}, 224211 (2020).

\bibitem{Li116} Y. Li and K. D. Zhu, High-order sideband optical properties of a DNA-quantum dot hybrid system [Invited], Photon. Res. \textbf{1}, 16 (2013).

\bibitem{Liu712521} Z. C. Zhang, Y. P. Wang, and X. G. Wang, PT-symmetry-breaking-enhanced cavity optomechanical magnetometry, Phys. Rev. A \textbf{102}, 023512 (2020); Z. X. Liu, B. Wang, C. Kong, L. G. Si, H. Xiong, and Y. Wu, A proposed method to measure weak magnetic field based on a hybrid optomechanical system, Sci. Rep. \textbf{7}, 12521 (2017).

\bibitem{Liu99033822} S. P. Liu, B. Liu, J. F. Wang, T. T. Sun, and W. X. Yang, Realization of a highly sensitive mass sensor in a quadratically coupled optomechanical system, Phys. Rev. A \textbf{99}, 033822 (2019).

\bibitem{Wang106803908} B. Wang, Z. X. Liu, H. Xiong, and Y. Wu, Highly sensitive mass sensing by means of the optomechanical nonlinearity, IEEE Photonics J. \textbf{10}, 6803908 (2018).

\bibitem{Liu439} S. P. Liu, W. X. Yang, Z. H. Zhu, T. Shui, and L. Li, Quadrature squeezing of a higher-order sideband spectrum in cavity optomechanics, Opt. Lett. \textbf{43}, 9 (2018).

\bibitem{Boyd3261074} R. W. Boyd and D. J. Gauthier, Controlling the velocity of light pulses, Science \textbf{326}, 1074 (2009).

\bibitem{Jiao97013843} Y. F. Jiao, T. X. Lu, and H. Jing, Optomechanical second-order sidebands and group delays in a Kerr resonator, Phys. Rev. A \textbf{97}, 013843 (2018).
%%%%%%%%%%qunyanchi
\bibitem{He351649} Q. He, F. Badshah, R. U. Din, H. Y. Zhang, Y. Hu, and G. Q. Ge, Optomechanically induced transparency and the long-lived slow light in a nonlinear system, J. Opt. Soc. Am. B \textbf{35}, 1649 (2018).

\bibitem{Li635090} L. Li, W. J. Nie, and A. X. Chen, Transparency and tunable slow and fast light in a nonlinear optomechanical cavity, Sci. Rep. \textbf{6}, 35090 (2016).

\bibitem{Mirza2725515} I. M. Mirza, W. C. Ge, and H. Jing, Optical nonreciprocity and slow light in coupled spinning optomechanical resonators, Opt. Express \textbf{27}, 25515 (2019).

\bibitem{Liao11698} Q. H. Liao, W. D. Bao, X. Xiao, W. J. Nie, and Y. C. Liu, Optomechanically induced transparency and slow-fast light effect in hybrid cavity optomechanical systems, Crystals, \textbf{11}, 698 (2021).

%
\bibitem{Zimmer92253201} F. Zimmer and M. Fleischhauer, Sagnac interferometry based on ultraslow polaritons in cold atomic vapors, Phys. Rev. Lett. \textbf{92}, 253201 (2004).

\bibitem{Shahriar75053807} M. S. Shahriar, G. S. Pati, R. Tripathi, V. Gopal, M. Messall, and K. Salit, Ultrahigh enhancement in absolute and relative rotation sensing using fast and slow light, Phys. Rev. A \textbf{75}, 053807 (2007).

%%%%%%%%%%%%%%%%%%%%%%%%%%%%55555markf
\bibitem{breuer2002} H. P. Breuer and F. Petruccione, \emph{The Theory of Open Quantum Systems}, (Oxford University Press, Oxford, 2002).

\bibitem{Weiss2008} D. F. Walls and G. J. Milburn, \textit{Quantum Optics} (Springer, Berlin, 1994).

\bibitem{Chang052105}  K. W. Chang and C. K. Law, Non-Markovian master equation for a damped oscillator with time-varying parameters, Phys. Rev. A \textbf{81}, 052105 (2010).

\bibitem{Tan032102}   H. T. Tan and W. M. Zhang, Non-Markovian dynamics of an open quantum system with initial system-reservoir correlations: A nanocavity coupled to a coupled-resonator optical waveguide, Phys. Rev. A \textbf{83}, 032102 (2011).

\bibitem{Longhi063826}   S. Longhi, Non-Markovian decay and lasing condition in an optical microcavity coupled to a structured reservoir, Phys. Rev. A \textbf{74}, 063826 (2006).



\bibitem{Leggett5911987}   A. J. Leggett, S. Chakravarty, A. T. Dorsey, M. P. A. Fisher, A. Garg, and W. Zwerger, Dynamics of the dissipative two-state system, Rev. Mod. Phys. \textbf{59}, 1 (1987).

%%%%%%%%%%%%%%%%%%%%%%%%%%%%%%%%%%%%%%%%%%%%%%%%%%%%%%%%%%%%%%%%%%%%%%%%%%%%%%%%%%







\bibitem{breuer1032104012009} H. P. Breuer, E. M. Laine, and J. Piilo, Measure for the Degree of Non-Markovian Behavior of Quantum Processes in Open Systems, Phys. Rev. Lett. \textbf{103}, 210401 (2009).
%https://doi.org/10.1103/PhysRevLett.103.210401

\bibitem{laine810621152010} E. M. Laine, J. Piilo, and H. P. Breuer, Measure for the non-Markovianity of quantum processes, Phys. Rev. A \textbf{81}, 062115 (2010).
%https://doi.org/10.1103/PhysRevA.81.062115

\bibitem{addis900521032014} C. Addis, B. Bylicka, D. Chru\'{s}ci\'{n}ski, and S. Maniscalco, Comparative study of non-Markovianity measures in exactly solvable one- and two-qubit models, Phys. Rev. A \textbf{90}, 052103 (2014).
%https://doi.org/10.1103/PhysRevA.90.052103

\bibitem{wibmann860621082012} S. Wi{\ss}mann, A. Karlsson, E. M. Laine, J. Piilo, and H. P. Breuer, Optimal state pairs for non-Markovian quantum dynamics, Phys. Rev. A \textbf{86}, 062108 (2012).
%https://doi.org/10.1103/PhysRevA.86.062108

\bibitem{wibmann920421082015} S. Wi{\ss}mann, H. P. Breuer, and B. Vacchini, Generalized trace-distance measure connecting quantum and classical non-Markovianity, Phys. Rev. A \textbf{92}, 042108 (2015); L. Xin, S. Xu, X. X. Yi, and H. Z. Shen, Tunable nonMarkovian dynamics with a three-level atom mediated by the classical laser in a semi-infinite photonic waveguide, Phys.
Rev. A \textbf{105}, 053706 (2022).
%https://doi.org/10.1103/PhysRevA.92.042108

\bibitem{shen960338052017} H. Z. Shen, D. X. Li, S. L. Su, Y. H. Zhou, and X. X. Yi, Exact non-Markovian dynamics of qubits coupled to two interacting environments, Phys. Rev. A \textbf{96}, 033805 (2017); H. Z. Shen, Y. Chen, T. Z. Luan, and X. X. Yi, Multiple single-photon generations in three-level atoms coupled to a cavity with non-Markovian effects, Phys. Rev. A \textbf{107}, 053705 (2023).
%https://doi.org/10.1103/PhysRevA.96.033805


\bibitem{lorenzo880201022013} S. Lorenzo, F. Plastina, and M. Paternostro, Geometrical characterization of non-Markovianity, Phys. Rev. A \textbf{88}, 020102(R) (2013).
%https://doi.org/10.1103/PhysRevA.88.020102

\bibitem{rivas1050504032010} \'{A}. Rivas, S. F. Huelga, and M. B. Plenio, Entanglement and Non-Markovianity of Quantum Evolutions, Phys. Rev. Lett. \textbf{105}, 050403 (2010); H. Z. Shen, Q. Wang, and X. X. Yi, Dispersive readout with non-Markovian environments, Phys. Rev. A \textbf{105}, 023707
(2022).
%https://doi.org/10.1103/PhysRevLett.105.050403

\bibitem{luo860441012012} S. L. Luo, S. S. Fu, and H. T. Song, Quantifying non-Markovianity via correlations, Phys. Rev. A \textbf{86}, 044101 (2012).
%https://doi.org/10.1103/PhysRevA.86.044101

\bibitem{wolf1011504022008} M. M. Wolf, J. Eisert, T. S. Cubitt, and J. I. Cirac, Assessing Non-Markovian Quantum Dynamics, Phys. Rev. Lett. \textbf{101}, 150402 (2008).
%https://doi.org/10.1103/PhysRevLett.101.150402

\bibitem{lu820421032010} X. M. Lu, X. G. Wang, and C. P. Sun, Quantum Fisher information flow and non-Markovian processes of open systems, Phys. Rev. A \textbf{82}, 042103 (2010).
%https://doi.org/10.1103/PhysRevA.82.042103

\bibitem{chruscinski1121204042014} D. Chru\'{s}ci\'{n}ski and S. Maniscalco, Degree of Non-Markovianity of Quantum Evolution, Phys. Rev. Lett. \textbf{112}, 120404 (2014).
%https://doi.org/10.1103/PhysRevLett.112.120404

%%%%%%%%%guangji
\bibitem{Zhang063853} H. Z. Shen, S. Xu, H. T. Cui, and X. X. Yi, Non-Markovian dynamics of a system of two-level atoms coupled to a structured environment, Phys. Rev. A \textbf{99}, 032101 (2019); W. Z. Zhang, J. Cheng, W. D. Li, and L. Zhou, Optomechanical cooling in the non-Markovian regime, Phys. Rev. A \textbf{93}, 063853 (2016).

\bibitem{Zhang19083022} W. Z. Zhang, Y. Han, B. Xiong, and L. Zhou, Optomechanical force sensor in a non-Markovian regime, New J. Phys. \textbf{19}, 083022 (2017).

\bibitem{Xiong436053} B. Xiong, X. Li, S. L. Chao, and L. Zhou, Optomechanical quadrature squeezing in the non-Markovian regime, Opt. Lett. \textbf{43}, 6053 (2018).

\bibitem{Zhao2729082} H. Z. Shen, S. L. Su, Y. H. Zhou, and X. X. Yi, Non-Markovian quantum Brownian motion in one dimension in electric fields,
Phys. Rev. A \textbf{97}, 042121 (2018); X. Y. Zhao, Macroscopic entanglement in optomechanical system induced by non-Markovian environment, Opt. Express \textbf{27}, 29082 (2019).
%https://opg.optica.org/abstract.cfm?uri=oe-27-20-29082
\bibitem{Triana116183602} J. F. Triana, A. F. Estrada, and L. A. Pach\'{o}n, Ultrafast optimal sideband cooling under non-Markovian evolution, Phys. Rev. Lett. \textbf{116}, 183602 (2016).

\bibitem{Cheng430385} J. Cheng, X. T. Liang, W. Z. Zhang, and X. M. Duan, Optomechanical state transfer in the presence of non-Markovian environments, Opt. Commun. \textbf{430}, 385 (2019).

\bibitem{Cheng623678} H. Z. Shen, S. Xu, S. Yi, and X. X. Yi, Controllable dissipation of a qubit coupled to an engineering reservoir, Phys. Rev. A \textbf{98}, 062106 (2018); J. Cheng, W. Z. Zhang, L. Zhou, and W. P. Zhang, Preservation macroscopic entanglement of optomechanical systems in non-Markovian environment, Sci. Rep. \textbf{6}, 23678 (2016).

\bibitem{Mu46270} Q. X. Mu, H. Li, X. Huang, and X. Y. Zhao, Microscopic-macroscopic entanglement transfer in optomechanical system: Non-Markovian effects, Opt. Commun. \textbf{462}, 70 (2018).

\bibitem{Li361363} X. Li, B. Xiong, S. L. Chao, and L. Zhou, Improving the sensitivity of weak microwave signal detection with optomechanical system under non-Markovian regime, J. Opt. Soc. Am. B \textbf{36}, 1363 (2019).

\bibitem{Ding111814} Q. Z. Ding, P. Zhao, Y. H. Ma, and Y. S. Chen, Impact of the central frequency of environment on non-Markovian dynamics in piezoelectric optomechanical devices, Sci. Rep. \textbf{11}, 1814 (2021).


%%%%%%%%%%%%%%%%555555


\bibitem{Sinha124043603} K. Sinha, P. Meystre, E. A. Goldschmidt, F. K. Fatemi, S. L. Rolston, and P. Solano, Non-Markovian collective emission from macroscopically separated emitters, Phys. Rev. Lett. \textbf{124}, 043603 (2020).

\bibitem{Wu103010601} W. Wu, S. Y. Bai, and J. H. An, Non-Markovian sensing of a quantum reservoir, Phys. Rev. A \textbf{103}, L010601 (2021).

\bibitem{Mu94012334} H. Z. Shen, X. Q. Shao, G. C. Wang, X. L. Zhao, and X. X.
Yi, Quantum phase transition in a coupled two-level system
embedded in anisotropic three-dimensional photonic crystals,
Phys. Rev. E \textbf{93}, 012107 (2016); Q. X. Mu, X. Y. Zhao, and T. Yu, Memory-effect-induced macroscopic-microscopic entanglement, Phys. Rev. A \textbf{94}, 012334 (2016).


%%%%%%%%%%%%%%55
\bibitem{caruso8612032014} F. Caruso, V. Giovannetti, C. Lupo, and S. Mancini, Quantum channels and memory effects, Rev. Mod. Phys. \textbf{86}, 1203 (2014).

\bibitem{darrigo3502112014} A. D'Arrigo, R. Lo Franco, G. Benenti, E. Paladino, and G. Falci, Recovering entanglement by local operations, Ann. Phys. (NY) \textbf{350}, 211 (2014).

\bibitem{lofranco900543042014} R. Lo Franco, A. D'Arrigo, G. Falci, G. Compagno, and E. Paladino, Preserving entanglement and nonlocality in solid-state qubits by dynamical decoupling, Phys. Rev. B \textbf{90}, 054304 (2014).
%https://doi.org/10.1103/PhysRevB.90.054304

\bibitem{bylicka457202014} B. Bylicka, D. Chru\'{s}ci\'{n}ski, and S. Maniscalco, Non-Markovianity and reservoir memory of quantum channels: a quantum information theory perspective, Sci. Rep. \textbf{4}, 5720 (2014).
%https://doi.org/10.1038/srep05720

\bibitem{xue860523042012} S. B. Xue, R. B. Wu, W. M. Zhang, J. Zhang, C. W. Li, and T. J. Tarn, Decoherence suppression via non-Markovian coherent feedback control, Phys. Rev. A \textbf{86}, 052304 (2012).
%https://doi.org/10.1103/PhysRevA.86.052304

\bibitem{Liu20117}  B. H. Liu, L. Li, Y. F. Huang, C. F. Li, G. C. Guo, E. M. Laine, H. P. Breuer, and J. Piilo, Experimental control of the transition from Markovian to non-Markovian dynamics of open quantum systems, Nat. Phys. \textbf{7}, 931 (2011).

\bibitem{Groblacher20156} S. Gr{\"o}blacher, A. Trubarov, N. Prigge, G. D. Cole, M. Aspelmeyer, and J. Eisert, Observation of non-Markovian micromechanical Brownian motion, Nat. Commun. \textbf{6}, 7606 (2015).

\bibitem{Xiong2019100} S. J. Xiong, Q. W. Hu, Z. Sun, L. Yu, Q. P. Su, J. M. Liu, and C. P. Yang, Non-Markovianity in experimentally simulated quantum channels: Role of counterrotating-wave terms, Phys. Rev. A \textbf{100}, 032101 (2019).

\bibitem{Cialdi2019100} S. Cialdi, C. Benedetti, D. Tamascelli, S. Olivares, M. G. A. Paris, and B. Vacchini, Experimental investigation of the effect of classical noise on quantum non-Markovian dynamics, Phys. Rev. A \textbf{100}, 052104 (2019).

\bibitem{Khurana201999} D. Khurana, B. K. Agarwalla, and T. S. Mahesh, Experimental emulation of quantum non-Markovian dynamics and coherence protection in the presence of information backflow, Phys. Rev. A \textbf{99}, 022107 (2019).

\bibitem{Madsen2011106} K. H. Madsen, S. Ates, T. Lund-Hansen, A. L{\"o}ffler, S. Reitzenstein, A. Forchel, and P. Lodahl, Observation of Non-Markovian Dynamics of a Single Quantum Dot in a Micropillar Cavity, Phys. Rev. Lett. \textbf{106}, 233601 (2011).

\bibitem{Guo2021126} Y. Guo, P. Taranto, B. H. Liu, X. M. Hu, Y. F. Huang, C. F. Li, and G. C. Guo, Experimental Demonstration of Instrument-Specific Quantum Memory Effects and Non-Markovian Process Recovery for Common-Cause Processes, Phys. Rev. Lett. \textbf{126}, 230401 (2021).

\bibitem{Li2022129} B. W. Li, Q. X. Mei, Y. K. Wu, M. L. Cai, Y. Wang, L. Yao, Z. C. Zhou, and L. M. Duan, Observation of Non-Markovian Spin Dynamics in a Jaynes-Cummings-Hubbard Model Using a Trapped-Ion Quantum Simulator, Phys. Rev. Lett. \textbf{129}, 140501 (2022).

\bibitem{Hoeppe2012108} U. Hoeppe, C. Wolff, J. K{\"u}chenmeister, J. Niegemann, M. Drescher, H. Benner, and K. Busch, Direct Observation of Non-Markovian Radiation Dynamics in 3D Bulk Photonic Crystals, Phys. Rev. Lett. \textbf{108}, 043603 (2012).

\bibitem{Xu201082} J. S. Xu, C. F. Li, C. J. Zhang, X. Y. Xu, Y. S. Zhang, and G. C. Guo, Experimental investigation of the non-Markovian dynamics of classical and quantum correlations, Phys. Rev. A \textbf{82}, 042328 (2010).

\bibitem{Tang201297} J. S. Tang, C. F. Li, Y. L. Li, X. B. Zou, G. C. Guo, H. P. Breuer, E. M. Laine, and J. Piilo, Measuring non-Markovianity of processes with controllable system-environment interaction, Europhys. Lett. \textbf{97}, 10002 (2012).

\bibitem{Uriri2020101} S. A. Uriri, F. Wudarski, I. Sinayskiy, F. Petruccione, and M. S.Tame, Experimental investigation of Markovian and non-Markovian channel addition, Phys. Rev. A \textbf{101}, 052107 (2020).

\bibitem{Anderson199347} M. H. Anderson, G. Vemuri, J. Cooper, P. Zoller, and S. J. Smith, Experimental study of absorption and gain by two-level atoms in a time-delayed non-Markovian optical field, Phys. Rev. A \textbf{47}, 3202 (1993).

\bibitem{Liu2020102} Z. D. Liu, Y. N. Sun, B. H. Liu, C. F. Li, G. C. Guo, S. Hamedani Raja, H. Lyyra, and J. Piilo, Experimental realization of high-fidelity teleportation via a non-Markovian open quantum system, Phys. Rev. A \textbf{102}, 062208 (2020).

\bibitem{breuer880210022016} H. P. Breuer, E. M. Laine, J. Piilo, and B. Vacchini, \textit{Colloquium}: Non-Markovian dynamics in open quantum systems, Rev. Mod. Phys. \textbf{88}, 021002 (2016).
%https://doi.org/10.1103/RevModPhys.88.021002

%5



\bibitem{Vega015001}   I. de Vega and D. Alonso, Dynamics of non-Markovian open quantum systems, Rev. Mod. Phys. \textbf{89}, 015001 (2017).


%%%%%%%%%%%%%%%%%%%%%%%%%%%%%%%%%%%%model

\bibitem{Gerry} C. C. Gerry and P. L. Knight, \textit{Introductory Quantum Optics} (Cambridge University Press, Cambridge, 2005).

\bibitem{Li100023838} W. A. Li and G. Y. Huang, Enhancement of optomechanically induced sum sidebands using parametric interactions, Phys. Rev. A \textbf{100}, 023838 (2019).

\bibitem{Nation841} P. D. Nation, J. R. Johansson, M. P. Blencowe, and F. Nori, \textit{Colloquium}: Stimulating uncertainty: Amplifying the quantum vacuum with superconducting circuits, Rev. Mod. Phys. \textbf{84}, 1 (2012).

\bibitem{Leghtas347853} Z. Leghtas, S. Touzard, I. M. Pop, A. Kou, B. Vlastakis, A. Petrenko, K. M. Sliwa, A. Narla, S. Shankar, M. J. Hatridge, M. Reagor, L. Frunzio, R. J. Schoelkopf, M. Mirrahimi, and M. H. Devoret, Confining the state of light to a quantum manifold by engineered two-photon loss, Science \textbf{347}, 853 (2015).

\bibitem{Clerk821155} A. A. Clerk, M. H. Devoret, S. M. Girvin, F. Marquardt, and R. J. Schoelkopf, Introduction to quantum noise, measurement, and amplification, Rev. Mod. Phys. \textbf{82}, 1155 (2010).

\bibitem{Shen100023814} S. T. Shen, Y. Qu, J. H. Li, and Y. Wu, Tunable photon statistics in parametrically amplified photonic molecules, Phys. Rev. A \textbf{100}, 023814 (2019).
%
\bibitem{Post39475} E. J. Post, Sagnac effect, Rev. Mod. Phys. \textbf{39}, 475 (1967).

\bibitem{Malykin431229} G. B. Malykin, The Sagnac effect: Correct and incorrect explanations, Phys. Usp. \textbf{43}, 1229 (2000).

\bibitem{Huang95023844} S. M. Huang and G. S. Agarwal, Robust force sensing for a free particle in a dissipative optomechanical system with a parametric amplifier, Phys. Rev. A \textbf{95}, 023844 (2017).

\bibitem{Scully} M. O. Scully and M. S. Zubairy, \textit{Quantum Optics} (Cambridge University Press, Cambridge, 1997).

\bibitem{Davuluri11264002} S. Davuluri and S. Y. Zhu, Controlling optomechanically induced transparency through rotation, Europhys. Lett. \textbf{112}, 64002 (2015).



\bibitem{Shahidani88053813} S. Shahidani, M. H. Naderi, and M. Soltanolkotabi, Control and manipulation of electromagnetically induced transparency in a nonlinear optomechanical system with two movable mirrors, Phys. Rev. A \textbf{88}, 053813 (2013).
%
\bibitem{Adiyatullin81329} A. F. Adiyatullin, M. D. Anderson, H. Flayac, M. T. Portella-Oberli, F. Jabeen, C. Ouellet-Plamondon, G. C. Sallen, and B. Deveaud, Periodic squeezing in a polariton Josephson junction, Nat. Commun. \textbf{8}, 1329 (2017).

\bibitem{Teufel471204} J. D. Teufel, D. Li, M. S. Allman, K. Cicak, A. J. Sirois, J. D. Whittaker, and R. W. Simmonds, Circuit cavity electromechanics in the strong-coupling regime, Nature \textbf{471}, 204 (2011).

\bibitem{Lu1}  T. X. Lu, X. Xiao, L. S. Chen, Q. Zhang, and H. Jing, Magnon-squeezing-enhanced slow light and second-order sideband in cavity magnomechanics, Phys. Rev. A \textbf{107}, 063714 (2023).

\bibitem{Lu2}  S. N. Huai, Y. L. Liu, J. Zhang, L. Yang, and Y. X. Liu, Enhanced sideband responses in a PT-symmetric-like cavity magnomechanical system, Phys. Rev. A \textbf{99}, 043803 (2019).

\bibitem{Milonni} P. W. Milonni, \textit{Fast light, slow light and lefthanded light}, (Institute of Physics Publishing, Bristol, 2005).

\bibitem{Zhang1321} X. Y. Zhang, Q. T. Cao, Z. Wang, Y. X. Liu, C. W. Qiu, L. Yang, Q. H. Gong, and Y. F. Xiao, Symmetry-breaking-induced nonlinear optics at a microcavity surface, Nat. Photonics \textbf{13}, 21 (2019).

\bibitem{Righini34435} G. C. Righini, Y. Dumeige, P. F\'{e}ron, M. Ferrari, G. N. Conti, D. Ristic, and S. Soria, Whispering gallery mode microresonators: Fundamentals and applications, Riv. Nuovo Cimento \textbf{34}, 435 (2011).






%%%%%%%%%%%%%%%%%%%%%%%%%%%%%%%%%%%%%%%%%%%%%%%%%%%%%%%%%%%%%%%%%%%%%%%%%%%%%%%%%%%%%%%%%%%%%%%%

%%%%%%%%%%%%%%%%%%%%%%%%%%%%%%%%%%%%%%%%%%%%%%%%%%%%%%%%%%%%%%%%%%%%%%%%%%%%%%%%%%%%%%%%%%%%%%%%

\bibitem{shen880338352013} H. Z. Shen, M. Qin, and X. X. Yi, Single-photon storing in coupled non-Markovian atom-cavity system, Phys. Rev. A \textbf{88}, 033835 (2013).

\bibitem{zhang870321172013} J. Zhang, Y. X. Liu, R. B. Wu, K. Jacobs, and F. Nori, Non-Markovian quantum input-output networks, Phys. Rev. A \textbf{87}, 032117 (2013).
%https://doi.org/10.1103/PhysRevA.87.032117


\bibitem{diosi850341012012} L. Di\'{o}si, Non-Markovian open quantum systems: Input-output fields, memory, and monitoring, Phys. Rev. A \textbf{85}, 034101 (2012).

%%%%%%%%%%%%%%%%%%%%%% %effective spectral density of environment



%https://doi.org/10.1103/PhysRevA.81.052103

\bibitem{xiong860321072012} H. N. Xiong, W. M. Zhang, M. W. Y. Tu, and D. Braun, Dynamically stabilized decoherence-free states in non-Markovian open fermionic systems, Phys. Rev. A \textbf{86}, 032107 (2012).
%https://doi.org/10.1103/PhysRevA.86.032107


%https://doi.org/10.1103/PhysRevA.88.033835

\bibitem{shen950121562017} H. Z. Shen, D. X. Li, and X. X. Yi, Non-Markovian linear response theory for quantum open systems and its applications, Phys. Rev. E \textbf{95}, 012156 (2017); H. Z. Shen, M. Qin, X. Q. Shao,
and X. X. Yi, General response formula and application to
topological insulator in quantum open system, Phys. Rev. E
\textbf{92}, 052122 (2015).
%https://doi.org/10.1103/PhysRevE.95.012156

%\bibitem{haikka810521032010} P. Haikka and S. Maniscalco, Non-Markovian dynamics of a damped driven two-state system, Phys. Rev. A \textbf{81}, 052103 (2010).

%%%%%Gaussian Ornstein-Uhlenbeck process
\bibitem{uhlenbeck368231930} G. E. Uhlenbeck and L. S. Ornstein, On the theory of the brownian motion, Phys. Rev. \textbf{36}, 823 (1930).
%https://doi.org/10.1103/PhysRev.36.823

\bibitem{gillespie5420841996} D. T. Gillespie, Exact numerical simulation of the Ornstein-Uhlenbeck process and its integral, Phys. Rev. E \textbf{54}, 2084 (1996).
%https://doi.org/10.1103/PhysRevE.54.2084

\bibitem{jing1052404032010} J. Jing and T. Yu, Non-Markovian relaxation of a three-level system: Quantum trajectory approach, Phys. Rev. Lett. \textbf{105}, 240403 (2010).
%https://doi.org/10.1103/PhysRevLett.105.240403





\bibitem{zz2} A. Majumdar and D. Gerace, Single-photon blockade in
doubly resonant nanocavities with second-order nonlinearity,
Phys. Rev. B \textbf{87}, 235319 (2013).

\bibitem{zz1} H. Z. Shen, Y. H. Zhou, and X. X. Yi, Quantum optical diode with semiconductor microcavities, Phys. Rev. A \textbf{90}, 023849 (2014); Y. H. Zhou, H. Z. Shen, and X. X. Yi, Unconventional photon
blockade with second-order nonlinearity, Phys. Rev. A \textbf{92},
023838 (2015).


\bibitem{zz3}S. Ferretti and D. Gerace, Single-photon nonlinear optics with Kerr-type nanostructured materials, Phys. Rev. B \textbf{85}, 033303 (2012).

\bibitem{zz4} H. Z. Shen, Y. H. Zhou, and X. X. Yi, Tunable photon blockade
in coupled semiconductor cavities, Phys. Rev. A \textbf{91}, 063808
(2015).


\bibitem{jack630438032001zs}J. C. Sankey, C. Yang, B. M. Zwickl, A. M. Jayich, and J. G. E. Harris, Strong and tunable nonlinear optomechanical coupling in a low-loss system, Nat. Phys. \textbf{6}, 707 (2010).


\bibitem{jack630438032001zzz} M. Bhattacharya, H. Uys, and P. Meystre, Optomechanical trapping and cooling of partially reflective mirrors, Phys. Rev. A \textbf{77}, 033819 (2008).


\bibitem{jack630438032001z1s} J. Q. Liao and F. Nori, Photon blockade in quadratically coupled optomechanical systems, Phys. Rev. A \textbf{88}, 023853 (2013).


\bibitem{jack630438032001zs3} H. K. Cheung and C. K. Law, Nonadiabatic optomechanical Hamiltonian of a moving dielectric membrane in a cavity, Phys. Rev. A \textbf{84}, 023812 (2011).

%\bibitem{Hovhannisyan98045101} K. V. Hovhannisyan and L. A. Correa, Measuring the temperature of cold many-body quantum systems, Phys. Rev. B \textbf{98}, 045101 (2018).

%\bibitem{Correa96062103} L. A. Correa, M. Perarnau-Llobet, K. V. Hovhannisyan, S. Hern\'{a}ndez-Santana, M. Mehboudi, and A. Sanpera, Enhancement of low-temperature thermometry by strong coupling, Phys. Rev. A \textbf{96}, 062103 (2017).

%\bibitem{Halliwell532012} J. J. Halliwell and T. Yu, Alternative derivation of the Hu-Paz-Zhang master equation of quantum Brownian motion, Phys. Rev. B \textbf{53}, 2012 (1996).

%\bibitem{Boyanovsky96062108} D. Boyanovsky and D. Jasnow, Heisenberg-Langevin versus quantum master equation, Phys. Rev. A \textbf{96}, 062108 (2017).

%\bibitem{Ferialdi96012109} L. Ferialdi and A. Smirne, Momentum coupling in non-Markovian quantum Brownian motion, Phys. Rev. A \textbf{96}, 012109 (2017).


\bibitem{Weiss1999} U. Weiss, \textit{Quantum Dissipative Systems}, 2nd ed. (World Scientific, Singapore, 1999).

\bibitem{Caldeira1983347} A. O. Caldeira and A. J. Leggett, Quantum tunnelling in a dissipative system, Ann. Phys. (NY) \textbf{149}, 374 (1983).

\bibitem{Spiechowicz052107} J. Spiechowicz, P. Bialas, and J. {\L}uczka, Quantum partition of energy for a free Brownian particle: Impact of dissipation, Phys. Rev. A \textbf{98}, 052107 (2018).

\bibitem{Einsiedler0222282020} S. Einsiedler, A. Ketterer, and H. P. Breuer, Non-Markovianity of quantum Brownian motion, Phys. Rev. A \textbf{102}, 022228 (2020).

\bibitem{Sinha051111} S. Sinha and P. A. Sreeram, Nonperturbative approach to quantum Brownian motion, E \textbf{79}, 051111 (2009).

\bibitem{Sun5118451995} C. P. Sun and L. H. Yu, Exact dynamics of a quantum dissipative system in a constant external field, Phys. Rev. A \textbf{51}, 1845 (1995).


%    [26] G. W. Ford and M. Kac, On the quantum langevin equation, J. Stat. Phys. \textbf{46}, 803 (1987).

%[27] P. De Smedt, D. D\"{u}rr, and J. L. Lebowitz, Quantum system in contact with a thermal environment: Rigorous treatment of a simple model, Commun. Math. Phys. \textbf{120}, 195 (1988).

\bibitem{Grabert1151988} H. Grabert, P. Schramm, and G. L. Ingold, Quantum Brownian motion: The functional integral approach, Phys. Rep. \textbf{168}, 115 (1988).

\bibitem{Caldeira5871983} A. O. Caldeira and A. J. Leggett, Path integral approach to quantum Brownian motion, Phys. A \textbf{121}, 587 (1983).




\bibitem{Gardiner1711022027} C. W. Gardiner and P. Zoller, \textit{Quantum Noise} (Springer,
Berlin, 2000).

\end{references}
\end{document}